\documentclass[pra,reprint,superscriptaddress,nofootinbib]{revtex4-1}
\usepackage{rotating}
\usepackage{floatpag}
\rotfloatpagestyle{empty}
\usepackage{amsthm}
\usepackage{amsmath}
\usepackage{amssymb}
\usepackage{graphics}
\usepackage{graphicx}
\usepackage{color}
\usepackage{hyperref}

\definecolor{ao(english)}{rgb}{0.0, 0.5, 0.0}
\definecolor{applegreen}{rgb}{0.55, 0.71, 0.0}
\definecolor{cadetblue}{rgb}{0.37, 0.62, 0.63}
\definecolor{cadet}{rgb}{0.33, 0.41, 0.47}
\definecolor{byzantine}{rgb}{0.74, 0.2, 0.64}
\definecolor{orange}{rgb}{1.0, 0.5, 0.0}
\definecolor{magenta}{rgb}{1.0, 0.0, 1.0}

\renewcommand{\i}{\mathrm{i}}

\newcommand{\cmag}[1]{{\color{magenta}{#1}}}

\newcommand{\TW}[1]{{\cmag{#1}}}

\begin{document}
\title{Formation of Bose--Einstein condensates}

\author{Matthew J. Davis}
\email{mdavis@physics.uq.edu.au}
\affiliation{School of Mathematics and Physics, The University of Queensland, St Lucia QLD 4072, Australia}
\affiliation{JILA, 440 UCB, University of Colorado, Boulder, Colorado 80309, USA}

\author{Tod M. Wright}
\affiliation{School of Mathematics and Physics, The University of Queensland, St Lucia QLD 4072, Australia}

\author{Thomas Gasenzer}
\affiliation{Kirchhoff-Institut f\"ur Physik, Universit\"at Heidelberg, Im Neuenheimer Feld 227, 69120 Heidelberg, Germany}

\author{Simon A. Gardiner}
\affiliation{Joint Quantum Centre (JQC) Durham-Newcastle, Department of Physics, Durham University, Durham DH1 3LE, United Kingdom}

\author{Nick P. Proukakis}
\affiliation{Joint Quantum Centre (JQC) Durham-Newcastle, School of Mathematics and Statistics, Newcastle University, Newcastle upon Tyne NE1 7RU, United Kingdom}

\begin{abstract}
The problem of understanding how a coherent, macroscopic Bose--Einstein condensate (BEC) emerges from the cooling of a thermal Bose gas has attracted significant theoretical and experimental interest over several decades. 
The pioneering achievement of BEC in weakly-interacting dilute atomic gases in 1995 was followed by a number of experimental studies examining the growth of the BEC number, as well as the development of its coherence.  
More recently there has been interest in connecting such experiments to universal aspects of nonequilibrium phase transitions, in terms of both static and dynamical critical exponents.
Here, the spontaneous formation of topological structures such as vortices and solitons in quenched cold-atom experiments has enabled the verification of the Kibble--Zurek mechanism predicting the density of topological defects in continuous phase transitions, first proposed in the context of the evolution of the early universe. 
This chapter reviews progress in the understanding of BEC formation, and discusses open questions and future research directions in the dynamics of phase transitions in quantum gases.
\end{abstract}

\maketitle
\section{Introduction}
The equilibrium phase diagram of the dilute Bose gas exhibits a continuous phase transition between condensed and noncondensed phases.  The order parameter characteristic of the condensed phase vanishes above some critical temperature $T_\mathrm{c}$ and grows continuously with decreasing temperature below this critical point.  
However, the dynamical process of condensate formation has proved to be a challenging phenomenon to address both theoretically and experimentally. This formation process is a crucial aspect of Bose systems and of direct relevance to all condensates discussed in this book, despite their evident system-specific properties. Important questions leading to  intense discussions in the early literature include the timescale for condensate formation, and the role of inhomogeneities and finite-size effects in 
``closed'' systems.  These issues are related to the concept of spontaneous symmetry breaking, its causes, and implications for physical systems (see, for example, the chapter by Snoke and Daley in this volume). 

In this chapter we give an overview of the dynamics of condensate formation and describe the present understanding provided by increasingly well controlled cold-atom experiments and corresponding theoretical advances over the past twenty years.  We focus on the growth of BECs in cooled Bose gases, which, from a theoretical standpoint, requires a suitable nonequilibrium formalism. A recent book provides a more complete introduction to a number of different theoretical approaches to the description of nonequilibrium and non-zero-temperature quantum gases~\cite{FINESS_book}. 

We note that the past decade has seen the observation of BEC in a number of diverse experimental systems beyond ultracold atoms, including exciton-polaritons, magnons, and phonons, which are covered in other chapters of this volume.  Many of the universal aspects of condensate formation  also apply to these systems.

\section{The physics of BEC formation}
\label{sec:Davis:BECFormation}
The essential character of the excitations and collective response of a condensed Bose gas is well described by perturbative approaches that take as their starting point the breaking of the $U(1)$ gauge symmetry of the Bose quantum field.  This approach can be extended further to provide a kinetic description of excitations in a condensed gas weakly perturbed away from equilibrium~\cite{griffin_nikuni_book_09}.  The description of the process of formation of a Bose-Einstein condensate in a closed system begins, however, in the opposite regime of kinetics of a non-condensed gas.  Over the past decades, there have been many studies using methods of kinetic theory to investigate the initiation of Bose--Einstein condensation.  It is now well established that these descriptions break down near the critical point, and in particular in any situation in which the formation process is far from adiabatic. 
A number of different theoretical methodologies have been applied to the issue of condensate formation, but most have converged to a similar description of the essential physics.  The prevailing view is  that a classical non-linear wave description ---  a form of  Gross--Pitaevskii equation --- can describe the nonequilibrium dynamics of the condensation process, which involves in general aspects of weak-wave turbulence and, in more aggressive cooling scenarios, strong turbulence.  The classical field describes the highly occupied modes of the gas at finite temperature and out of equilibrium.

A summary of the consensus picture of condensate formation in a Bose gas cooled from above the critical temperature is as follows.  Well above the critical point the coherences between particles in distinct eigenstates of the appropriate single-particle Hamiltonian are negligible and the system is well described by a quantum Boltzmann kinetic equation for the occupation numbers of these single-particle modes.  As cooling of the gas proceeds due to inter-particle collisions and interactions with an external bath, if one is present, the occupation numbers of lower-energy modes increase.  Once phase correlations between these modes become significant, the system is best described in terms of an emergent quasiclassical field, which may in general exhibit large phase fluctuations, topological structures and turbulent dynamics, the nature of which may vary over time and depend on the specific details of the system --- including its dimensionality, density, and strength of interactions.  
This regime is sometimes referred to as a nonequilibrium \emph{quasicondensate}, in analogy to the phase-fluctuating equilibrium regimes of low-dimensional Bose systems~\cite{Popov1972,popov_book_83}.  The eventual relaxation of this quasicondensate establishes phase coherence across the sample, producing the state that we routinely call a Bose--Einstein condensate.

\subsection{The pre-condensation kinetic  regime}
\label{Pre-condensationKinetic}
Early investigations of the kinetics of condensation of a gas of massive bosons began with studies of such a system coupled to a thermal bath with infinite heat capacity, consisting of phonons \cite{Inoue1976a} or fermions \cite{levich_yakhot_77,levich_yakhot_78a,levich_yakhot_78b}.  These works inherited ideas from earlier studies of condensation of photons in cosmological scenarios~\cite{Zeldovich1968a}.  In a homogeneous system, condensation is signified by a delta-function singularity of the momentum distribution at zero momentum (see, e.g., Ref.~\cite{pitaevskii_stringari_book_03}).  Levich and Yakhot found~\cite{levich_yakhot_77} that an initially non-degenerate equilibrium ideal Bose gas brought in contact with a bath with a temperature below 
$T_\mathrm{c}$ would develop such a singularity at zero momentum only in the limit of an infinite evolution time (see also Ref.~\cite{Tikhodeev1990a}).  These same authors subsequently found that the introduction of collisions between the bosons lead to the ``explosive'' development of a singular peak at zero momentum after a finite evolution time~\cite{levich_yakhot_78a,levich_yakhot_78b}.  They were careful to point out, however, the approximations involved in their treatment of interactions, and indeed that the development of such coherence invalidates the assumptions underlying the quantum Boltzmann description, conjecturing that ``the system in the course of phase transition passes through a stage which may be identified as a period of strong turbulence'' \cite{levich_yakhot_78a}. 

Experimental attempts in the 1980s to achieve Bose condensation of spin-polarized hydrogen (see the chapter by Greytak and Kleppner for an overview and recent developments), and excitons in semiconductors such as Cu$_2$O, inspired renewed theoretical interest in Bose-gas kinetics.  Eckern developed a kinetic theory~\cite{eckern_84} for Hartree--Fock--Bogoliubov quasiparticles appropriate to the relaxation of the system on the condensed side of the transition.  Snoke and Wolfe revisited the question of the kinetics of approach to the condensation transition by undertaking numerical calculations of the quantum Boltzmann equation~\cite{snoke_wolfe_89}.  They found in particular that the bosonic enhancement of scattering rates in the degenerate regime offset the increased number of scattering events required for rethermalisation in this regime, such that re-equilibration of a shock-cooled thermal distribution takes place on the order of three to four kinetic collision times, $\tau_\mathrm{kin}=(\rho\sigma v_{T})^{-1}$, where $\rho$ is the particle density, $\sigma$ is the collisional cross section, and the mean thermal velocity 
$v_{T} = (3 k_\mathrm{B} T / m)^{1/2}$.
These results imply that a Boltzmann-equation description of this early kinetic regime is valid even for short-lived particles such as excitons, as the particle lifetime is long compared to this equilibration timescale.

Over time a comprehensive picture of the process of condensation of a quench-cooled gas has emerged, and comprises three distinct stages of nonequilibrium dynamics:
a kinetic redistribution of population towards lower energy modes in the non-condensed phase, development of an instability that leads to nucleation of the condensate and a subsequent build-up of coherence, and finally condensate growth and phase ordering. 
In the midst of increasingly intensive efforts to achieve Bose condensation in dilute atomic gases, by then including the new system of alkali-metal vapours, these stages were analysed in more detail in the early 1990s, beginning with a series of papers by Stoof \cite{stoof_91,stoof_92, stoof_95,stoof_97,stoof_99}, and by Svistunov, Kagan, and Shlyapnikov \cite{svistunov_91,kagan_svistunov_92,kagan_svistunov_94,kagan_95}.

In Ref.~\cite{svistunov_91}, Svistunov discussed condensate formation in a weakly interacting, dilute Bose gas, with so-called gas parameter $\zeta= \rho^{1/3} a\ll1$, where $a$ is the scattering length.  In a closed system, a cooling quench generically leads to a particle distribution which, below some energy scale $\varepsilon_{0}$,  exceeds the equilibrium occupation number corresponding to the total energy and particle content.
Energy and momentum conservation then imply that a few particles scattered to high-momentum modes carry away a large fraction of the excess energy associated with this over-occupation, allowing the momentum of a majority of the particles to decrease.  
Should the characteristic energy scale of the overpopulated regime be sufficiently small, 
$\varepsilon_{0}\ll\hbar^{2}\rho^{2/3}/m\sim k_\mathrm{B}T_\mathrm{c}$, mode-occupation numbers in this regime will be much larger than unity, and the subsequent particle transport in momentum space towards lower energies is described by the quantum Boltzmann equation in the 
classical-wave limit~\cite{svistunov_91,kagan_svistunov_92,kagan_svistunov_94}.
This is valid for modes with energies above the scale set by the chemical potential 
$\mu=g\rho\sim\zeta k_\mathrm{B} T_\mathrm{c}$ of the ultimate equilibrium state, where $g=4\pi \hbar^{2}a/m$ is the interaction constant for particles of mass $m$.  At lower energies, the phase correlations between momentum modes become significant, and a description beyond the quantum Boltzmann equation is required.  

We note that for open systems such as exciton-polariton condensates,  the quasi-coherent dynamics of such low-energy modes will in general be  sensitive to the driving and dissipation corresponding to the continual  decay and replenishment of the bosons.
Such external coupling can dramatically alter the behaviour of the system, and its effects on condensate formation dynamics are a subject of current research --- 
see, e.g., Refs.~\cite{sieberer_huber_13,altman_sieberer_15,dagvadorj_fellows_14} and the chapter by Altman {\em et al.}.  Hereafter, unless otherwise specified, the theoretical developments we discuss pertain to closed systems in which the bosons undergoing condensation are conserved in number during the formation process.

By assuming the scattering matrix elements in the wave Boltzmann equation to be independent of the mode energies, Svistunov discussed several different transport scenarios within the framework of weak-wave turbulence, in analogy to similar processes underlying Langmuir-wave turbulence in plasmas \cite{Zakharov1992a}.
He concluded that the initial kinetic transport stage of the condensation process evolves as a weakly non-local particle wave in momentum space.  
Specifically, he proposed that the particle-flux wave followed the self-similar form $n(\varepsilon,t)\sim\varepsilon_{1}(t)^{-7/6}f(\varepsilon/\varepsilon_{1}(t))$, with $\varepsilon_{1}(t)\sim(t-t_{*})^{3}$, and scaling function $f$ falling off as $f(x)\propto x^{-\alpha}$ for $x\gg1$, with $\alpha=7/6$.
Following the arrival of this wave at time $t_*\simeq t_0 +\hbar\varepsilon_0/\mu^2$, a quasi-stationary wave-turbulent cascade forms in which particles are transported locally, from momentum shell to momentum shell, from the scale $\varepsilon_{0}$ of the energy concentration in the initial state to the low-energy regime $\varepsilon\lesssim\mu$ where coherence formation sets in.

The wave-kinetic (or weak-wave turbulence) stage of condensate formation following a cooling quench was investigated in more detail by Semikoz and Tkachev \cite{Semikoz1995a.PhysRevLett.74.3093,Semikoz1997a}\TW{,} who solved the wave Boltzmann equation numerically and found results consistent with the above scenario, albeit with a slightly shifted power-law exponent $\alpha\simeq1.24$ for the wave-turbulence spectrum.  Later dynamical classical-field simulations of the condensation formation process by Berloff and Svistunov \cite{berloff_svistunov_02} further corroborated the above picture.

\subsection{The formation of coherence}
It has been known for some time that a kinetic Boltzmann equation model is unable to describe the development of a macroscopic zero-momentum occupation in the absence of seeding or other modifications ~\cite{levich_yakhot_77,snoke_wolfe_89,svistunov_91}.  In any event, the quantum Boltzmann equation ceases to be valid in the high-density, low-energy regime in which condensation occurs.
The two-body scattering receives significant many-body corrections once the interaction energy $ g\int_{k\lesssim p}\mathrm{d}\mathbf{k}\,n_{k}$ of particles with momenta below a given scale $p$ exceeds the kinetic energy at that scale, and these are indeed the prevailing conditions when phase coherence emerges and the condensate begins to grow~\cite{svistunov_91}.

In a series of papers~\cite{stoof_91,stoof_92,stoof_97,stoof_99}, Stoof took  account of these many-body corrections and developed a theory of condensate nucleation
 resting on kinetic equations incorporating a ladder-resummed many-body $T$-matrix determined from a one-particle-irreducible (1PI) effective action or free-energy functional. 
In the 1PI formalism the propagators appearing in the effective action are taken as fixed, determined in this case by the initial thermal Bose number 
distribution and the spectral properties of a free gas.

Constructed within the Schwinger--Keldysh closed-time-path framework, the method allows the determination of the time evolution of the self-energy and thus of an effective chemical potential for the 
zero-momentum mode through the phase transition. 
During the kinetic stage, once the system has reached temperatures 
below the interaction-renormalized critical temperature, the self-energy renders the 
vacuum state of the zero-momentum mode metastable.
Stoof found that this modification of the self-energy occurs on a time scale 
$\sim\hbar/k_\mathrm{B}T_\mathrm{c}$ and gives rise to a small seed population in the zero mode, $n_{0}\sim\zeta^{2}\rho$, within the kinetic time scale 
$\tau_\mathrm{kin}\sim \hbar/(\zeta^{2}k_\mathrm{B}T_\mathrm{c})$.	
He argued that, following this seeding, the system undergoes an unstable semi-classical evolution of the low-energy modes. 
Taking interactions between quasiparticles into account he found that the 
squared dispersion $\omega(\mathbf{p})^{2}$ becomes negative for $p\lesssim\hbar\sqrt{an_{0}(t)}$, i.e., below a momentum scale of the order of the inverse healing length associated with the density $n_{0}(t)$ of the existing condensed fraction.
As a result, the condensate grows linearly in time
over the kinetic time scale $\tau_\mathrm{kin}$.
The growth process eventually ceases due to the conservation of total particle number, whereafter the final kinetic equilibration of quasi-particles takes place over a time scale 
$\sim\hbar/(\zeta^{3}k_\mathrm{B}T_\mathrm{c})$ as discussed 
previously by Eckern \cite{eckern_84}, and by Semikoz and Tkachev  \cite{Semikoz1997a}.

\subsection{Turbulent condensation}
The semi-classical scenario 
of Stoof is built on the assumptions that the cooling quench has driven the system to the critical point in a quasi-adiabatic fashion, and that the neglect of thermal fluctuations and nonequilibrium over-occupations in the self-energy is justified \cite{stoof_99}.  
However, as previously pointed out in Ref.~\cite{levich_yakhot_78a}, a more vigorous quench may drive the system into an intermediate stage of strong turbulence, where the coherences between wave frequencies lead to the formation of coherent structures, such as vortices, that have a significant influence on the subsequent dynamics.
The main processes and scales governing this stage were discussed in detail by Kagan and Svistunov~\cite{kagan_svistunov_94,kagan_95}. 
As a result of excess particles being transported kinetically into the coherent regime (wave numbers below the final inverse healing length, $k\lesssim\xi^{-1}\sim\sqrt{a\rho}$),  the density and phase of the Bose field fluctuate strongly  on length scales 
shorter than $\xi$.
The growing population at even smaller wave numbers then implies, according to Refs.~\cite{svistunov_91,kagan_svistunov_92,kagan_svistunov_94}, the formation of a \emph{quasicondensate} over the respective length scales, as the coherent evolution of the field according to the Gross--Pitaevskii equation causes the density fluctuations to strongly decrease at the expense of phase fluctuations.
This short-range phase-ordering occurs on a time scale 
$\tau_\mathrm{c}\sim\hbar/\mu\sim\hbar/(\zeta k_\mathrm{B}T_\mathrm{c})$.  Depending on the flux of excess particles entering the coherent regime, this leads to quasicondensate formation over a minimum length scale $l_\mathrm{v}>\xi$ (see Sects.~\ref{sec:Davis:KibbleZurek} and \ref{sec:Davis:NTFP}) \cite{Nowak:2012gd}.
The phase, however, remains strongly fluctuating on larger length scales due to the formation of topological defects --- vortex lines and rings.
These vortices appear in the form of clumps of strongly tangled filaments \cite{Schwarz1988a} with an average distance between filaments of order $l_\mathrm{v}$.
If the cooling quench is sufficiently strong to drive the system near a non-thermal fixed point, cf.~Sect.~\ref{sec:Davis:NTFP}, this quasicondensate is characterised by new universal scaling laws in space and time.

The work of Kagan and Svistunov  laid the foundations for studying the role of superfluid turbulence in the process of Bose--Einstein condensation.
Kozik and Svistunov have subsequently elucidated the decay of the vortex tangle 
via the transport of Kelvin waves created on the vortex filaments through their reconnections, which 
can itself assume a wave-turbulent structure \cite{Kozik2004a.PhysRevLett.92.035301,Kozik2005a.PhysRevLett.94.025301,Kozik2005a.PhysRevB.72.172505,Kozik2009a}.


\section{Condensate formation experiments}
\subsection{Growth of condensate number}

We now provide a historical overview of both experiments and theory related to condensate formation in ultracold atomic gases. 
The first experiments to achieve Bose--Einstein condensation in 1995~\cite{anderson_ensher_95,davis_mewes_95} reached the phase space density necessary for quantum degeneracy using the technique of  evaporative cooling~\cite{ketterle_vandruten_96} --- the steady removal of the most energetic atoms, followed by rethermalisation to a lower temperature via atomic collisions.  These experiments, which concentrated on the BEC atom number as the conceptionally simplest observable, provided an indication of the time scale for condensation in trapped atomic gases, in the range of milliseconds to seconds.  This gave the impetus for the development of
a quantum kinetic theory  by Gardiner and Zoller using the techniques of open quantum systems.  They first considered the homogeneous Bose gas~\cite{gardiner_zoller_97a}, before extending the formalism to trapped gases~\cite{gardiner_zoller_98, gardiner_zoller_00}. Their methodology split the system into a ``condensate band'', containing modes significantly affected by the presence of a BEC, and a ``non-condensate band'' containing all other levels.  A master equation was derived for the condensate band using standard techniques~\cite{gardiner_zoller_book_04}, yielding equations of motion for the occupations of the condensate mode and the low-lying excited states contained in the condensate band.   A simple BEC growth equation derived from this approach  provided a reasonable first estimate of the time of formation for the   $^{87}$Rb and $^{23}$Na BECs of the JILA~\cite{anderson_ensher_95} and MIT~\cite{davis_mewes_95} groups, respectively.

The first experiment to explicitly study the formation dynamics of a BEC in a dilute weakly interacting gas was performed by the Ketterle group at MIT, using their newly developed technique of non-destructive imaging~\cite{miesner_stamper-kurn_98}.  Beginning with an equilibrium gas just above the critical temperature, they performed a sudden evaporative cooling ``quench'' by removing all atoms above a certain energy.  The subsequent evolution led to the formation of a condensate, with a characteristic S-shaped curve for the growth in condensate number.  This was interpreted as 
evidence of bosonic stimulation in the growth process, and they fitted the simple BEC growth equation of Ref.~\cite{gardiner_zoller_97b} to their experimental observations.  However, the measured growth rates did not fit the theory all that well.  

Gardiner and co-workers subsequently developed an expanded rate-equa\-tion approach incorporating the dynamics of a number of quasiparticle levels~\cite{gardiner_lee_98,lee_gardiner_00}.  This formalism predicted faster growth rates, mostly due to the enhancement of collision rates by bosonic stimulation, but still failed to agree with the experimental data.  One limitation of this approach was that it neglected the evaporative cooling dynamics of the thermal cloud, instead treating it as being in a supersaturated thermal equilibrium.

The details of the evaporative cooling were simulated in two closely related works by Davis \emph{et al.}~\cite{davis_gardiner_00} and Bijlsma \emph{et al.}~\cite{bijlsma_zaremba_00}.  The former was based on the quantum kinetic theory of Gardiner and Zoller, while the latter
emerged as a limit of the 
field-theoretical approach of Stoof \cite{stoof_97,stoof_99} and
the ``ZNG'' formalism previously developed for nonequilibrium trapped Bose gases~\cite{zaremba_nikuni_99} by Zaremba, Nikuni, and Griffin.
The latter authors used a broken-symmetry approach to derive a quantum Boltzmann equation for non-condensed atoms coupled to a Gross--Pitaevskii equation for the condensate~\cite{zaremba_griffin_98,zaremba_nikuni_99}, thereby extending their two-fluid model for trapped BECs~\cite{zaremba_griffin_98}, which was based on the pioneering work of Kirkpatrick and Dorfman~\cite{kirkpatrick_dorfman_83,kirkpatrick_dorfman_85a,kirkpatrick_dorfman_85b,kirkpatrick_dorfman_85c}.  
The ZNG methodology has since been used successfully and extensively  to study a variety of nonequilibrium phenomena in partially condensed 
Bose gases, such as the temperature dependence of collective excitations, as reviewed in Ref.~\cite{griffin_nikuni_book_09}.  As 
this methodology is explicitly based on 
symmetry breaking, it cannot address the initial seeding of a BEC, or any critical physics arising from fluctuations. However, it can model continued growth once a BEC is present.

The works of Davis~\emph{et al.}~\cite{davis_gardiner_00} and Bijlsma \emph{et al.}~\cite{bijlsma_zaremba_00} both introduced approximations to the formalisms they were built on, assuming that the condensate grew adiabatically in its ground state, and 
treating all non-condensed atoms in a Boltzmann-like approach.  Both papers boiled down to simulating the quantum Boltzmann equation in the ergodic approximation, in which the 
phase-space distribution depends on the phase-space variables only through the energy~\cite{luiten_reynolds_96}. Despite the different approaches, the calculations were in excellent agreement with one another --- yet still quantitatively disagreed with the MIT experimental data~\cite{miesner_stamper-kurn_98}.  This disagreement has remained unexplained.

\begin{figure*}
\centering
\includegraphics[height=6cm]{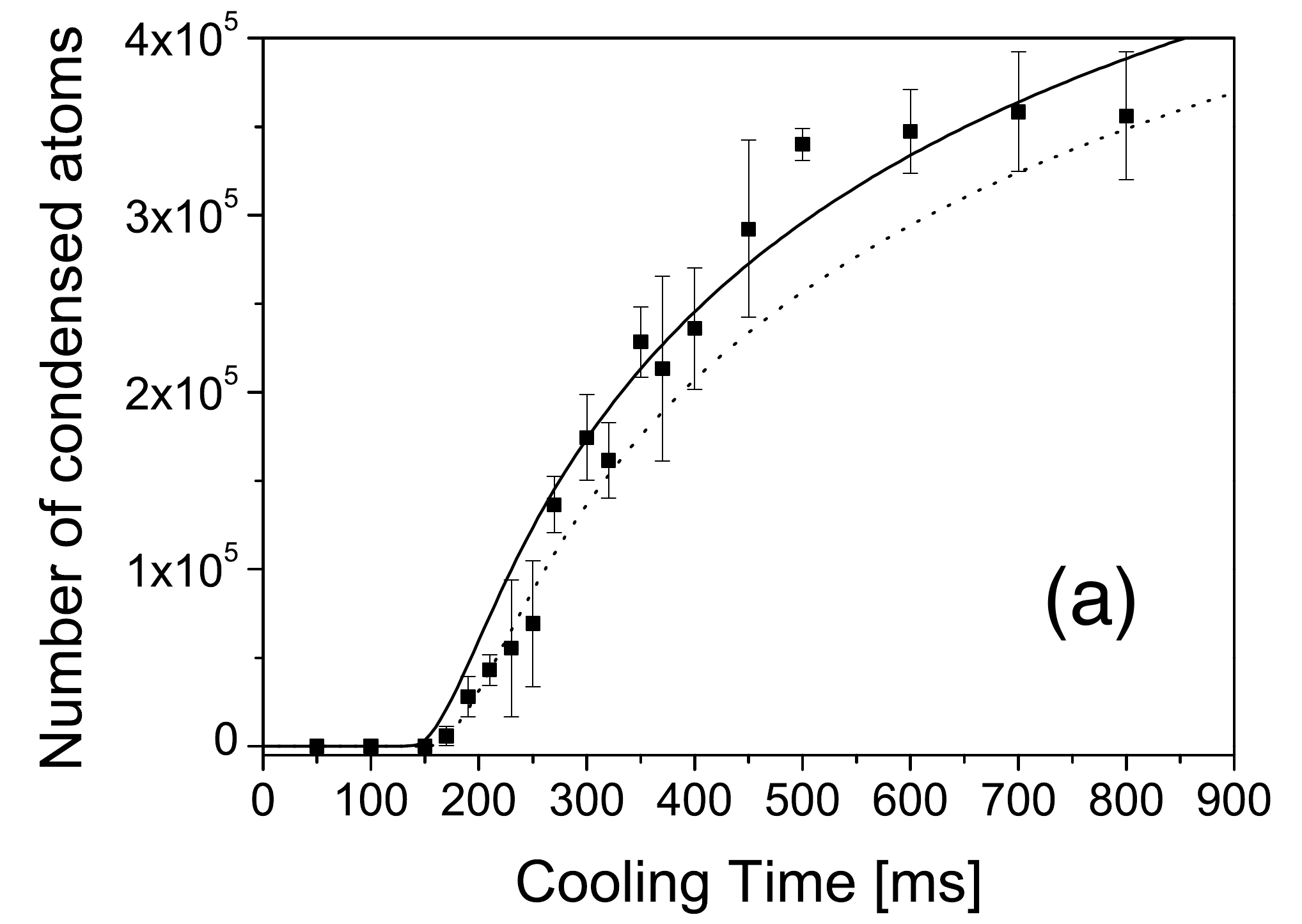}
\includegraphics[height=6cm]{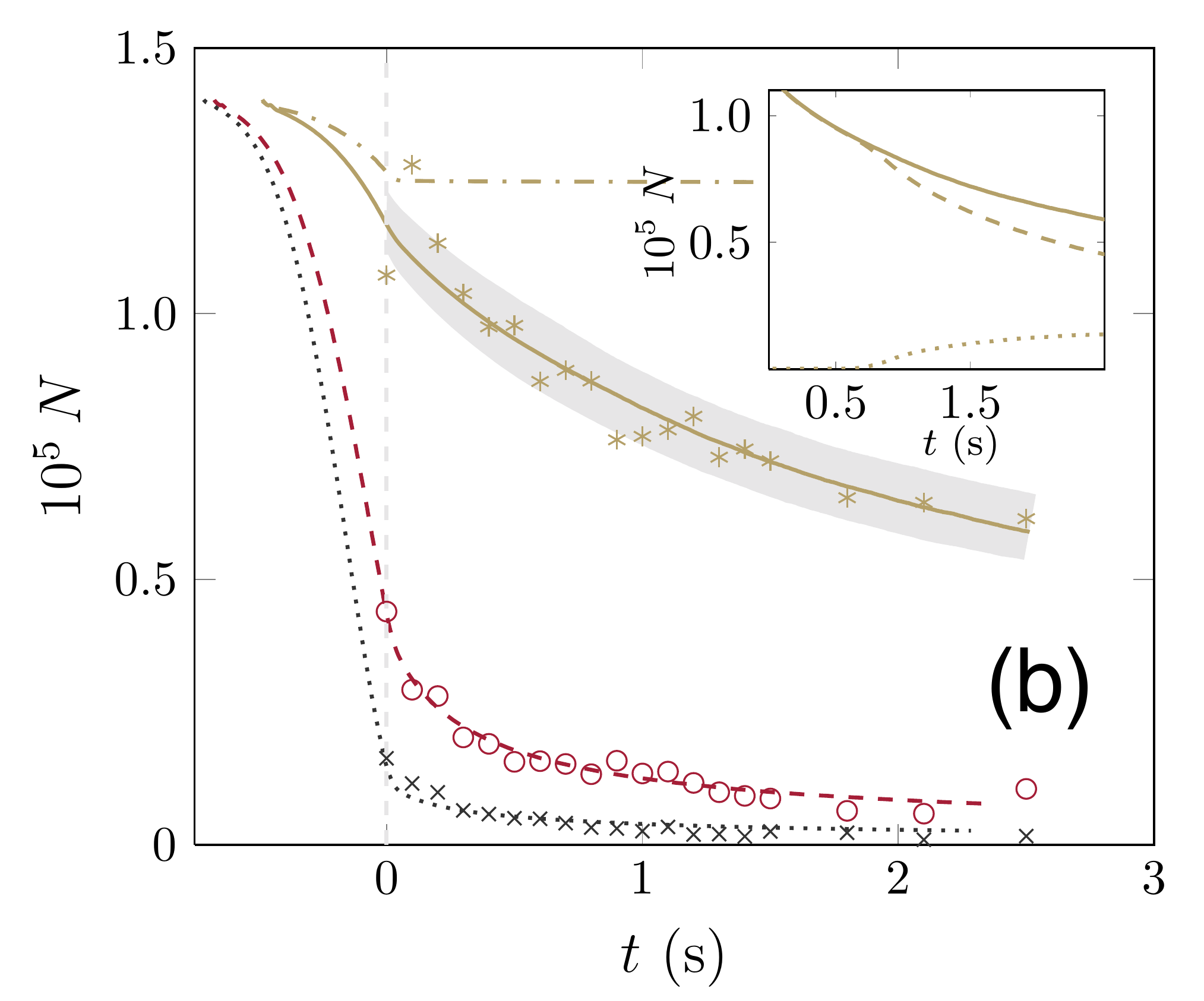}
\caption{(a) Growth of an atomic Bose--Einstein condensate modelled with the quantum Boltzmann equation.  The experiment began with a $^{87}$Rb Bose gas in an elongated harmonic trap with $N_{\i} = (4.2\pm0.2) \times 10^6$ atoms at a temperature of $T_\mathrm{i} = (640 \pm 30)$ nK, before turning on rf evaporative cooling with a truncation energy of 
$1.4 k_\mathrm{B} T$.  The solid and dotted lines show the theoretical calculations with a starting number of $N_\i = 4.2\times10^6$ and $N_\i = 4.4\times10^6$ atoms respectively.  Taken from Ref.~\cite{kohl_davis_02}. (b) Surface evaporation leading to the formation of a BEC, showing the total atom number for three different cloud-surface distances.  The lines are the results of ZNG simulations, the points are from experiment.  The dot-dash gold line is for a ZNG simulation neglecting collisions in the thermal cloud, demonstrating that modelling the full dynamics of the thermal cloud is necessary for a quantitative understanding of the experiment.  The inset shows the total number, thermal cloud number, and condensate number, from top to bottom respectively, as a function of time. Taken from Ref.~\cite{markle_allen_15}.}
\label{fig:Davis1}
\end{figure*}

A second  study of evaporative cooling to BEC in a dilute gas was performed by the group of Esslinger and H\"{a}nsch in Munich~\cite{kohl_davis_02}.  In this experiment the Bose cloud, which was again initially prepared in an equilibrium state slightly above 
$T_\mathrm{c}$, was subjected to a continuous rf field inducing the ejection of 
high-energy atoms from the sample.  By adjusting the frequency of the applied field and thus the energies of the removed atoms, these authors were able to investigate the growth of the condensate for varying rates of evaporative cooling.  
Davis and Gardiner extended their earlier approach~\cite{davis_gardiner_00} to include the effects of three-body loss and gravitational sag on the cooling of the $^{87}$Rb cloud in this experiment~\cite{davis_gardiner_02}.  Their calculations yielded excellent agreement with the experimental data of Ref.~\cite{kohl_davis_02} within its statistical uncertainty for all but the slowest cooling scenarios considered.  An example is shown in Fig.~\ref{fig:Davis1}(a). 

In 1997 Pinkse~\emph{et al.}~\cite{pinkse_mosk_97} experimentally demonstrated that adiabatically changing the trap shape could increase the 
phase-space density of an atomic gas by up to a factor of two and conjectured that this effect could be exploited to cross the BEC transition in a thermodynamically reversible fashion.  This scenario was subsequently realised in the MIT group by Stamper-Kurn \emph{et al.}~\cite{stamper-kurn_miesner_98} by slowly ramping on a tight ``dimple'' trap formed from an optical dipole potential on top of a weaker harmonic magnetic trap.  
This experiment 
was the setting for the first application of a stochastic Gross--Pitaevskii methodology~\cite{stoof_01}, previously developed from a nonequilibrium formalism for Bose gases  by Stoof~\cite{stoof_99}.   This is based on the many-body T-matrix approximation, and uses the Schwinger--Keldysh path integral formulation of nonequilibrium quantum field theory to derive a Fokker-Planck equation for both the coherent and incoherent dynamics of a Bose gas.  The classical modes of the 
gas were represented by a Gross--Pitaevskii equation, with additional dissipative and noise terms resulting from a collisional coupling to a thermal bath with a temperature $T$ and chemical potential $\mu$.

 Proukakis~\emph{et al.}~\cite{proukakis_schmiedmayer_06} subsequently used this methodology to study the formation of quasicondensates in a one-dimensional dimple trap.  A much later experiment~\cite{garrett_ratnapala_11} 
investigated the dynamics of condensate formation following the sudden introduction of a dimple trap, and included quantum-kinetic simulations that were in good agreement with the data.

A novel method of cooling a bosonic cloud to condensation was introduced in 2003 by the Cornell group at JILA~\cite{harber_mcguirk_03}, who demonstrated the evaporative cooling of an atomic Bose cloud brought in close proximity to a dielectric surface, due to the selective adsorption of high-energy atoms.
More recently, similar experiments have been undertaken by the Durham \cite{marchant_handel_11} and T\"ubingen groups~\cite{markle_allen_15}, with the observed rates of loss in the latter case explained accurately by non-ergodic ZNG-method calculations of the evaporative cooling dynamics.  
Example results are shown in Fig.~\ref{fig:Davis1}(b).

\subsection{Other theories for condensate formation}
For completeness, here we briefly outline other theoretical methods that can be applied to condensate formation.
A generalised kinetic equation for thermally excited Bogoliubov quasiparticles was obtained by Imamovic-To\-ma\-so\-vic and Griffin~\cite{imamovic-tomasovic_griffin_01} based on the application of the Kada\-noff--Baym nonequilibrium Green's function approach~\cite{kadanoff_baym_book_62} to a trapped Bose gas. 
This kinetic equation reduces to that of Eckern~\cite{eckern_84} in the homogenous limit and to that of ZNG~\cite{zaremba_nikuni_99} when the quasiparticle character of the excitation spectrum is neglected. 
Walser \emph{et al.}~\cite{walser_williams_99,walser_cooper_01} derived a kinetic theory for a weakly interacting condensed Bose gas in terms of a coarse graining of the $N$-particle density operator over configurational variables.  Neglecting short-lived correlations between colliding atoms in a Markov approximation, they obtained kinetic equations for the condensate and noncondensate mean fields which were subsequently shown to be microscopically equivalent~\cite{wachter_walser_01a} to the nonequilibrium Green's function approach of Ref.~\cite{imamovic-tomasovic_griffin_01}.  
Exactly the same kinetic equations were derived by Proukakis \cite{proukakis_2001a}, within the formalism of his earlier quantum kinetic formulation \cite{proukakis_burnett_1996,proukakis_burnett_1998}, based on the adiabatic elimination of rapidly-evolving averages of non-condensate operators, ideas which fed into the development of the ZNG kinetic model \cite{shi_griffin_98}.
Although elegant, these formalisms have not provided a tractable computational methodology for modelling condensate formation away from the quasistatic limit.

A non-perturbative method for the many-body dynamics of the Bose gas far from equilibrium has been developed by Berges, Gasenzer, and co-workers~\cite{gasenzer_berges_05,berges_gasenzer_07,branschadel_gasenzer_08,Bodet2011a}.  This two-particle irreducible (2PI) effective-action approach provides a systematic way to derive approximate Kadanoff--Baym equations consistent with conservation laws such as those for energy and particle number. 
In contrast to 1PI methods, single-particle correlators are 
determined self-consistently by these equations. 
This approach allows the description of strongly correlated systems, and has been exploited in the context of turbulent condensation \cite{Berges:2008wm,Berges:2008sr,Scheppach:2009wu} where it provides a self-consistently determined many-body $T$~matrix. 
This 2PI effective-action approach is useful for studying strongly interacting systems such as 1D gases with large coupling constant~\cite{Kronenwett:2010ic}, or relaxation and (pre-)thermalization of strongly correlated spinor gases \cite{Babadi2015a.PhysRevX.5.041005}.

\subsection{Other pioneering condensate-formation experiments}

There are a number of experimental methods  other than evaporative cooling to increase the 
phase-space density of a quantum gas and form a condensate.   We briefly mention them here for completeness.

An experimental technique that has proved to be extremely useful for multi-component quantum gases is the method of sympathetic cooling, in which an atomic gas is cooled by virtue of its collisional interaction with a second gas of atoms, distinguished from the first either isotopically or by internal quantum numbers, which is itself subject to, e.g., evaporative cooling.  This technique was first demonstrated by Myatt \emph{et al.}~\cite{myatt_burt_97} in a gas comprising two distinct spin states of $^{87}$Rb, and was subsequently employed to cool a single-component Fermi gas to degeneracy by Schreck \emph{et al.}~\cite{schreck_ferrari_01}.

In a similar spirit, in 2009 the Inguscio group in Florence used entropy exchange between components of a two-species $^{87}$Rb-$^{41}$K Bose gas mixture to induce BEC in one of the components~\cite{catani_barontini_09}.  The two gases were brought close to degeneracy by cooling, after which the strength of the  $^{41}$K  trapping potential was adiabatically increased, by introducing an optical dipole potential 
to which the $^{87}$Rb component 
was largely insensitive.  In a single-component system this would lead to an increase in the temperature and leave the phase-space density unaffected.  However, in the dual-species setup the $^{87}$Rb cloud acted as a thermal reservoir, suppressing the temperature increase of the $^{41}$K component and causing it to cross the BEC threshold.

In 2004, the Sengstock group observed the formation of a BEC at constant temperature~\cite{Erhard2004a}.  Working with a spin-1 system, they prepared a partially condensed gas consisting of  $m_F = \pm1$ states.  Spin collisions within the BEC components populated the $m_F = 0$ state, which then quickly thermalised.  When the population of the $m_F =  0$ component reached the critical number a new BEC emerged.  The experiment was modelled with a simple rate equation.

In the same year, Ketterle's MIT group  performed an experiment 
in which they distilled a BEC from one trap minimum to another~\cite{shin_saba_04}.  A non-zero-temperature BEC was formed in an optical dipole trap, before a second trap with a greater potential depth was brought nearby.  Atoms of sufficient thermal energy were able to cross the barrier between the two potential minima, populating the second trap.  Eventually the first condensate evaporated, and a second condensate formed in the new global trap minimum.

Finally, we mention a recent experiment by the group of Schreck at Innsbruck, 
who demonstrated the first experimental production of a BEC 
solely by laser cooling~\cite{stellmer_pasquiou_13}.  This feat was made possible by laser cooling on a narrow-linewidth transition of $^{84}$Sr, resulting in a low 
Doppler-limit temperature of just 350 nK.  A ``light-shift'' laser beam was introduced at the centre of the trap so that the atoms in that region no longer responded to the laser cooling, 
after which an additional dimple trap was introduced to confine the atoms.  Repeatedly cycling the dimple trap on and off resulted in the formation of several condensates~\cite{stellmer_pasquiou_13}.

\subsection{Low-dimensional Bose systems and phase fluctuations}

The experiments described above were in the three-dimensional (3D) realm, in which long-wavelength phase fluctuations are strongly suppressed away from the vicinity of the phase transition.  In lower dimensional systems such fluctuations are enhanced, leading to dramatic modifications to the physics of the degenerate regime.  In a two-dimensional (2D) system, thermal fluctuations of the phase erode the long-range order associated with true condensation, leaving only so-called quasi-long-range order characterised by correlation functions that decay algebraically with spatial separation~\cite{Popov1972}.  A more complete analysis reveals the importance of vortex-antivortex pairs in this phase-fluctuating ``quasi-condensed'' regime~\cite{kosterlitz_thouless_73}.  Such pairs undergo a so-called Berezinskii--Kosterlitz--Thouless (BKT) deconfinement transition at some finite temperature, above which even quasi-long-range order is lost and superfluidity is extinguished.  
Two-dimensional Bose systems are of particular interest due to their natural realisation in systems such as liquid helium films and the fact that the degenerate Bose quasiparticles such as excitons and polaritons in semiconductor systems are typically confined in a planar geometry.  An insightful overview of BKT physics can be found in the chapter by Kim, Nitsche and Yamamoto in this volume. 

There have been numerous experimental 
realisations of (quasi-)2D Bose gases in cold-atom experiments~\cite{hadzibabic_kruger_06,schweikhard_tung_07,kruger_hadzibabic_07,clade_ryu_09,tung_lamporesi_10,hung_zhang_11}, with most notable the observations of 
thermally activated vortices 
via interferometric measurements \cite{hadzibabic_kruger_06} and the direct probing of the equation of state and 
scale invariance of the 2D system \cite{hung_zhang_11} (see the chapter by Chin and Refs.~\cite{prokofev_ruebenacker_01,simula_blakie_06,holzmann_krauth_08,bisset_davis_09,cockburn_proukakis_12} for related theoretical considerations). 
Further details and a lengthy discussion of the interplay between BKT and BEC in homogeneous and trapped systems can be found in 
Ref.~\cite{hadzibabic_dalibard_12}.  Although theoretical works on the dynamics of such systems have existed for some time, little experimental work on the formation dynamics of condensates in these systems has been undertaken (aside from the quasi-2D Kibble-Zurek works discussed in the following section). 
Considerable discussion is currently taking place 
regarding the emergence and nature of the BKT transition in driven-dissipative polariton condensates: experimentalists have observed evidence for quasi-long-range order \cite{roumpos_lohse_12,nitsche_kim_14} (see 
Kim \emph{et al.}'s chapter), but the 
nature of the transition and its ``nonequilibrium" features are 
topics of current debate \cite{dagvadorj_fellows_14,altman_sieberer_15} (see also the chapter by Keeling {\em et al.}). 

In one dimension, the effects of phase fluctuations are even more pronounced, leading to the complete destruction of long-range order and superfluidity at any finite temperature.  Many experiments with cold atoms in elongated 
``cigar-shaped'' traps have investigated the physics of such (quasi-) one-dimensional systems, though again, little work has been done on the formation dynamics of these degenerate samples.  We note, however, that quasicondensate regimes somewhat analogous to those of (quasi-) one-dimensional systems can be realized in elongated 3D traps~\cite{petrov_shlyapnikov_01}.  In such a regime, the Bose gas behaves much as a conventional three-dimensional Bose condensate, except that the coherence length of the gas is shorter than the system extent along the long axis of the trap.  A study of condensate formation 
in this regime was performed by the Amsterdam group of Walraven~\cite{shvarchuck_buggle_02} in 2002 in 
an elongated $^{23}$Na 
cloud.  Similarly to the MIT experiment~\cite{miesner_stamper-kurn_98}, they performed rapid quench cooling of their sample from just above the critical temperature.  However, the system was in the hydrodynamic regime in the 
weakly trapped dimension, i.e., the mean distance between collisions was much shorter than the system length.  

 It was argued that the system rapidly came to a local thermal equilibrium in the radial direction, resulting in cooling of the cloud below the \emph{local} degeneracy temperature over a large spatial region and generating an elongated quasicondensate.  However, 
the extent of this quasicondensate along the long axis of the trap was larger than that expected at equilibrium, leading to large amplitude oscillations.  The momentum distribution of the 
cloud was imaged via ``condensate focussing'', with the breadth of the focal point giving an indication of the magnitude of the phase fluctuations present in the 
sample.  This interesting experiment was somewhat ahead of its time, with theoretical techniques unable to address many of the nonequilibrium aspects of the problem.  

In 2007 the group of Aspect from Institut d'Optique also studied the formation of a quasicondensate in an elongated 
three-dimensional trap via continuous evaporative cooling~\cite{hugbart_retter_07} in a similar fashion to the earlier work by K\"{o}hl \emph{et al.}~\cite{kohl_davis_02}.  As well as measuring the condensate number, they also performed Bragg spectroscopy during the growth to determine the momentum width and hence the coherence length of the system.  They found that the momentum width they measured rapidly decreased with time to the 
width expected in equilibrium for the instantaneous value of the condensate number. 
Modelling of the growth of the condensate population 
using the methodology of Ref.~\cite{davis_gardiner_02}
produced results in good agreement with the experimental data, apart from an unexplained delay of 10--50 ms, depending on the rate of evaporation.

\section{Criticality and nonequilibrium dynamics}

As Bose--Einstein condensation is a continuous phase transition, the theory of critical phenomena~\cite{binney_dowrick_92} predicts that in the vicinity of the critical point the correlations of the Bose field obey universal scaling relations.  In particular, the scaling of correlations at and near equilibrium is governed by a set of universal critical exponents and scaling functions, independent of the microscopic parameters of the gas.  
For a homogeneous system close to criticality, 
standard theory predicts that the correlation length $\xi$, relaxation time $\tau$, and first-order correlation function $G(x)=\langle\psi^{\dagger}(x)\psi(0)\rangle$ obey scaling laws
\begin{equation}
\xi = \frac{\xi_0}{|\epsilon|^\nu}, 
\quad 
\tau = \frac{\tau_0}{|\epsilon|^{\nu z}},
\quad
G(x) = \epsilon^{\nu(d-2+\eta)}\mathcal{F}(\epsilon^{\nu} x),
\label{davis_crit_exponents}
\end{equation}
with $\epsilon = T/T_\mathrm{c} -1$ the reduced temperature, $\nu$ and $z$ the correlation length and dynamical critical exponents, $\eta$ the scaling dimension of the Bose field, and $\mathcal{F}$ a universal scaling function.  The static  Bose gas belongs to the $XY$ (or $O(2)$) universality class, and is thus expected to have the same critical exponents as superfluid helium, i.e., in 3D, $\nu \simeq 0.67$ and $\eta=0.038(4)$~\cite{Zinn-Justin2002}.  
The critical dynamics of the system are expected to conform to those of the diffusive model denoted by $F$ in the classification of Ref.~\cite{Hohenberg1977a}, implying a value $z=3/2$ for the dynamical critical exponent.

The influence of critical physics is significantly reduced in 
the conditions of harmonic confinement typical of experimental Bose-gas systems, as compared to homogeneous systems.  Within a local-density approximation, the inhomogeneous thermodynamic parameters of the system imply that only a small fraction of atoms in the gas enter the critical regime, and so global observables are relatively insensitive to the effects of criticality.  Nevertheless, a few experiments have attempted to observe aspects of the critical physics of trapped Bose gases.  

In a homogeneous gas the introduction of interparticle interactions has no effect on the critical temperature at the mean-field level, but the magnitude and even the sign of the shift due to critical fluctuations was debated for several decades (see Ref.~\cite{baym_blaizot_99} and references therein) before being settled by classical-field Monte-Carlo calculations~\cite{kashurnikov_prokofev_01,arnold_moore_01}.  An experiment by the Aspect group carefully measured a shift in critical temperature of the trapped gas, but was unable to unambiguously infer any 
beyond-mean-field contribution to this shift~\cite{gerbier_thywissen_04,davis_blakie_06}.  A later experiment by the group of Hadzibabic made use of a Feshbach resonance to control the interaction strength in $^{41}$K, and found clear evidence of a positive beyond-mean-field shift~\cite{smith_campbell_11} (see also the chapter by Smith in this volume).  

\begin{figure*}
\centering
\includegraphics[height=5.5cm]{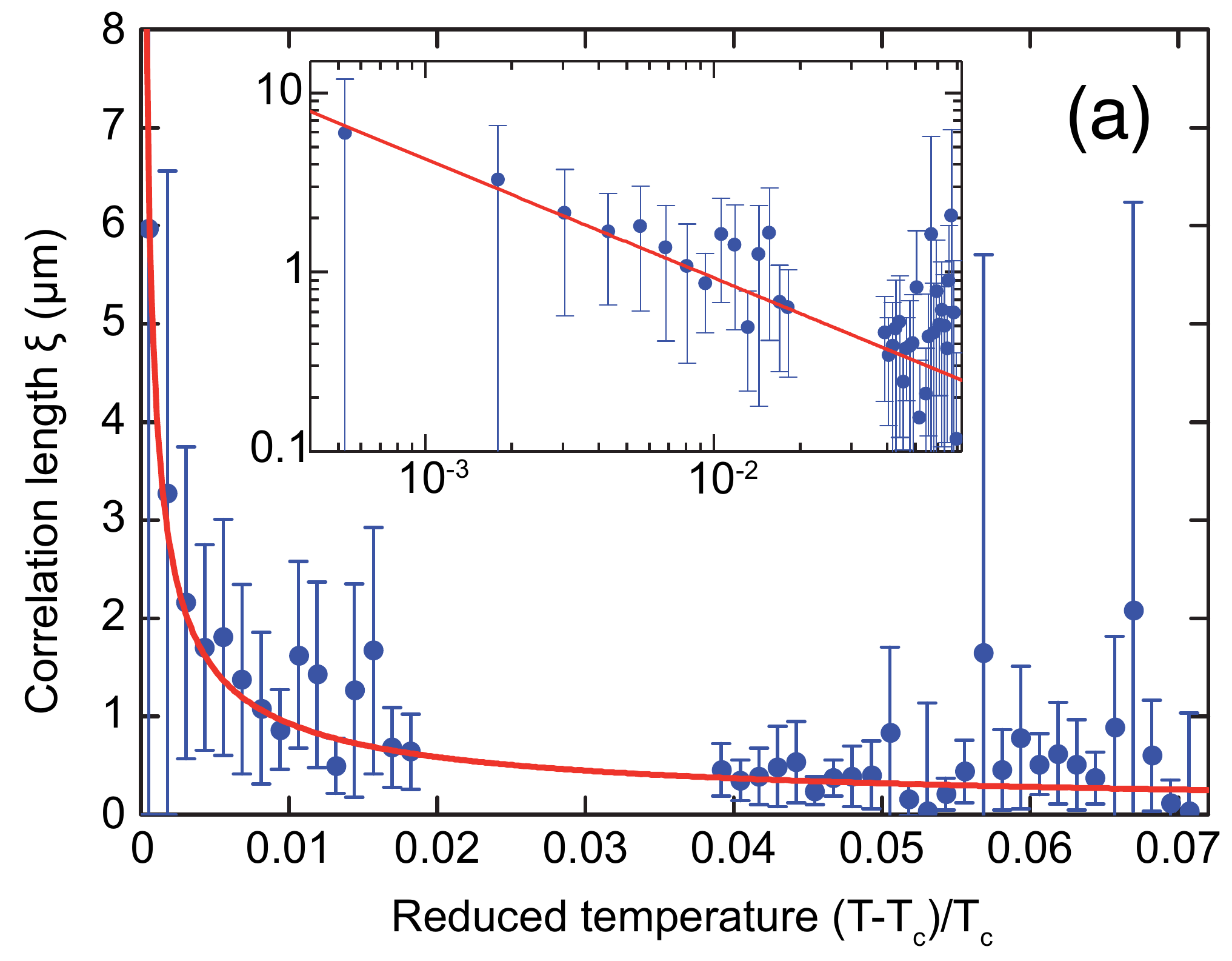}
\includegraphics[height=5.5cm]{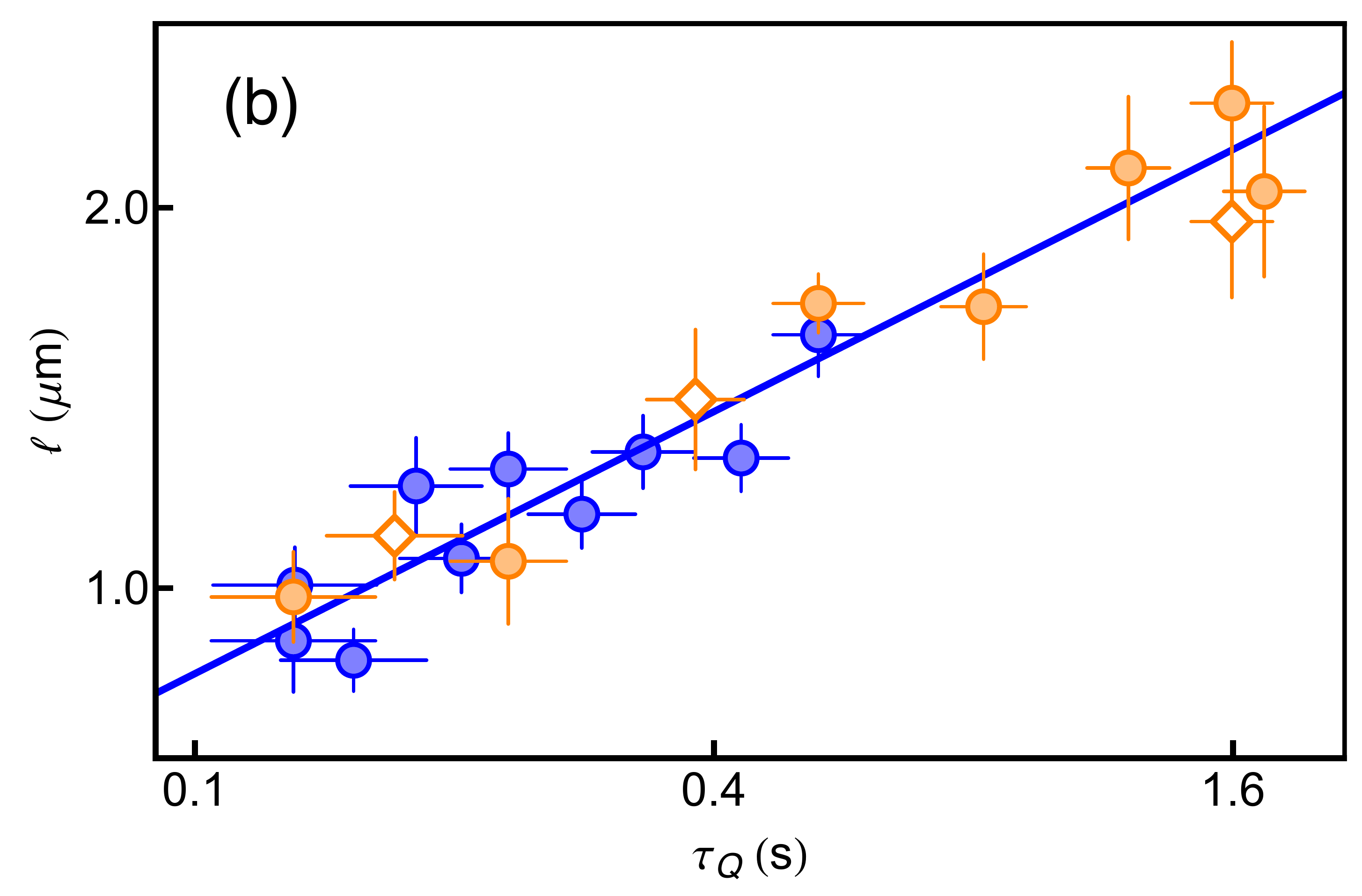}
\caption{Critical phenomena in BECs. (a) Divergence of the equilibrium correlation length $\xi$ as a function of the reduced temperature, and the fitting of the critical exponent, giving the result $\nu = 0.67\pm 0.13$. Inset: Double logarithmic plot of the same data. Taken from Ref.~\cite{donner_ritter_07}. (b) Log-log plot of the dependence of the correlation length, here labelled $\ell$, as a function of the characteristic time 
$\tau_\mathrm{Q}$ of the quench through the BEC phase transition.  The solid line corresponds to a Kibble--Zurek power-law scaling 
$\ell \propto \tau_\mathrm{Q}^b$ with $b = 0.35\pm0.04$, in agreement with the beyond-mean-field prediction $b=1/3$ of the so-called $F$ model~\cite{Hohenberg1977a} and inconsistent with the mean-field value $b = 1/4$.  This in turn implies a value $z = 1.4 \pm 0.2$ for the dynamical critical exponent. Taken from Ref.~\cite{navon_gaunt_15}.
}
\label{fig:Davis2}
\end{figure*}

In 2007 the ETH Z\"urich group of Esslinger revisited their experiments on condensate formation and the coherence of a three-dimensional BEC with a new tool: the ability to count single atoms passing through an optical cavity below their ultra-cold gas~\cite{ritter_ottl_07}.  They outcoupled atoms from two different vertical locations from their sample as it was cooled, realising interference in the falling matter waves.  By monitoring the visibility of the fringes, they were able to measure the growth of the coherence length as a function of time.  Using the same optical cavity setup, the Esslinger group subsequently measured the coherence length of their Bose gas as it was driven through the critical temperature by a small background heating rate, and determined the correlation-length critical exponent to be $\nu = 0.67 \pm 0.13$~\cite{donner_ritter_07}. Their results are shown in Fig.~\ref{fig:Davis2}(a).  Classical-field simulations of their experiment were in reasonable agreement, determining $\nu = 0.80 \pm 0.12$~\cite{bezett_blakie_09}.  

Although an important topic in its own right, the greatest significance of the equilibrium theory of critical fluctuations to studies of condensate formation is that it provides a basis for generalisations of concepts such as critical scaling laws and universality classes to the domain of nonequilibrium physics.  In the remainder of this section we discuss two such extensions: the Kibble--Zurek mechanism (KZM), and the theory of non-thermal fixed points.

\subsection{The Kibble--Zurek mechanism}
\label{sec:Davis:KibbleZurek}

The theory of the Kibble--Zurek mechanism leverages the well-established results of the equilibrium theory of criticality to make immediate predictions for universal scaling behaviour in the nonequilibrium dynamics of passage through a second-order phase transition.
The underlying idea --- that causally disconnected regions of space break symmetry independently, leading to the formation of topological defects --- was first discussed by Kibble~\cite{kibble_76}, who predicted that the distribution of defects following the transition would be determined by the instantaneous correlation length of the system as it passes through the Ginzburg temperature~\cite{landau_lifshitz_book_80}.  Zurek later emphasised~\cite{zurek_85} the importance of dynamic critical phenomena~\cite{Hohenberg1977a} in such a scenario. 
In particular, the scaling relations~\eqref{davis_crit_exponents} imply that both the correlation length and the characteristic relaxation time of the system diverge as the critical point is approached ($\epsilon \rightarrow 0$), imposing a limit to the size of spatial regions over which order can be established during the transition.  Topological defects will thus be seeded, with a density determined by the correlation length at the time the system ``freezes'' during the transition, and will subsequently decay in the symmetry-broken phase.  The more rapidly the system passes through the critical point, the shorter the correlation length that is frozen in, and therefore more topological defects will form.  
A dimensional analysis predicts that a linear ramp 
$\epsilon(t) = -t/\tau_\mathrm{Q}$ of the reduced temperature through the critical point on a characteristic time scale 
$\tau_\mathrm{Q}$ results in a distribution of spontaneously formed defects with a density
$n_d$
that scales as~\cite{zurek_96}
\begin{equation}
n_d
\propto 
\tau_\mathrm{Q}^{(p-d)\nu/(1 + \nu z)},
\end{equation}
where $d$ is the dimensionality of the sample and $p$ is the intrinsic dimensionality of the defects.%

Zurek initially described the KZM in the context of vortices in the $\lambda$-transition of superfluid $^4$He~\cite{zurek_85}.  Although vortices are observed in the wake of this transition, it is difficult to identify them as having formed due to the KZM rather than being induced by, e.g., inadvertent stirring~\cite{zurek_96} (see also the chapter by Pickett in this volume).  
The prospect of generating vorticity in atomic BECs by means of the KZM was first discussed by Anglin and Zurek in 1999~\cite{anglin_zurek_99}.    However, it was not until the 2008 experiment of the Anderson group at the University of Arizona~\cite{weiler_neely_08} that spontaneously formed vortices were first observed in such a system (see also Ref.~\cite{freilich_bianchi_10}).  
 
\begin{figure*}
\centering
\includegraphics[height=5.5cm]{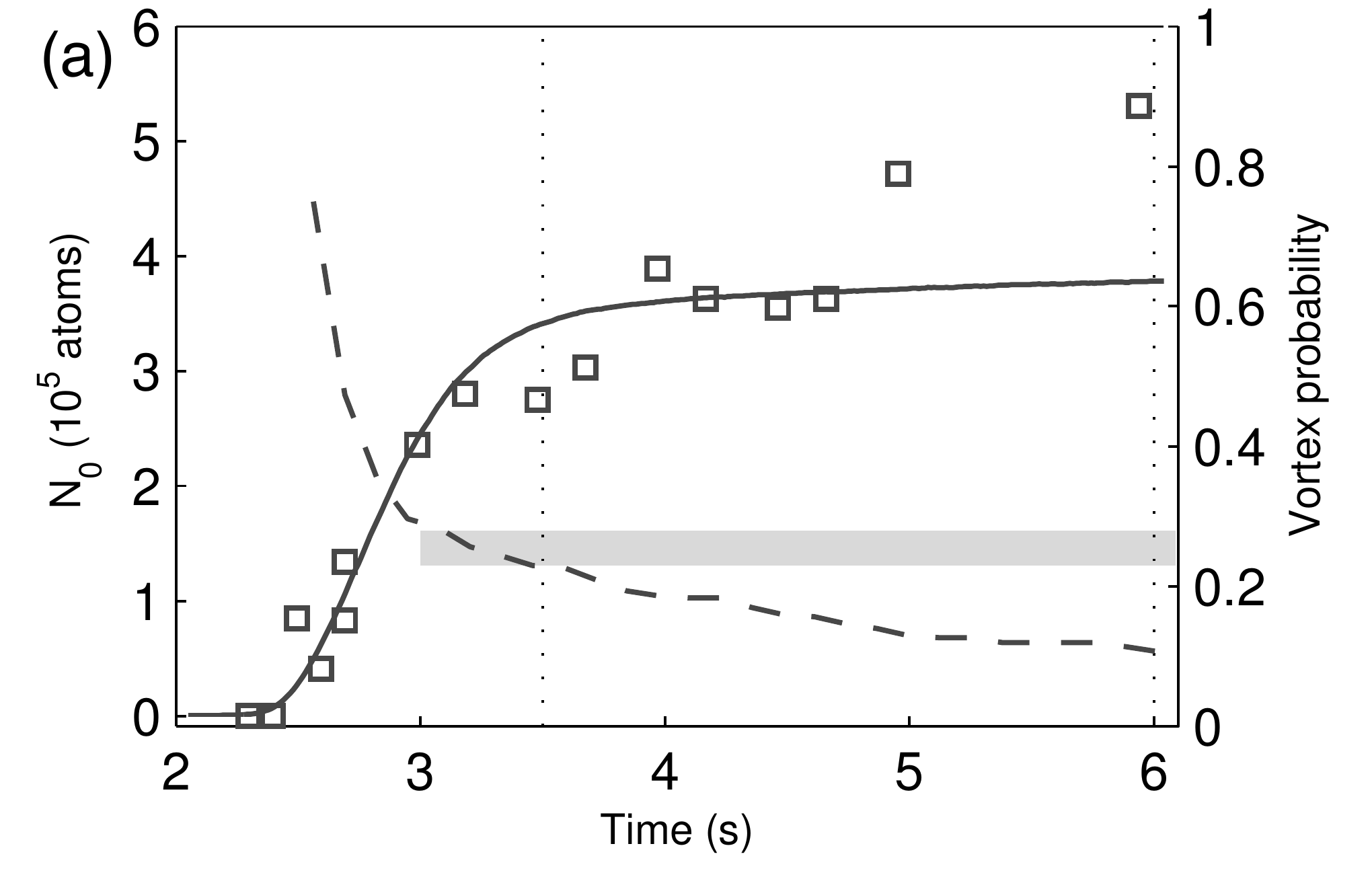}
\includegraphics[height=6cm]{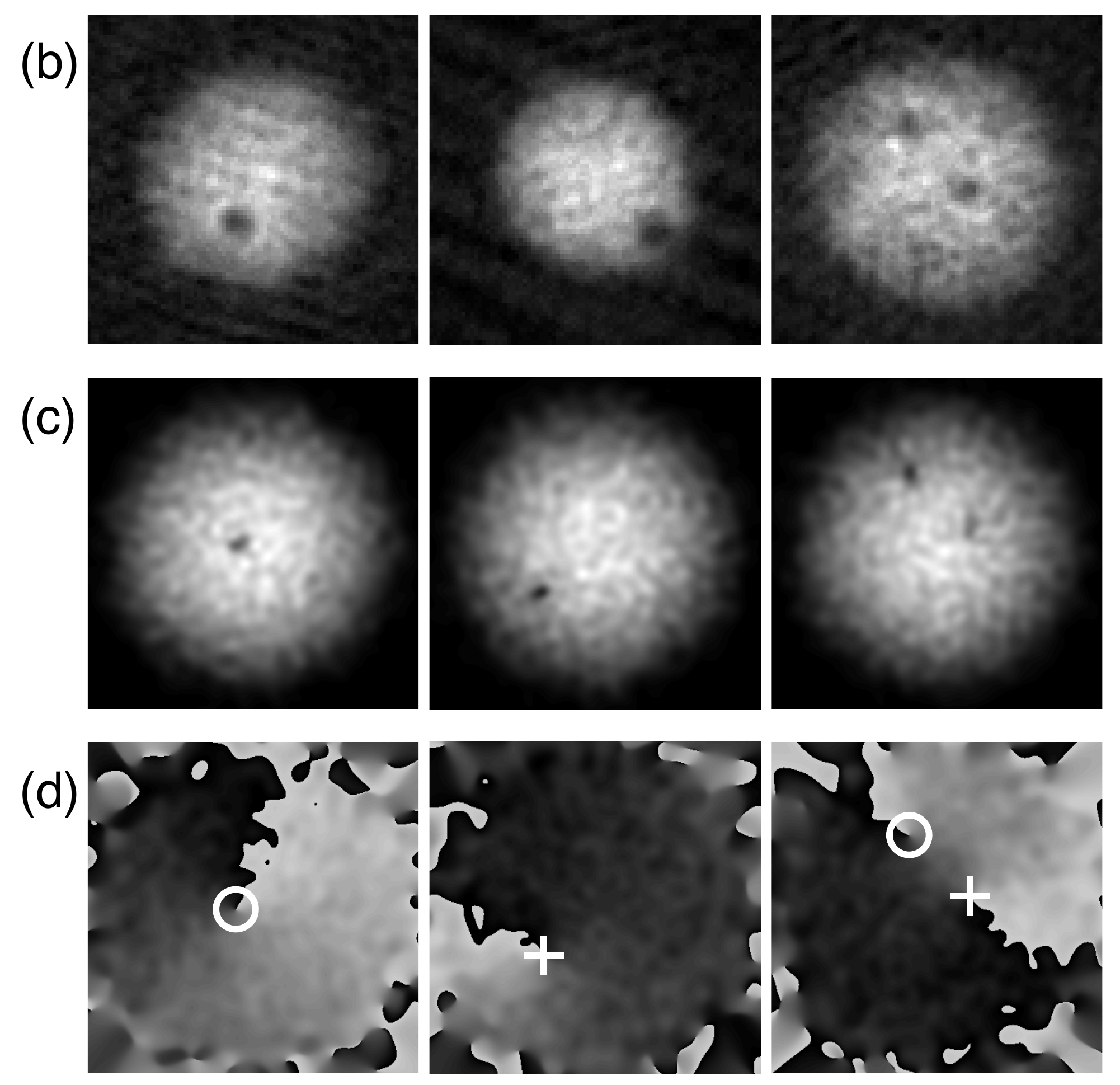}
\caption{Spontaneous vortices in the formation of a Bose--Einstein condensate. (a) Squares: Experimentally measured condensate population as a function of time.  Solid line: Condensate number from stochastic Gross--Pitaevskii simulations.  Dashed line: probability of finding one or more vortices in the simulations as a function of time, averaged over 298 trajectories.  The shaded area indicates the statistical uncertainty in the experimentally measured vortex probability at $t= 6.0$ s.  It was observed in experiment that there was no discernible vortex decay between 3.5 s and 6.0 s  (b) Experimental absorption images taken after 59 ms time-of-flight showing the presence of vortices.  (c) Simulated in-trap column densities  at $t=3.5$ s [indicated by the left vertical dotted line in (a).] (d) Phase images through the $z=0$ plane, with plusses (open circles) representing vortices with positive (negative) circulation. 
Adapted from C.~N.~Weiler \emph{et al.}~\cite{weiler_neely_08}.}
\label{fig:Davis3}
\end{figure*}

The observations of spontaneous vortices in Ref.~\cite{weiler_neely_08} were supported by numerical simulations using the stochastic projected Gross-Pitaevskii equation description of Gardiner and Davis~\cite{gardiner_davis_03}.  Their results are shown in Fig.~\ref{fig:Davis3}.
This formalism is essentially a variant of the Gardiner-Zoller quantum kinetic theory, obtained by making a high-temperature approximation to the condensate-band master equation and then exploiting the quantum-classical correspondence of the Wigner representation to obtain a stochastic classical-field description of the condensate band~\cite{gardiner_davis_03, blakie_bradley_08}. 
Although derived using different theoretical techniques, the resulting description is similar to the stochastic Gross-Pitaevskii equation of Stoof~\cite{stoof_99,stoof_01}, both in terms of its physical content and its computational implementation
--- see, e.g., discussion in Refs.~\cite{proukakis_jackson_08,cockburn_proukakis_09}.
A related phase-space method originating in quantum optics known as the positive-P representation has also been applied to ultracold gases~\cite{steel_olsen_98}.  This has been used to investigate cooling 
of a small system towards BEC by Drummond and Corney~\cite{drummond_corney_99}, 
who observed features consistent with spontaneously formed vortices.  Despite formally being a statistically exact method, for interacting systems it tends to suffer from numerical divergences after a relatively short evolution time.

It seems likely that spontaneously formed vortices and other defects were present in earlier BEC-formation experiments, but not observed due to the practical difficulties inherent in resolving these defects in experimental imaging --- and indeed the fact that these experiments were not attempting to investigate whether such structures were present. 
Another difficulty in identifying quantitative signatures of the KZM in experimental BECs is the inhomogeneity of the system in the experimental trapping potential, which is typically harmonic.  From the point of view of a local-density approximation, this inhomogeneity implies that the instantaneous coherence length and relaxation timescale are spatially varying quantities, and that the transition occurs at different times in different regions of space as the system is cooled. 
Following preliminary reports of the experimental observation of dark solitons following the formation of a quasi-one-dimensional BEC by the group of Engels at the University of Washington~\cite{chiang_hamner_09}, Zurek applied the framework of the KZM to a quasi-1D BEC in a cigar-shaped trap to estimate the scaling of the number of spontaneously generated solitons as a function of the quench time~\cite{zurek_09,damski_zurek_10}.  Witkowska \emph{et al.}~\cite{witkowska_deuar_11} numerically studied cooling leading to solitons in a comparable one-dimensional geometry.  Zurek's methodology for inhomogeneous systems was applied by del~Campo \emph{et al.}~\cite{delcampo_retzker_11} to strongly oblate geometries in which vortex filaments behave approximately as point vortices in the plane, an idealisation of the geometry of the experiment of Weiler \emph{et al.} \cite{weiler_neely_08}.

Lamporesi \emph{et al.}~\cite{lamporesi_donadello_13} recently reported the spontaneous creation of Kibble--Zurek dark solitons in the formation of a BEC in 
an elongated trap, and found the scaling of the number of observed defects with cooling rate in good agreement with the predictions of Zurek~\cite{zurek_09}.  It was later realised that the apparent solitons were actually solitonic vortices \cite{trento_prl}. 
The effects of inhomogeneity in such experiments can be mitigated by the realisation of ``box-like'' 
flat-bottomed trapping geometries.  The Dalibard group in Paris has observed the formation of spontaneous vortices in 
a quasi-2D box-like geometry, and found scaling of the vortex number with quench rate in good agreement with the predictions of the KZM~\cite{chomaz_corman_15}.   
We also note further work by the Dalibard group~\cite{corman_chomaz_14} verifying the production of quench-induced supercurrents in a toroidal or ``ring-trap'' geometry~\cite{das_sabbatini_12} analogous to the annular sample of superfluid helium considered in Zurek's original proposal~\cite{zurek_85}.

Experimental investigations of the KZM in dilute atomic gases have largely 
focused on the imaging of defects in the wake of the phase transition --- either following time-of-flight expansion~\cite{weiler_neely_08,lamporesi_donadello_13,chomaz_corman_15} or {\em in situ} \cite{trento_prl}.  However, the accurate extraction of critical scaling behaviour from such observations is hampered by the large background excitation of the field near the transition, and the 
relaxation (or ``coarsening'') dynamics of defects in the symmetry-broken phase.  An alternative approach is to make quantitative measurements of global properties of the system following the quench.  Performing quench experiments in a three-dimensional box-like geometry, the Hadzibabic group in Cambridge~\cite{navon_gaunt_15} made careful measurements of the scaling of the correlation length with quench time.  From the measured scaling law, these authors were able to infer a beyond-mean-field value $z = 1.4 \pm 0.2$ for the dynamical critical exponent for this universality class.  Some of the results of Ref.~\cite{navon_gaunt_15} are displayed in Fig.~\ref{fig:Davis2}(b).

The possibilities for the 
trapping and cooling of multicomponent systems in atomic physics experiments have naturally lead to investigations of the spontaneous formation of more complicated topological defects during a phase transition. Although such experiments have so-far largely focused on the formation of defects following a quench of Hamiltonian parameters~\cite{sadler_higbie_06, de_campbell_14}, the formation of nontrivial domain structures following gradual sympathetic cooling in immiscible $^{85}$Rb-$^{87}$Rb~\cite{papp_pino_08} and $^{87}$Rb-$^{133}$Cs~\cite{mccarron_cho_11} Bose-Bose mixtures has also been observed.  
The competing growth dynamics of the two immiscible components in the formation of such a binary condensate have recently been investigated theoretically in the limit of a sudden temperature quench~\cite{liu_pattinson_15} (see also Refs.~\cite{sabbatini_zurek_11,swislocki_witkowska_13,hofmann_natu_14} for related critical scaling in other Hamiltonian quenches).  These investigations indicate the rich nonequilibrium dynamics possible in these systems, including strong memory effects on the coarsening of spontaneously formed defects and the potential ``microtrapping'' of one component in spontaneous defects formed in the other.

\subsection{Non-thermal fixed points}
\label{sec:Davis:NTFP}

A general characterisation of the relaxation dynamics of quantum many-body systems quenched far out of equilibrium remains a largely open problem.  In particular, it is interesting to ask to what extent analogues of the universal descriptions arising from the equilibrium theory of critical fluctuations may exist for nonequilibrium systems.  A recent advance towards answering such questions has been made in the development of the theory of non-thermal fixed points: universal nonequilibrium configurations showing scaling in space and (evolution) time, characterised by a small number of fundamental properties.  The theory of such fixed points transposes the concepts of equilibrium and diffusive near-equilibrium renormalisation-group theory to the real-time evolution of nonequilibrium systems.  These developments provide, for example, a framework within which to understand the turbulent, coarsening, and relaxation dynamics following the creation of various kinds of defects and nonlinear patterns in a Kibble-Zurek quench.

The existence and significance of non-thermal scaling solutions in space and time was discussed by Berges  and collaborators in the context of reheating after early-universe inflation \cite{Berges:2008wm,Berges:2008sr} and then 
generalised by Berges, Gasenzer, and coworkers to scenarios of strong matter-wave turbulence~\cite{Scheppach:2009wu,Mathey2014a.PhysRevA.92.023635}.
For the condensation dynamics of the dilute Bose gas discussed here, the presence of a non-thermal fixed point can exert a significant  influence in the case of a strong cooling quench  \cite{Nowak:2012gd,Orioli:2015dxa,Berges:2012us}.

\begin{figure}
\centering
\includegraphics[width=8cm]{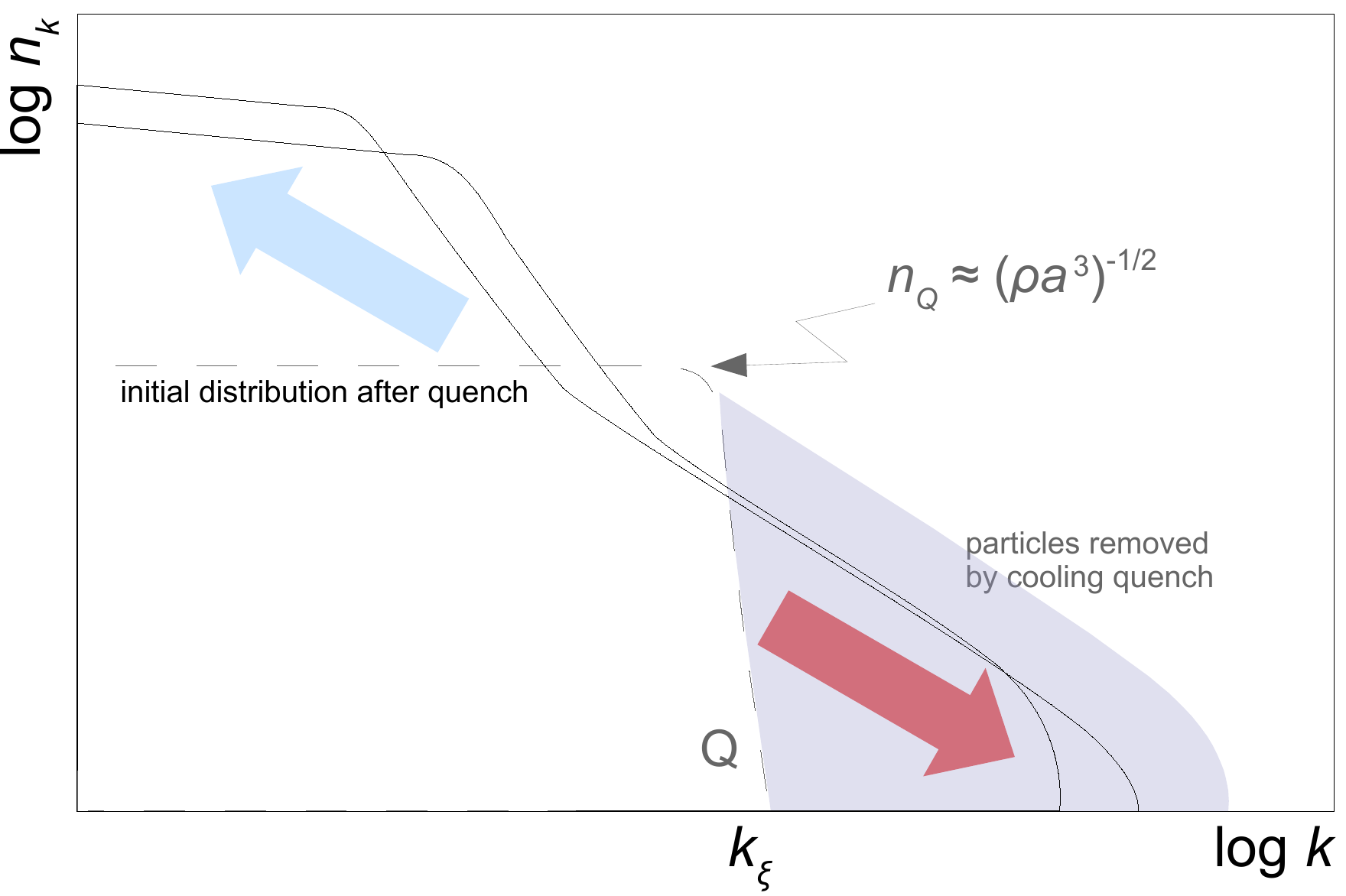}
\caption{Sketch of the evolution of the single-particle momentum distribution $n_{k}(t)$ of a Bose gas close to a non-thermal fixed point (after Ref.~\cite{Orioli:2015dxa}). Starting from the extreme initial distribution (dashed line, see main text for details) produced, e.g., by a strong cooling quench, a bidirectional redistribution of particles in momentum space (arrows) builds up a quasicondensate in the infrared while refilling the thermal tail at large momenta. The particle transport towards zero momentum is characterised by a self-similar scaling evolution in space and time, $n(k,t)=(t/t_{0})^{\alpha}n([t/t_{0}]^{\beta}k,t_{0})$, with characteristic scaling exponents $\alpha$, $\beta$. Note the double-logarithmic scale.}
\label{fig:NTFP}
\end{figure}

As an illustration, we consider a particle distribution that drops abruptly above the healing-length scale $k_{\xi}=\sqrt{8\pi a\rho}$, as depicted on a double-logarith\-mic scale in Fig.~\ref{fig:NTFP} (dashed line).  In order for the influence of the non-thermal fixed point to be observed, the decay of $n(k)$ above $Q\simeq k_{\xi}$ is assumed to be much steeper than the quasi-thermal scaling that develops in the kinetic stage of condensation following a weak quench~\cite{Semikoz1995a.PhysRevLett.74.3093,Semikoz1997a}, as discussed in Sect.~\ref{sec:Davis:BECFormation}.
Such a distribution would, e.g., result from a severe cooling quench of a thermal Bose gas initially just above the critical temperature where $T>|\mu|/k_\mathrm{B}$ such that the Bose-Einstein distribution has developed a Rayleigh--Jeans scaling regime where $n(k)\sim 2mk_\mathrm{B}T/(\hbar k)^{2}$. 
The modulus of the chemical potential of this state determines the momentum scale $Q$ where the flat infrared scaling of the distribution goes over to the Rayleigh--Jeans scaling at larger $k$.
If this chemical potential is of the order of the ground-state energy of the post-quench fully condensed gas, $(\hbar Q)^2/2m\simeq|\mu|\simeq g\rho$, with $g=4\pi\hbar^{2} a/m$, then the energy of the entire gas is concentrated at the scale $Q\simeq k_{\xi}$ after the quench.
This is a key feature of the extreme nonequilibrium initial state from which a non-thermal fixed point can be approached.
We note that, if in this state there is no significant zero-mode occupation $n_{0}$, the respective occupation number at $Q$ is on the order of the inverse of the diluteness parameter, $n_{Q}\sim \zeta^{-3/2}$. 

In analogy to the weak-wave-turbulence scenario \cite{svistunov_91,kagan_svistunov_92,Semikoz1995a.PhysRevLett.74.3093,Semikoz1997a} discussed in Sect.~\ref{Pre-condensationKinetic}, the initial overpopulation of modes with energies $\sim (\hbar Q)^{2}/2m$ leads to inverse particle transport while energy is transported to higher wavenumbers, as indicated by the arrows in Fig.~\ref{fig:NTFP} \cite{Nowak:2012gd,Orioli:2015dxa,Berges:2012us}. 
However, the inverse transport involves non-local scattering and thus does not represent a cascade.
Furthermore, in contrast to the case of a weak quench \cite{stoof_91,stoof_92,stoof_99,svistunov_91,kagan_svistunov_92,Semikoz1997a,berloff_svistunov_02}, in which weak-wave turbulence produces a quasi-thermal momentum distribution that  relaxes quickly to a thermal equilibrium distribution, here the inverse transport is characterised by a strongly non-thermal power-law scaling in the infrared.  Specifically, the momentum distribution $n(k)\sim k^{-d-2}\sim k^{-5}$ in $d=3$ dimensions~\cite{Scheppach:2009wu} provides the ``smoking-gun''  of the influence of the non-thermal fixed point. 
Semiclassical simulations by Nowak, Gasenzer and collaborators \cite{Nowak:2010tm,Nowak:2011sk,Nowak:2012gd,Nowak:2013juc} showed that this scaling is associated with the creation, dilution, coarsening and relaxation of a complex vortex tangle, as predicted on phenomenological grounds in Ref.~\cite{kagan_svistunov_94}, and other types of (quasi-)topological excitations in low-dimensional, spinor and gauge systems~\cite{Schole:2012kt,Schmidt:2012kw,Karl:2013mn, Karl:2013kua, Gasenzer:2013era}.  

The dynamics in the vicinity of the fixed point are characterised by an anomalously slow relaxation of the total vortex line length, which exhibits an algebraic decay $\sim t^{-0.88}$ (see Ref.~\cite{Schole:2012kt} for analogous results in the 2D case). At the same time, the condensate population grows as $n_{0}(t)\sim t^{2}$ \cite{Nowak:2012gd,Orioli:2015dxa}, a significant slowing compared to the $\sim t^{3}$ behaviour observed for weakly nonequilibrium condensate formation \cite{damle_majumdar_96,Nowak:2012gd}. 

In the vicinity of the fixed point 
the momentum distribution is expected to follow a self-similar scaling behaviour in space and time in the infrared, $n(k,t)=(t/t_{0})^{\alpha}n([t/t_{0}]^{\beta}k,t_{0})$.
For a 3D Bose gas, these scaling exponents have recently been numerically determined to be $\alpha=1.66(12)$, $\beta=0.55(3)$, in agreement with the analytically predicted values $\alpha=\beta d$, $\beta=1/2$~\cite{Orioli:2015dxa}.
This behaviour, here corresponding  to the dilution and relaxation of vortices leading to a build-up of the condensate population, represents the generalisation of critical slowing-down to real-time evolution far away from thermal equilibrium.
At very late times, the system leaves  the vicinity of the non-thermal fixed point, typically when the last topological patterns decay, and finally approaches thermal equilibrium \cite{Kozik2004a.PhysRevLett.92.035301,Kozik2005a.PhysRevLett.94.025301,Kozik2005a.PhysRevB.72.172505,Kozik2009a,Schole:2012kt,connaughton_josserand_05}.
This equilibrium state corresponds to a fully established condensate superimposed with weak sound excitations.

In summary, non-thermal fixed points are nonequilibrium field configurations, exhibiting universal scaling in time and space, to which the system is attracted if suitably forced --- e.g., in the case of Bose condensation, following a sufficiently strong cooling quench.
In the vicinity of such fixed points, the relaxation of the field is critically slowed down and the dynamics exhibit self-similar time evolution, governed by new critical exponents and scaling functions.
The possibility of categorising systems into 
generalised ``universality classes'' associated with the new critical exponents is a fascinating prospect and the subject of current research~\cite{sieberer_huber_13,altman_sieberer_15}.

Finally in this section we note that prethermalisation or ``pre-Gibbsiani\-sa\-tion'', i.e., the approach of a state characterised by a Generalised Gibbs ensemble, usually in near-integrable systems (see the chapter by Schmiedmayer), represents a special case of a Gaussian non-thermal fixed point, meaning that the effective coupling of the prethermalised modes vanishes.
It is expected that the exponents $\alpha$ and $\beta$ in such a situation can become very small compared to unity.
Only at very late times the remaining effects of interactions may eventually drive the system away from the fixed point towards a thermal state. 

\section{Conclusions and outlook}

In this chapter we have provided a brief introduction to the scenario of the formation of a Bose--Einstein condensate, and physics related to the dynamics of the BEC phase transition.  We have 
given a fairly comprehensive 
review of the experiments studying the formation of simple, single-component BECs in three-dimensional atomic gases,
with brief mentions of how such features are affected by reduced effective dimensionality or in cases where more than one condensate may co-exist. However,  the underlying physics described here is relevant to several other systems, most notably exciton-polaritons confined in strictly two-dimensional geometries featuring pumping and decay, where experiments on condensate formation have also been performed~\cite{nardin_lagoudakis_09,belykh_sibeldin_13,lagoudakis_manni_11}. 

An interesting question is what are the similarities and differences between these systems, and others such as BECs of photons~\cite{klaers_schmitt_10} and magnons~\cite{demokritov_demidov_06}.  Furthermore, what can phase transitions in quantum gases teach us about phase transitions that cannot be accessed experimentally, such as inflationary scenarios of early-universe evolution?  This was one of the motivating questions in the formulation of the Kibble--Zurek mechanism, as well as in the development of the theory of non-thermal fixed points.  It remains to be seen what we can learn about such matters as the formation of cosmological topological defects and (possibly)  
baryon asymmetry by studying nanokelvin gases here on earth.

\bibliography{Davis_CUP_References_combined2}

\begin{thebibliography}{169}
\expandafter\ifx\csname natexlab\endcsname\relax\def\natexlab#1{#1}\fi
\expandafter\ifx\csname bibnamefont\endcsname\relax
  \def\bibnamefont#1{#1}\fi
\expandafter\ifx\csname bibfnamefont\endcsname\relax
  \def\bibfnamefont#1{#1}\fi
\expandafter\ifx\csname citenamefont\endcsname\relax
  \def\citenamefont#1{#1}\fi
\expandafter\ifx\csname url\endcsname\relax
  \def\url#1{\texttt{#1}}\fi
\expandafter\ifx\csname urlprefix\endcsname\relax\def\urlprefix{URL }\fi
\providecommand{\bibinfo}[2]{#2}
\providecommand{\eprint}[2][]{\url{#2}}

\bibitem[{\citenamefont{Proukakis et~al.}(2013)\citenamefont{Proukakis,
  Gardiner, Davis, and Szyma\'{n}ska}}]{FINESS_book}
\bibinfo{editor}{\bibfnamefont{N.~P.} \bibnamefont{Proukakis}},
  \bibinfo{editor}{\bibfnamefont{S.~A.} \bibnamefont{Gardiner}},
  \bibinfo{editor}{\bibfnamefont{M.~J.} \bibnamefont{Davis}}, \bibnamefont{and}
  \bibinfo{editor}{\bibfnamefont{M.}~\bibnamefont{Szyma\'{n}ska}}, eds.,
  \emph{\bibinfo{title}{Quantum Gases: Finite Temperature and Non-Equilibrium
  Dynamics}} (\bibinfo{publisher}{Imperial College Press},
  \bibinfo{address}{London, UK}, \bibinfo{year}{2013}).

\bibitem[{\citenamefont{Griffin et~al.}(2009)\citenamefont{Griffin, Nikuni, and
  Zaremba}}]{griffin_nikuni_book_09}
\bibinfo{author}{\bibfnamefont{A.}~\bibnamefont{Griffin}},
  \bibinfo{author}{\bibfnamefont{T.}~\bibnamefont{Nikuni}}, \bibnamefont{and}
  \bibinfo{author}{\bibfnamefont{E.}~\bibnamefont{Zaremba}},
  \emph{\bibinfo{title}{{B}ose-{C}ondensed {G}ases at {F}inite {T}emperatures}}
  (\bibinfo{publisher}{Cambridge University Press},
  \bibinfo{address}{Cambridge, UK}, \bibinfo{year}{2009}).

\bibitem[{\citenamefont{Popov}(1972)}]{Popov1972}
\bibinfo{author}{\bibfnamefont{V.~N.} \bibnamefont{Popov}},
  \bibinfo{journal}{Theor. Math. Phys.} \textbf{\bibinfo{volume}{11}},
  \bibinfo{pages}{565} (\bibinfo{year}{1972}).

\bibitem[{\citenamefont{Popov}(1983)}]{popov_book_83}
\bibinfo{author}{\bibfnamefont{V.~N.} \bibnamefont{Popov}},
  \emph{\bibinfo{title}{Functional Integrals in Quantum Field Theory and
  Statistical Physics}} (\bibinfo{publisher}{Reidel},
  \bibinfo{address}{Dordrecht}, \bibinfo{year}{1983}).

\bibitem[{\citenamefont{Inoue and Hanamura}(1976)}]{Inoue1976a}
\bibinfo{author}{\bibfnamefont{A.}~\bibnamefont{Inoue}} \bibnamefont{and}
  \bibinfo{author}{\bibfnamefont{E.}~\bibnamefont{Hanamura}},
  \bibinfo{journal}{J. Phys. Soc. Jpn.} \textbf{\bibinfo{volume}{41}},
  \bibinfo{pages}{771} (\bibinfo{year}{1976}).

\bibitem[{\citenamefont{Levich and Yakhot}(1977)}]{levich_yakhot_77}
\bibinfo{author}{\bibfnamefont{E.}~\bibnamefont{Levich}} \bibnamefont{and}
  \bibinfo{author}{\bibfnamefont{V.}~\bibnamefont{Yakhot}},
  \bibinfo{journal}{Physical Review B} \textbf{\bibinfo{volume}{15}},
  \bibinfo{pages}{243} (\bibinfo{year}{1977}).

\bibitem[{\citenamefont{Levich and
  Yakhot}(1978{\natexlab{a}})}]{levich_yakhot_78a}
\bibinfo{author}{\bibfnamefont{E.}~\bibnamefont{Levich}} \bibnamefont{and}
  \bibinfo{author}{\bibfnamefont{V.}~\bibnamefont{Yakhot}},
  \bibinfo{journal}{J. Phys. A} \textbf{\bibinfo{volume}{11}},
  \bibinfo{pages}{2237} (\bibinfo{year}{1978}{\natexlab{a}}).

\bibitem[{\citenamefont{Levich and
  Yakhot}(1978{\natexlab{b}})}]{levich_yakhot_78b}
\bibinfo{author}{\bibfnamefont{E.}~\bibnamefont{Levich}} \bibnamefont{and}
  \bibinfo{author}{\bibfnamefont{V.}~\bibnamefont{Yakhot}},
  \bibinfo{journal}{J. Low. Temp. Phys.} \textbf{\bibinfo{volume}{27}},
  \bibinfo{pages}{107} (\bibinfo{year}{1978}{\natexlab{b}}).

\bibitem[{\citenamefont{Zeldovich and Levich}(1969)}]{Zeldovich1968a}
\bibinfo{author}{\bibfnamefont{Y.~B.} \bibnamefont{Zeldovich}}
  \bibnamefont{and} \bibinfo{author}{\bibfnamefont{E.~V.}
  \bibnamefont{Levich}}, \bibinfo{journal}{[Zh. Eksp. Teor. Fiz. 55, 2423
  (1968)] Sov. Phys. JETP} \textbf{\bibinfo{volume}{28}}, \bibinfo{pages}{1287}
  (\bibinfo{year}{1969}).

\bibitem[{\citenamefont{Pitaevskii and
  Stringari}(2003)}]{pitaevskii_stringari_book_03}
\bibinfo{author}{\bibfnamefont{L.~P.} \bibnamefont{Pitaevskii}}
  \bibnamefont{and}
  \bibinfo{author}{\bibfnamefont{S.}~\bibnamefont{Stringari}},
  \emph{\bibinfo{title}{{B}ose--{E}instein Condensation}}
  (\bibinfo{publisher}{Clarendon Press}, \bibinfo{address}{Oxford, UK},
  \bibinfo{year}{2003}).

\bibitem[{\citenamefont{Tikhodeev}(1990)}]{Tikhodeev1990a}
\bibinfo{author}{\bibfnamefont{S.~G.} \bibnamefont{Tikhodeev}},
  \bibinfo{journal}{[Zh. Eksp. Teor. Fiz. 97, 681 (1990)] Sov. Phys. JETP}
  \textbf{\bibinfo{volume}{70}}, \bibinfo{pages}{380} (\bibinfo{year}{1990}).

\bibitem[{\citenamefont{Eckern}(1984)}]{eckern_84}
\bibinfo{author}{\bibfnamefont{U.}~\bibnamefont{Eckern}}, \bibinfo{journal}{J.
  Low. Temp. Phys.} \textbf{\bibinfo{volume}{54}}, \bibinfo{pages}{333}
  (\bibinfo{year}{1984}).

\bibitem[{\citenamefont{Snoke and Wolfe}(1989)}]{snoke_wolfe_89}
\bibinfo{author}{\bibfnamefont{D.~W.} \bibnamefont{Snoke}} \bibnamefont{and}
  \bibinfo{author}{\bibfnamefont{J.~P.} \bibnamefont{Wolfe}},
  \bibinfo{journal}{Phys. Rev. B} \textbf{\bibinfo{volume}{39}},
  \bibinfo{pages}{4030} (\bibinfo{year}{1989}).

\bibitem[{\citenamefont{Stoof}(1991)}]{stoof_91}
\bibinfo{author}{\bibfnamefont{H.~T.~C.} \bibnamefont{Stoof}},
  \bibinfo{journal}{Phys. Rev. Lett.} \textbf{\bibinfo{volume}{66}},
  \bibinfo{pages}{3148} (\bibinfo{year}{1991}).

\bibitem[{\citenamefont{Stoof}(1992)}]{stoof_92}
\bibinfo{author}{\bibfnamefont{H.~T.~C.} \bibnamefont{Stoof}},
  \bibinfo{journal}{Phys. Rev. A} \textbf{\bibinfo{volume}{45}},
  \bibinfo{pages}{8398} (\bibinfo{year}{1992}).

\bibitem[{\citenamefont{Stoof}(1995)}]{stoof_95}
\bibinfo{author}{\bibfnamefont{H.~T.~C.} \bibnamefont{Stoof}},
  \emph{\bibinfo{title}{Bose-{E}instein Condensation}}
  (\bibinfo{publisher}{Cambridge University Press}, \bibinfo{year}{1995}),
  chap. \bibinfo{chapter}{Condensate Formation in a {B}ose Gas}, p.
  \bibinfo{pages}{226}.

\bibitem[{\citenamefont{Stoof}(1997)}]{stoof_97}
\bibinfo{author}{\bibfnamefont{H.~T.~C.} \bibnamefont{Stoof}},
  \bibinfo{journal}{Phys. Rev. Lett.} \textbf{\bibinfo{volume}{78}},
  \bibinfo{pages}{768} (\bibinfo{year}{1997}).

\bibitem[{\citenamefont{Stoof}(1999)}]{stoof_99}
\bibinfo{author}{\bibfnamefont{H.~T.~C.} \bibnamefont{Stoof}},
  \bibinfo{journal}{J. Low Temp. Phys.} \textbf{\bibinfo{volume}{114}},
  \bibinfo{pages}{11} (\bibinfo{year}{1999}).

\bibitem[{\citenamefont{Svistunov}(1991)}]{svistunov_91}
\bibinfo{author}{\bibfnamefont{B.~V.} \bibnamefont{Svistunov}},
  \bibinfo{journal}{J.~Mosc.~Phys.~Soc.} \textbf{\bibinfo{volume}{1}},
  \bibinfo{pages}{373} (\bibinfo{year}{1991}).

\bibitem[{\citenamefont{Kagan et~al.}(1992)\citenamefont{Kagan, Svistunov, and
  Shlyapnikov}}]{kagan_svistunov_92}
\bibinfo{author}{\bibfnamefont{Y.}~\bibnamefont{Kagan}},
  \bibinfo{author}{\bibfnamefont{B.~V.} \bibnamefont{Svistunov}},
  \bibnamefont{and} \bibinfo{author}{\bibfnamefont{G.~V.}
  \bibnamefont{Shlyapnikov}}, \bibinfo{journal}{Zh.~\'Eksp.~Teor.~Fiz.}
  \textbf{\bibinfo{volume}{101}}, \bibinfo{pages}{528} (\bibinfo{year}{1992}),
  \bibinfo{note}{[Sov. Phys. JETP \textbf{75}, 387 (1992)]}.

\bibitem[{\citenamefont{Kagan and Svistunov}(1994)}]{kagan_svistunov_94}
\bibinfo{author}{\bibfnamefont{Y.}~\bibnamefont{Kagan}} \bibnamefont{and}
  \bibinfo{author}{\bibfnamefont{B.~V.} \bibnamefont{Svistunov}},
  \bibinfo{journal}{Zh.~\'Eksp.~Teor.~Fiz.} \textbf{\bibinfo{volume}{105}},
  \bibinfo{pages}{353} (\bibinfo{year}{1994}), \bibinfo{note}{[Sov. Phys. JETP
  \textbf{78}, 187 (1994)]}.

\bibitem[{\citenamefont{Kagan}(1995)}]{kagan_95}
\bibinfo{author}{\bibfnamefont{Y.}~\bibnamefont{Kagan}},
  \emph{\bibinfo{title}{Bose-{E}instein Condensation}}
  (\bibinfo{publisher}{Cambridge University Press}, \bibinfo{year}{1995}),
  chap. \bibinfo{chapter}{Kinetics of {B}ose-{E}instein Condensate Formation in
  an Interacting {B}ose Gas}, p. \bibinfo{pages}{202}.

\bibitem[{\citenamefont{Sieberer et~al.}(2013)\citenamefont{Sieberer, Huber,
  Altman, and Diehl}}]{sieberer_huber_13}
\bibinfo{author}{\bibfnamefont{L.~M.} \bibnamefont{Sieberer}},
  \bibinfo{author}{\bibfnamefont{S.~D.} \bibnamefont{Huber}},
  \bibinfo{author}{\bibfnamefont{E.}~\bibnamefont{Altman}}, \bibnamefont{and}
  \bibinfo{author}{\bibfnamefont{S.}~\bibnamefont{Diehl}},
  \bibinfo{journal}{Phys. Rev. Lett.} \textbf{\bibinfo{volume}{110}},
  \bibinfo{pages}{195301} (\bibinfo{year}{2013}).

\bibitem[{\citenamefont{Altman et~al.}(2015)\citenamefont{Altman, Sieberer,
  Chen, Diehl, and Toner}}]{altman_sieberer_15}
\bibinfo{author}{\bibfnamefont{E.}~\bibnamefont{Altman}},
  \bibinfo{author}{\bibfnamefont{L.~M.} \bibnamefont{Sieberer}},
  \bibinfo{author}{\bibfnamefont{L.}~\bibnamefont{Chen}},
  \bibinfo{author}{\bibfnamefont{S.}~\bibnamefont{Diehl}}, \bibnamefont{and}
  \bibinfo{author}{\bibfnamefont{J.}~\bibnamefont{Toner}},
  \bibinfo{journal}{Phys. Rev. X} \textbf{\bibinfo{volume}{5}},
  \bibinfo{pages}{011017} (\bibinfo{year}{2015}).

\bibitem[{\citenamefont{Dagvadorj et~al.}(2015)\citenamefont{Dagvadorj,
  Fellows, Matyjaskiewicz, Marchetti, Carusotto, and
  Szymanska}}]{dagvadorj_fellows_14}
\bibinfo{author}{\bibfnamefont{G.}~\bibnamefont{Dagvadorj}},
  \bibinfo{author}{\bibfnamefont{J.~M.} \bibnamefont{Fellows}},
  \bibinfo{author}{\bibfnamefont{S.}~\bibnamefont{Matyjaskiewicz}},
  \bibinfo{author}{\bibfnamefont{F.~M.} \bibnamefont{Marchetti}},
  \bibinfo{author}{\bibfnamefont{I.}~\bibnamefont{Carusotto}},
  \bibnamefont{and} \bibinfo{author}{\bibfnamefont{M.~H.}
  \bibnamefont{Szymanska}}, \bibinfo{journal}{Phys. Rev. X}
  \textbf{\bibinfo{volume}{5}}, \bibinfo{pages}{041028} (\bibinfo{year}{2015}).

\bibitem[{\citenamefont{Zakharov et~al.}(1992)\citenamefont{Zakharov, {L'vov},
  and Falkovich}}]{Zakharov1992a}
\bibinfo{author}{\bibfnamefont{V.~E.} \bibnamefont{Zakharov}},
  \bibinfo{author}{\bibfnamefont{V.~S.} \bibnamefont{{L'vov}}},
  \bibnamefont{and}
  \bibinfo{author}{\bibfnamefont{G.}~\bibnamefont{Falkovich}},
  \emph{\bibinfo{title}{Kolmogorov Spectra of Turbulence I: Wave Turbulence}}
  (\bibinfo{publisher}{Springer, Berlin}, \bibinfo{year}{1992}).

\bibitem[{\citenamefont{Semikoz and
  Tkachev}(1995)}]{Semikoz1995a.PhysRevLett.74.3093}
\bibinfo{author}{\bibfnamefont{D.~V.} \bibnamefont{Semikoz}} \bibnamefont{and}
  \bibinfo{author}{\bibfnamefont{I.~I.} \bibnamefont{Tkachev}},
  \bibinfo{journal}{Phys. Rev. Lett.} \textbf{\bibinfo{volume}{74}},
  \bibinfo{pages}{3093} (\bibinfo{year}{1995}).

\bibitem[{\citenamefont{Semikoz and Tkachev}(1997)}]{Semikoz1997a}
\bibinfo{author}{\bibfnamefont{D.~V.} \bibnamefont{Semikoz}} \bibnamefont{and}
  \bibinfo{author}{\bibfnamefont{I.~I.} \bibnamefont{Tkachev}},
  \bibinfo{journal}{Phys. Rev. D} \textbf{\bibinfo{volume}{55}},
  \bibinfo{pages}{489} (\bibinfo{year}{1997}).

\bibitem[{\citenamefont{Berloff and Svistunov}(2002)}]{berloff_svistunov_02}
\bibinfo{author}{\bibfnamefont{N.~G.} \bibnamefont{Berloff}} \bibnamefont{and}
  \bibinfo{author}{\bibfnamefont{B.~V.} \bibnamefont{Svistunov}},
  \bibinfo{journal}{Phys. Rev. A} \textbf{\bibinfo{volume}{66}},
  \bibinfo{pages}{013603} (\bibinfo{year}{2002}).

\bibitem[{\citenamefont{{Nowak} et~al.}(2014)\citenamefont{{Nowak}, {Schole},
  and {Gasenzer}}}]{Nowak:2012gd}
\bibinfo{author}{\bibfnamefont{B.}~\bibnamefont{{Nowak}}},
  \bibinfo{author}{\bibfnamefont{J.}~\bibnamefont{{Schole}}}, \bibnamefont{and}
  \bibinfo{author}{\bibfnamefont{T.}~\bibnamefont{{Gasenzer}}},
  \bibinfo{journal}{New J. Phys.} \textbf{\bibinfo{volume}{16}},
  \bibinfo{pages}{093052} (\bibinfo{year}{2014}).

\bibitem[{\citenamefont{Schwarz}(1988)}]{Schwarz1988a}
\bibinfo{author}{\bibfnamefont{K.~W.} \bibnamefont{Schwarz}},
  \bibinfo{journal}{Phys. Rev. B} \textbf{\bibinfo{volume}{38}},
  \bibinfo{pages}{2398} (\bibinfo{year}{1988}).

\bibitem[{\citenamefont{Kozik and
  Svistunov}(2004)}]{Kozik2004a.PhysRevLett.92.035301}
\bibinfo{author}{\bibfnamefont{E.}~\bibnamefont{Kozik}} \bibnamefont{and}
  \bibinfo{author}{\bibfnamefont{B.}~\bibnamefont{Svistunov}},
  \bibinfo{journal}{Phys. Rev. Lett.} \textbf{\bibinfo{volume}{92}},
  \bibinfo{pages}{035301} (\bibinfo{year}{2004}).

\bibitem[{\citenamefont{Kozik and
  Svistunov}(2005{\natexlab{a}})}]{Kozik2005a.PhysRevLett.94.025301}
\bibinfo{author}{\bibfnamefont{E.}~\bibnamefont{Kozik}} \bibnamefont{and}
  \bibinfo{author}{\bibfnamefont{B.}~\bibnamefont{Svistunov}},
  \bibinfo{journal}{Phys. Rev. Lett.} \textbf{\bibinfo{volume}{94}},
  \bibinfo{pages}{025301} (\bibinfo{year}{2005}{\natexlab{a}}).

\bibitem[{\citenamefont{Kozik and
  Svistunov}(2005{\natexlab{b}})}]{Kozik2005a.PhysRevB.72.172505}
\bibinfo{author}{\bibfnamefont{E.}~\bibnamefont{Kozik}} \bibnamefont{and}
  \bibinfo{author}{\bibfnamefont{B.}~\bibnamefont{Svistunov}},
  \bibinfo{journal}{Phys. Rev. B} \textbf{\bibinfo{volume}{72}},
  \bibinfo{pages}{172505} (\bibinfo{year}{2005}{\natexlab{b}}).

\bibitem[{\citenamefont{Kozik and Svistunov}(2009)}]{Kozik2009a}
\bibinfo{author}{\bibfnamefont{E.~V.} \bibnamefont{Kozik}} \bibnamefont{and}
  \bibinfo{author}{\bibfnamefont{B.~V.} \bibnamefont{Svistunov}},
  \bibinfo{journal}{J. Low Temp. Phys.} \textbf{\bibinfo{volume}{156}},
  \bibinfo{pages}{215} (\bibinfo{year}{2009}).

\bibitem[{\citenamefont{Anderson et~al.}(1995)\citenamefont{Anderson, Ensher,
  Matthews, Wieman, and Cornell}}]{anderson_ensher_95}
\bibinfo{author}{\bibfnamefont{M.~H.} \bibnamefont{Anderson}},
  \bibinfo{author}{\bibfnamefont{J.~R.} \bibnamefont{Ensher}},
  \bibinfo{author}{\bibfnamefont{M.~R.} \bibnamefont{Matthews}},
  \bibinfo{author}{\bibfnamefont{C.~E.} \bibnamefont{Wieman}},
  \bibnamefont{and} \bibinfo{author}{\bibfnamefont{E.~A.}
  \bibnamefont{Cornell}}, \bibinfo{journal}{Science}
  \textbf{\bibinfo{volume}{269}}, \bibinfo{pages}{198} (\bibinfo{year}{1995}).

\bibitem[{\citenamefont{Davis et~al.}(1995)\citenamefont{Davis, Mewes, Andrews,
  van Druten, Durfee, Kurn, and Ketterle}}]{davis_mewes_95}
\bibinfo{author}{\bibfnamefont{K.~B.} \bibnamefont{Davis}},
  \bibinfo{author}{\bibfnamefont{M.~O.} \bibnamefont{Mewes}},
  \bibinfo{author}{\bibfnamefont{M.~R.} \bibnamefont{Andrews}},
  \bibinfo{author}{\bibfnamefont{N.~J.} \bibnamefont{van Druten}},
  \bibinfo{author}{\bibfnamefont{D.~S.} \bibnamefont{Durfee}},
  \bibinfo{author}{\bibfnamefont{D.~M.} \bibnamefont{Kurn}}, \bibnamefont{and}
  \bibinfo{author}{\bibfnamefont{W.}~\bibnamefont{Ketterle}},
  \bibinfo{journal}{Phys. Rev. Lett.} \textbf{\bibinfo{volume}{75}},
  \bibinfo{pages}{3969} (\bibinfo{year}{1995}).

\bibitem[{\citenamefont{Ketterle and {Van
  Druten}}(1996)}]{ketterle_vandruten_96}
\bibinfo{author}{\bibfnamefont{W.}~\bibnamefont{Ketterle}} \bibnamefont{and}
  \bibinfo{author}{\bibfnamefont{N.~J.} \bibnamefont{{Van Druten}}}
  (\bibinfo{publisher}{Academic Press}, \bibinfo{year}{1996}),
  vol.~\bibinfo{volume}{37} of \emph{\bibinfo{series}{Advances In Atomic,
  Molecular, and Optical Physics}}, p. \bibinfo{pages}{181}.

\bibitem[{\citenamefont{Gardiner and Zoller}(1997)}]{gardiner_zoller_97a}
\bibinfo{author}{\bibfnamefont{C.~W.} \bibnamefont{Gardiner}} \bibnamefont{and}
  \bibinfo{author}{\bibfnamefont{P.}~\bibnamefont{Zoller}},
  \bibinfo{journal}{Phys. Rev. A} \textbf{\bibinfo{volume}{55}},
  \bibinfo{pages}{2902} (\bibinfo{year}{1997}).

\bibitem[{\citenamefont{Gardiner and Zoller}(1998)}]{gardiner_zoller_98}
\bibinfo{author}{\bibfnamefont{C.~W.} \bibnamefont{Gardiner}} \bibnamefont{and}
  \bibinfo{author}{\bibfnamefont{P.}~\bibnamefont{Zoller}},
  \bibinfo{journal}{Phys. Rev. A} \textbf{\bibinfo{volume}{58}},
  \bibinfo{pages}{536} (\bibinfo{year}{1998}).

\bibitem[{\citenamefont{Gardiner and Zoller}(2000)}]{gardiner_zoller_00}
\bibinfo{author}{\bibfnamefont{C.~W.} \bibnamefont{Gardiner}} \bibnamefont{and}
  \bibinfo{author}{\bibfnamefont{P.}~\bibnamefont{Zoller}},
  \bibinfo{journal}{Phys. Rev. A} \textbf{\bibinfo{volume}{61}},
  \bibinfo{pages}{033601} (\bibinfo{year}{2000}).

\bibitem[{\citenamefont{Gardiner and Zoller}(2004)}]{gardiner_zoller_book_04}
\bibinfo{author}{\bibfnamefont{C.~W.} \bibnamefont{Gardiner}} \bibnamefont{and}
  \bibinfo{author}{\bibfnamefont{P.}~\bibnamefont{Zoller}},
  \emph{\bibinfo{title}{Quantum Noise}} (\bibinfo{publisher}{Springer-Verlag},
  \bibinfo{address}{Berlin Heidelberg}, \bibinfo{year}{2004}),
  \bibinfo{edition}{3rd} ed.

\bibitem[{\citenamefont{Miesner et~al.}(1998)\citenamefont{Miesner,
  Stamper-Kurn, Andrews, Durfee, Inouye, and
  Ketterle}}]{miesner_stamper-kurn_98}
\bibinfo{author}{\bibfnamefont{H.-J.} \bibnamefont{Miesner}},
  \bibinfo{author}{\bibfnamefont{D.~M.} \bibnamefont{Stamper-Kurn}},
  \bibinfo{author}{\bibfnamefont{M.~R.} \bibnamefont{Andrews}},
  \bibinfo{author}{\bibfnamefont{D.~S.} \bibnamefont{Durfee}},
  \bibinfo{author}{\bibfnamefont{S.}~\bibnamefont{Inouye}}, \bibnamefont{and}
  \bibinfo{author}{\bibfnamefont{W.}~\bibnamefont{Ketterle}},
  \bibinfo{journal}{Science} \textbf{\bibinfo{volume}{279}},
  \bibinfo{pages}{1005} (\bibinfo{year}{1998}).

\bibitem[{\citenamefont{Gardiner et~al.}(1997)\citenamefont{Gardiner, Zoller,
  Ballagh, and Davis}}]{gardiner_zoller_97b}
\bibinfo{author}{\bibfnamefont{C.~W.} \bibnamefont{Gardiner}},
  \bibinfo{author}{\bibfnamefont{P.}~\bibnamefont{Zoller}},
  \bibinfo{author}{\bibfnamefont{R.~J.} \bibnamefont{Ballagh}},
  \bibnamefont{and} \bibinfo{author}{\bibfnamefont{M.~J.} \bibnamefont{Davis}},
  \bibinfo{journal}{Phys. Rev. Lett.} \textbf{\bibinfo{volume}{79}},
  \bibinfo{pages}{1793} (\bibinfo{year}{1997}).

\bibitem[{\citenamefont{Gardiner et~al.}(1998)\citenamefont{Gardiner, Lee,
  Ballagh, Davis, and Zoller}}]{gardiner_lee_98}
\bibinfo{author}{\bibfnamefont{C.~W.} \bibnamefont{Gardiner}},
  \bibinfo{author}{\bibfnamefont{M.~D.} \bibnamefont{Lee}},
  \bibinfo{author}{\bibfnamefont{R.~J.} \bibnamefont{Ballagh}},
  \bibinfo{author}{\bibfnamefont{M.~J.} \bibnamefont{Davis}}, \bibnamefont{and}
  \bibinfo{author}{\bibfnamefont{P.}~\bibnamefont{Zoller}},
  \bibinfo{journal}{Phys. Rev. Lett.} \textbf{\bibinfo{volume}{81}},
  \bibinfo{pages}{5266} (\bibinfo{year}{1998}).

\bibitem[{\citenamefont{Lee and Gardiner}(2000)}]{lee_gardiner_00}
\bibinfo{author}{\bibfnamefont{M.~D.} \bibnamefont{Lee}} \bibnamefont{and}
  \bibinfo{author}{\bibfnamefont{C.~W.} \bibnamefont{Gardiner}},
  \bibinfo{journal}{Phys. Rev. A} \textbf{\bibinfo{volume}{62}},
  \bibinfo{pages}{033606} (\bibinfo{year}{2000}).

\bibitem[{\citenamefont{Davis et~al.}(2000)\citenamefont{Davis, Gardiner, and
  Ballagh}}]{davis_gardiner_00}
\bibinfo{author}{\bibfnamefont{M.~J.} \bibnamefont{Davis}},
  \bibinfo{author}{\bibfnamefont{C.~W.} \bibnamefont{Gardiner}},
  \bibnamefont{and} \bibinfo{author}{\bibfnamefont{R.~J.}
  \bibnamefont{Ballagh}}, \bibinfo{journal}{Phys. Rev. A}
  \textbf{\bibinfo{volume}{62}}, \bibinfo{pages}{063608}
  (\bibinfo{year}{2000}).

\bibitem[{\citenamefont{Bijlsma et~al.}(2000)\citenamefont{Bijlsma, Zaremba,
  and Stoof}}]{bijlsma_zaremba_00}
\bibinfo{author}{\bibfnamefont{M.~J.} \bibnamefont{Bijlsma}},
  \bibinfo{author}{\bibfnamefont{E.}~\bibnamefont{Zaremba}}, \bibnamefont{and}
  \bibinfo{author}{\bibfnamefont{H.~T.~C.} \bibnamefont{Stoof}},
  \bibinfo{journal}{Phys. Rev. A} \textbf{\bibinfo{volume}{62}},
  \bibinfo{pages}{063609} (\bibinfo{year}{2000}).

\bibitem[{\citenamefont{Zaremba et~al.}(1999)\citenamefont{Zaremba, Nikuni, and
  Griffin}}]{zaremba_nikuni_99}
\bibinfo{author}{\bibfnamefont{E.}~\bibnamefont{Zaremba}},
  \bibinfo{author}{\bibfnamefont{T.}~\bibnamefont{Nikuni}}, \bibnamefont{and}
  \bibinfo{author}{\bibfnamefont{A.}~\bibnamefont{Griffin}},
  \bibinfo{journal}{J. Low. Temp. Phys.} \textbf{\bibinfo{volume}{116}},
  \bibinfo{pages}{277} (\bibinfo{year}{1999}).

\bibitem[{\citenamefont{Zaremba et~al.}(1998)\citenamefont{Zaremba, Griffin,
  and Nikuni}}]{zaremba_griffin_98}
\bibinfo{author}{\bibfnamefont{E.}~\bibnamefont{Zaremba}},
  \bibinfo{author}{\bibfnamefont{A.}~\bibnamefont{Griffin}}, \bibnamefont{and}
  \bibinfo{author}{\bibfnamefont{T.}~\bibnamefont{Nikuni}},
  \bibinfo{journal}{Phys. Rev. A} \textbf{\bibinfo{volume}{57}},
  \bibinfo{pages}{4695} (\bibinfo{year}{1998}).

\bibitem[{\citenamefont{Kirkpatrick and
  Dorfman}(1983)}]{kirkpatrick_dorfman_83}
\bibinfo{author}{\bibfnamefont{T.~R.} \bibnamefont{Kirkpatrick}}
  \bibnamefont{and} \bibinfo{author}{\bibfnamefont{J.~R.}
  \bibnamefont{Dorfman}}, \bibinfo{journal}{Phys. Rev. A}
  \textbf{\bibinfo{volume}{28}}, \bibinfo{pages}{2576} (\bibinfo{year}{1983}).

\bibitem[{\citenamefont{Kirkpatrick and
  Dorfman}(1985{\natexlab{a}})}]{kirkpatrick_dorfman_85a}
\bibinfo{author}{\bibfnamefont{T.~R.} \bibnamefont{Kirkpatrick}}
  \bibnamefont{and} \bibinfo{author}{\bibfnamefont{J.~R.}
  \bibnamefont{Dorfman}}, \bibinfo{journal}{J. Low Temp. Phys.}
  \textbf{\bibinfo{volume}{58}}, \bibinfo{pages}{399}
  (\bibinfo{year}{1985}{\natexlab{a}}).

\bibitem[{\citenamefont{Kirkpatrick and
  Dorfman}(1985{\natexlab{b}})}]{kirkpatrick_dorfman_85b}
\bibinfo{author}{\bibfnamefont{T.~R.} \bibnamefont{Kirkpatrick}}
  \bibnamefont{and} \bibinfo{author}{\bibfnamefont{J.~R.}
  \bibnamefont{Dorfman}}, \bibinfo{journal}{J. Low Temp. Phys.}
  \textbf{\bibinfo{volume}{58}}, \bibinfo{pages}{301}
  (\bibinfo{year}{1985}{\natexlab{b}}).

\bibitem[{\citenamefont{Kirkpatrick and
  Dorfman}(1985{\natexlab{c}})}]{kirkpatrick_dorfman_85c}
\bibinfo{author}{\bibfnamefont{T.~R.} \bibnamefont{Kirkpatrick}}
  \bibnamefont{and} \bibinfo{author}{\bibfnamefont{J.~R.}
  \bibnamefont{Dorfman}}, \bibinfo{journal}{J. Low Temp. Phys.}
  \textbf{\bibinfo{volume}{59}}, \bibinfo{pages}{1}
  (\bibinfo{year}{1985}{\natexlab{c}}).

\bibitem[{\citenamefont{Luiten et~al.}(1996)\citenamefont{Luiten, Reynolds, and
  Walraven}}]{luiten_reynolds_96}
\bibinfo{author}{\bibfnamefont{O.~J.} \bibnamefont{Luiten}},
  \bibinfo{author}{\bibfnamefont{M.~W.} \bibnamefont{Reynolds}},
  \bibnamefont{and} \bibinfo{author}{\bibfnamefont{J.~T.~M.}
  \bibnamefont{Walraven}}, \bibinfo{journal}{Phys. Rev. A}
  \textbf{\bibinfo{volume}{53}}, \bibinfo{pages}{381} (\bibinfo{year}{1996}).

\bibitem[{\citenamefont{K\"{o}hl et~al.}(2002)\citenamefont{K\"{o}hl, Davis,
  Gardiner, H\"{a}nsch, and Esslinger}}]{kohl_davis_02}
\bibinfo{author}{\bibfnamefont{M.}~\bibnamefont{K\"{o}hl}},
  \bibinfo{author}{\bibfnamefont{M.~J.} \bibnamefont{Davis}},
  \bibinfo{author}{\bibfnamefont{C.~W.} \bibnamefont{Gardiner}},
  \bibinfo{author}{\bibfnamefont{T.~W.} \bibnamefont{H\"{a}nsch}},
  \bibnamefont{and} \bibinfo{author}{\bibfnamefont{T.~W.}
  \bibnamefont{Esslinger}}, \bibinfo{journal}{Phys. Rev. Lett.}
  \textbf{\bibinfo{volume}{88}}, \bibinfo{pages}{080402}
  (\bibinfo{year}{2002}).

\bibitem[{\citenamefont{Davis and Gardiner}(2002)}]{davis_gardiner_02}
\bibinfo{author}{\bibfnamefont{M.~J.} \bibnamefont{Davis}} \bibnamefont{and}
  \bibinfo{author}{\bibfnamefont{C.~W.} \bibnamefont{Gardiner}},
  \bibinfo{journal}{J. Phys. B: At. Mol. Opt. Phys.}
  \textbf{\bibinfo{volume}{35}}, \bibinfo{pages}{733} (\bibinfo{year}{2002}).

\bibitem[{\citenamefont{Pinske et~al.}(1997)\citenamefont{Pinske, Mosk,
  Weidem{\"u}ller, Reynolds, Hijmans, and Walraven}}]{pinkse_mosk_97}
\bibinfo{author}{\bibfnamefont{P.~W.~H.} \bibnamefont{Pinske}},
  \bibinfo{author}{\bibfnamefont{A.}~\bibnamefont{Mosk}},
  \bibinfo{author}{\bibfnamefont{M.}~\bibnamefont{Weidem{\"u}ller}},
  \bibinfo{author}{\bibfnamefont{M.~W.} \bibnamefont{Reynolds}},
  \bibinfo{author}{\bibfnamefont{T.~W.} \bibnamefont{Hijmans}},
  \bibnamefont{and} \bibinfo{author}{\bibfnamefont{J.~T.~M.}
  \bibnamefont{Walraven}}, \bibinfo{journal}{Phys. Rev. Lett.}
  \textbf{\bibinfo{volume}{78}}, \bibinfo{pages}{990} (\bibinfo{year}{1997}).

\bibitem[{\citenamefont{Stamper-Kurn et~al.}(1998)\citenamefont{Stamper-Kurn,
  Miesner, Chikkatur, Inouye, Stenger, and Ketterle}}]{stamper-kurn_miesner_98}
\bibinfo{author}{\bibfnamefont{D.~M.} \bibnamefont{Stamper-Kurn}},
  \bibinfo{author}{\bibfnamefont{H.-J.} \bibnamefont{Miesner}},
  \bibinfo{author}{\bibfnamefont{A.~P.} \bibnamefont{Chikkatur}},
  \bibinfo{author}{\bibfnamefont{S.}~\bibnamefont{Inouye}},
  \bibinfo{author}{\bibfnamefont{J.}~\bibnamefont{Stenger}}, \bibnamefont{and}
  \bibinfo{author}{\bibfnamefont{W.}~\bibnamefont{Ketterle}},
  \bibinfo{journal}{Phys. Rev. Lett.} \textbf{\bibinfo{volume}{81}},
  \bibinfo{pages}{2194} (\bibinfo{year}{1998}).

\bibitem[{\citenamefont{Stoof and Bijlsma}(2001)}]{stoof_01}
\bibinfo{author}{\bibfnamefont{H.~T.~C.} \bibnamefont{Stoof}} \bibnamefont{and}
  \bibinfo{author}{\bibfnamefont{M.~J.} \bibnamefont{Bijlsma}},
  \bibinfo{journal}{J. Low. Temp. Phys.} \textbf{\bibinfo{volume}{124}},
  \bibinfo{pages}{431} (\bibinfo{year}{2001}).

\bibitem[{\citenamefont{Proukakis et~al.}(2006)\citenamefont{Proukakis,
  Schmiedmayer, and Stoof}}]{proukakis_schmiedmayer_06}
\bibinfo{author}{\bibfnamefont{N.~P.} \bibnamefont{Proukakis}},
  \bibinfo{author}{\bibfnamefont{J.}~\bibnamefont{Schmiedmayer}},
  \bibnamefont{and} \bibinfo{author}{\bibfnamefont{H.~T.~C.}
  \bibnamefont{Stoof}}, \bibinfo{journal}{Phys. Rev. A}
  \textbf{\bibinfo{volume}{73}}, \bibinfo{pages}{053603}
  (\bibinfo{year}{2006}).

\bibitem[{\citenamefont{Garrett et~al.}(2011)\citenamefont{Garrett, Ratnapala,
  van Ooijen, Vale, Weegink, Schnelle, Vainio, Heckenberg, Rubinsztein-Dunlop,
  and Davis}}]{garrett_ratnapala_11}
\bibinfo{author}{\bibfnamefont{M.~C.} \bibnamefont{Garrett}},
  \bibinfo{author}{\bibfnamefont{A.}~\bibnamefont{Ratnapala}},
  \bibinfo{author}{\bibfnamefont{E.~D.} \bibnamefont{van Ooijen}},
  \bibinfo{author}{\bibfnamefont{C.~J.} \bibnamefont{Vale}},
  \bibinfo{author}{\bibfnamefont{K.}~\bibnamefont{Weegink}},
  \bibinfo{author}{\bibfnamefont{S.~K.} \bibnamefont{Schnelle}},
  \bibinfo{author}{\bibfnamefont{O.}~\bibnamefont{Vainio}},
  \bibinfo{author}{\bibfnamefont{N.~R.} \bibnamefont{Heckenberg}},
  \bibinfo{author}{\bibfnamefont{H.}~\bibnamefont{Rubinsztein-Dunlop}},
  \bibnamefont{and} \bibinfo{author}{\bibfnamefont{M.~J.} \bibnamefont{Davis}},
  \bibinfo{journal}{Phys. Rev. A} \textbf{\bibinfo{volume}{83}},
  \bibinfo{pages}{013630} (\bibinfo{year}{2011}).

\bibitem[{\citenamefont{Harber et~al.}(2003)\citenamefont{Harber, McGuirk,
  Obrecht, and Cornell}}]{harber_mcguirk_03}
\bibinfo{author}{\bibfnamefont{D.~M.} \bibnamefont{Harber}},
  \bibinfo{author}{\bibfnamefont{J.~M.} \bibnamefont{McGuirk}},
  \bibinfo{author}{\bibfnamefont{J.~M.} \bibnamefont{Obrecht}},
  \bibnamefont{and} \bibinfo{author}{\bibfnamefont{E.~A.}
  \bibnamefont{Cornell}}, \bibinfo{journal}{J. Low Temp. Phys.}
  \textbf{\bibinfo{volume}{133}}, \bibinfo{pages}{229} (\bibinfo{year}{2003}).

\bibitem[{\citenamefont{Marchant et~al.}(2011)\citenamefont{Marchant, H\:andel,
  Wiles, Hopkins, and Cornish}}]{marchant_handel_11}
\bibinfo{author}{\bibfnamefont{A.~L.} \bibnamefont{Marchant}},
  \bibinfo{author}{\bibfnamefont{S.}~\bibnamefont{H\:andel}},
  \bibinfo{author}{\bibfnamefont{T.~P.} \bibnamefont{Wiles}},
  \bibinfo{author}{\bibfnamefont{S.~A.} \bibnamefont{Hopkins}},
  \bibnamefont{and} \bibinfo{author}{\bibfnamefont{S.~L.}
  \bibnamefont{Cornish}}, \bibinfo{journal}{New J. Phys.}
  \textbf{\bibinfo{volume}{13}}, \bibinfo{pages}{125003}
  (\bibinfo{year}{2011}).

\bibitem[{\citenamefont{M\"arkle et~al.}(2014)\citenamefont{M\"arkle, Allen,
  Federsel, Jetter, G\"unther, Fort\'agh, Proukakis, and
  Judd}}]{markle_allen_15}
\bibinfo{author}{\bibfnamefont{J.}~\bibnamefont{M\"arkle}},
  \bibinfo{author}{\bibfnamefont{A.~J.} \bibnamefont{Allen}},
  \bibinfo{author}{\bibfnamefont{P.}~\bibnamefont{Federsel}},
  \bibinfo{author}{\bibfnamefont{B.}~\bibnamefont{Jetter}},
  \bibinfo{author}{\bibfnamefont{A.}~\bibnamefont{G\"unther}},
  \bibinfo{author}{\bibfnamefont{J.}~\bibnamefont{Fort\'agh}},
  \bibinfo{author}{\bibfnamefont{N.~P.} \bibnamefont{Proukakis}},
  \bibnamefont{and} \bibinfo{author}{\bibfnamefont{T.~E.} \bibnamefont{Judd}},
  \bibinfo{journal}{Phys. Rev. A} \textbf{\bibinfo{volume}{90}},
  \bibinfo{pages}{023614} (\bibinfo{year}{2014}).

\bibitem[{\citenamefont{Imamovic-Tomasovic and
  Griffin}(2001)}]{imamovic-tomasovic_griffin_01}
\bibinfo{author}{\bibfnamefont{M.}~\bibnamefont{Imamovic-Tomasovic}}
  \bibnamefont{and} \bibinfo{author}{\bibfnamefont{A.}~\bibnamefont{Griffin}},
  \bibinfo{journal}{J. Low Temp. Phys.} \textbf{\bibinfo{volume}{122}},
  \bibinfo{pages}{616} (\bibinfo{year}{2001}).

\bibitem[{\citenamefont{Kadanoff and Baym}(1962)}]{kadanoff_baym_book_62}
\bibinfo{author}{\bibfnamefont{L.~P.} \bibnamefont{Kadanoff}} \bibnamefont{and}
  \bibinfo{author}{\bibfnamefont{G.}~\bibnamefont{Baym}},
  \emph{\bibinfo{title}{Quantum Statistical Mechanics}}
  (\bibinfo{publisher}{W.A. Benjamin}, \bibinfo{address}{Menlo Park, CA, USA},
  \bibinfo{year}{1962}).

\bibitem[{\citenamefont{Walser et~al.}(1999)\citenamefont{Walser, Williams,
  Cooper, and Holland}}]{walser_williams_99}
\bibinfo{author}{\bibfnamefont{R.}~\bibnamefont{Walser}},
  \bibinfo{author}{\bibfnamefont{J.}~\bibnamefont{Williams}},
  \bibinfo{author}{\bibfnamefont{J.}~\bibnamefont{Cooper}}, \bibnamefont{and}
  \bibinfo{author}{\bibfnamefont{M.}~\bibnamefont{Holland}},
  \bibinfo{journal}{Phys. Rev. A} \textbf{\bibinfo{volume}{59}},
  \bibinfo{pages}{3878} (\bibinfo{year}{1999}).

\bibitem[{\citenamefont{Walser et~al.}(2001)\citenamefont{Walser, Cooper, and
  Holland}}]{walser_cooper_01}
\bibinfo{author}{\bibfnamefont{R.}~\bibnamefont{Walser}},
  \bibinfo{author}{\bibfnamefont{J.}~\bibnamefont{Cooper}}, \bibnamefont{and}
  \bibinfo{author}{\bibfnamefont{M.}~\bibnamefont{Holland}},
  \bibinfo{journal}{Phys. Rev. A} \textbf{\bibinfo{volume}{63}},
  \bibinfo{pages}{013607} (\bibinfo{year}{2001}).

\bibitem[{\citenamefont{Wachter et~al.}(2001)\citenamefont{Wachter, Walser,
  Cooper, and Holland}}]{wachter_walser_01a}
\bibinfo{author}{\bibfnamefont{J.}~\bibnamefont{Wachter}},
  \bibinfo{author}{\bibfnamefont{R.}~\bibnamefont{Walser}},
  \bibinfo{author}{\bibfnamefont{J.}~\bibnamefont{Cooper}}, \bibnamefont{and}
  \bibinfo{author}{\bibfnamefont{M.}~\bibnamefont{Holland}},
  \bibinfo{journal}{Phys. Rev. A} \textbf{\bibinfo{volume}{64}},
  \bibinfo{pages}{053612} (\bibinfo{year}{2001}).

\bibitem[{\citenamefont{Proukakis}(2001)}]{proukakis_2001a}
\bibinfo{author}{\bibfnamefont{N.~P.} \bibnamefont{Proukakis}},
  \bibinfo{journal}{J. Phys. B: At. Mol. Opt. Phys.}
  \textbf{\bibinfo{volume}{34}}, \bibinfo{pages}{4737} (\bibinfo{year}{2001}).

\bibitem[{\citenamefont{Proukakis and Burnett}(1996)}]{proukakis_burnett_1996}
\bibinfo{author}{\bibfnamefont{N.~P.} \bibnamefont{Proukakis}}
  \bibnamefont{and} \bibinfo{author}{\bibfnamefont{K.}~\bibnamefont{Burnett}},
  \bibinfo{journal}{J. Res. Natl. Inst. Stand. Technol.}
  \textbf{\bibinfo{volume}{101}}, \bibinfo{pages}{457} (\bibinfo{year}{1996}).

\bibitem[{\citenamefont{Proukakis et~al.}(1998)\citenamefont{Proukakis,
  Burnett, and Stoof}}]{proukakis_burnett_1998}
\bibinfo{author}{\bibfnamefont{N.~P.} \bibnamefont{Proukakis}},
  \bibinfo{author}{\bibfnamefont{K.}~\bibnamefont{Burnett}}, \bibnamefont{and}
  \bibinfo{author}{\bibfnamefont{H.~T.~C.} \bibnamefont{Stoof}},
  \bibinfo{journal}{Phys. Rev. A} \textbf{\bibinfo{volume}{57}},
  \bibinfo{pages}{1230} (\bibinfo{year}{1998}).

\bibitem[{\citenamefont{Shi and Griffin}(1998)}]{shi_griffin_98}
\bibinfo{author}{\bibfnamefont{H.}~\bibnamefont{Shi}} \bibnamefont{and}
  \bibinfo{author}{\bibfnamefont{A.}~\bibnamefont{Griffin}},
  \bibinfo{journal}{Physics Reports} \textbf{\bibinfo{volume}{304}},
  \bibinfo{pages}{1 } (\bibinfo{year}{1998}), ISSN \bibinfo{issn}{0370-1573}.

\bibitem[{\citenamefont{Gasenzer et~al.}(2005)\citenamefont{Gasenzer, Berges,
  Schmidt, and Seco}}]{gasenzer_berges_05}
\bibinfo{author}{\bibfnamefont{T.}~\bibnamefont{Gasenzer}},
  \bibinfo{author}{\bibfnamefont{J.}~\bibnamefont{Berges}},
  \bibinfo{author}{\bibfnamefont{M.~G.} \bibnamefont{Schmidt}},
  \bibnamefont{and} \bibinfo{author}{\bibfnamefont{M.}~\bibnamefont{Seco}},
  \bibinfo{journal}{Phys. Rev. A} \textbf{\bibinfo{volume}{72}},
  \bibinfo{pages}{063604} (\bibinfo{year}{2005}).

\bibitem[{\citenamefont{Berges and Gasenzer}(2007)}]{berges_gasenzer_07}
\bibinfo{author}{\bibfnamefont{J.}~\bibnamefont{Berges}} \bibnamefont{and}
  \bibinfo{author}{\bibfnamefont{T.}~\bibnamefont{Gasenzer}},
  \bibinfo{journal}{Phys. Rev. A} \textbf{\bibinfo{volume}{76}},
  \bibinfo{pages}{033604} (\bibinfo{year}{2007}).

\bibitem[{\citenamefont{Bransch{\"a}del and
  Gasenzer}(2008)}]{branschadel_gasenzer_08}
\bibinfo{author}{\bibfnamefont{A.}~\bibnamefont{Bransch{\"a}del}}
  \bibnamefont{and} \bibinfo{author}{\bibfnamefont{T.}~\bibnamefont{Gasenzer}},
  \bibinfo{journal}{J. Phys. B: At. Mol. Opt. Phys.}
  \textbf{\bibinfo{volume}{41}}, \bibinfo{pages}{135302}
  (\bibinfo{year}{2008}).

\bibitem[{\citenamefont{{Bodet} et~al.}(2012)\citenamefont{{Bodet},
  {Kronenwett}, {Nowak}, {Sexty}, and {Gasenzer}}}]{Bodet2011a}
\bibinfo{author}{\bibfnamefont{C.}~\bibnamefont{{Bodet}}},
  \bibinfo{author}{\bibfnamefont{M.}~\bibnamefont{{Kronenwett}}},
  \bibinfo{author}{\bibfnamefont{B.}~\bibnamefont{{Nowak}}},
  \bibinfo{author}{\bibfnamefont{D.}~\bibnamefont{{Sexty}}}, \bibnamefont{and}
  \bibinfo{author}{\bibfnamefont{T.}~\bibnamefont{{Gasenzer}}},
  \emph{\bibinfo{title}{Quantum Gases: Finite Temperature and Non-Equilibrium
  Dynamics}} (\bibinfo{publisher}{Imperial College Press, London},
  \bibinfo{year}{2012}), chap. \bibinfo{chapter}{{Non-equilibrium Quantum
  Many-Body Dynamics: Functional Integral Approaches}}.

\bibitem[{\citenamefont{Berges et~al.}(2008)\citenamefont{Berges, Rothkopf, and
  Schmidt}}]{Berges:2008wm}
\bibinfo{author}{\bibfnamefont{J.}~\bibnamefont{Berges}},
  \bibinfo{author}{\bibfnamefont{A.}~\bibnamefont{Rothkopf}}, \bibnamefont{and}
  \bibinfo{author}{\bibfnamefont{J.}~\bibnamefont{Schmidt}},
  \bibinfo{journal}{Phys. Rev. Lett.} \textbf{\bibinfo{volume}{101}},
  \bibinfo{pages}{041603} (\bibinfo{year}{2008}).

\bibitem[{\citenamefont{Berges and Hoffmeister}(2009)}]{Berges:2008sr}
\bibinfo{author}{\bibfnamefont{J.}~\bibnamefont{Berges}} \bibnamefont{and}
  \bibinfo{author}{\bibfnamefont{G.}~\bibnamefont{Hoffmeister}},
  \bibinfo{journal}{Nucl. Phys.} \textbf{\bibinfo{volume}{B813}},
  \bibinfo{pages}{383} (\bibinfo{year}{2009}).

\bibitem[{\citenamefont{Scheppach et~al.}(2010)\citenamefont{Scheppach, Berges,
  and Gasenzer}}]{Scheppach:2009wu}
\bibinfo{author}{\bibfnamefont{C.}~\bibnamefont{Scheppach}},
  \bibinfo{author}{\bibfnamefont{J.}~\bibnamefont{Berges}}, \bibnamefont{and}
  \bibinfo{author}{\bibfnamefont{T.}~\bibnamefont{Gasenzer}},
  \bibinfo{journal}{Phys. Rev. A} \textbf{\bibinfo{volume}{81}},
  \bibinfo{pages}{033611} (\bibinfo{year}{2010}).

\bibitem[{\citenamefont{Kronenwett and Gasenzer}(2011)}]{Kronenwett:2010ic}
\bibinfo{author}{\bibfnamefont{M.}~\bibnamefont{Kronenwett}} \bibnamefont{and}
  \bibinfo{author}{\bibfnamefont{T.}~\bibnamefont{Gasenzer}},
  \bibinfo{journal}{Appl. Phys. B} \textbf{\bibinfo{volume}{102}},
  \bibinfo{pages}{469} (\bibinfo{year}{2011}).

\bibitem[{\citenamefont{Babadi et~al.}(2015)\citenamefont{Babadi, Demler, and
  Knap}}]{Babadi2015a.PhysRevX.5.041005}
\bibinfo{author}{\bibfnamefont{M.}~\bibnamefont{Babadi}},
  \bibinfo{author}{\bibfnamefont{E.}~\bibnamefont{Demler}}, \bibnamefont{and}
  \bibinfo{author}{\bibfnamefont{M.}~\bibnamefont{Knap}},
  \bibinfo{journal}{Phys. Rev. X} \textbf{\bibinfo{volume}{5}},
  \bibinfo{pages}{041005} (\bibinfo{year}{2015}).

\bibitem[{\citenamefont{Myatt et~al.}(1997)\citenamefont{Myatt, Burt, Ghrist,
  Cornell, and Wieman}}]{myatt_burt_97}
\bibinfo{author}{\bibfnamefont{C.~J.} \bibnamefont{Myatt}},
  \bibinfo{author}{\bibfnamefont{E.~A.} \bibnamefont{Burt}},
  \bibinfo{author}{\bibfnamefont{R.~W.} \bibnamefont{Ghrist}},
  \bibinfo{author}{\bibfnamefont{E.~A.} \bibnamefont{Cornell}},
  \bibnamefont{and} \bibinfo{author}{\bibfnamefont{C.~E.}
  \bibnamefont{Wieman}}, \bibinfo{journal}{Phys. Rev. Lett.}
  \textbf{\bibinfo{volume}{78}}, \bibinfo{pages}{586} (\bibinfo{year}{1997}).

\bibitem[{\citenamefont{Schreck et~al.}(2001)\citenamefont{Schreck, Ferrari,
  Corwin, Cubizolles, Khaykovich, Mewes, and Salomon}}]{schreck_ferrari_01}
\bibinfo{author}{\bibfnamefont{F.}~\bibnamefont{Schreck}},
  \bibinfo{author}{\bibfnamefont{G.}~\bibnamefont{Ferrari}},
  \bibinfo{author}{\bibfnamefont{K.~L.} \bibnamefont{Corwin}},
  \bibinfo{author}{\bibfnamefont{J.}~\bibnamefont{Cubizolles}},
  \bibinfo{author}{\bibfnamefont{L.}~\bibnamefont{Khaykovich}},
  \bibinfo{author}{\bibfnamefont{M.-O.} \bibnamefont{Mewes}}, \bibnamefont{and}
  \bibinfo{author}{\bibfnamefont{C.}~\bibnamefont{Salomon}},
  \bibinfo{journal}{Phys. Rev. A} \textbf{\bibinfo{volume}{64}},
  \bibinfo{pages}{011402} (\bibinfo{year}{2001}).

\bibitem[{\citenamefont{Catani et~al.}(2009)\citenamefont{Catani, Barontini,
  Lamporesi, Rabatti, Thalhammer, Minardi, Stringari, and
  Inguscio}}]{catani_barontini_09}
\bibinfo{author}{\bibfnamefont{J.}~\bibnamefont{Catani}},
  \bibinfo{author}{\bibfnamefont{G.}~\bibnamefont{Barontini}},
  \bibinfo{author}{\bibfnamefont{G.}~\bibnamefont{Lamporesi}},
  \bibinfo{author}{\bibfnamefont{F.}~\bibnamefont{Rabatti}},
  \bibinfo{author}{\bibfnamefont{G.}~\bibnamefont{Thalhammer}},
  \bibinfo{author}{\bibfnamefont{F.}~\bibnamefont{Minardi}},
  \bibinfo{author}{\bibfnamefont{S.}~\bibnamefont{Stringari}},
  \bibnamefont{and} \bibinfo{author}{\bibfnamefont{M.}~\bibnamefont{Inguscio}},
  \bibinfo{journal}{Phys. Rev. Lett.} \textbf{\bibinfo{volume}{103}},
  \bibinfo{pages}{140401} (\bibinfo{year}{2009}).

\bibitem[{\citenamefont{Erhard et~al.}(2004)\citenamefont{Erhard, Schmaljohann,
  Kronj\"ager, Bongs, and Sengstock}}]{Erhard2004a}
\bibinfo{author}{\bibfnamefont{M.}~\bibnamefont{Erhard}},
  \bibinfo{author}{\bibfnamefont{H.}~\bibnamefont{Schmaljohann}},
  \bibinfo{author}{\bibfnamefont{J.}~\bibnamefont{Kronj\"ager}},
  \bibinfo{author}{\bibfnamefont{K.}~\bibnamefont{Bongs}}, \bibnamefont{and}
  \bibinfo{author}{\bibfnamefont{K.}~\bibnamefont{Sengstock}},
  \bibinfo{journal}{Phys. Rev. A} \textbf{\bibinfo{volume}{70}},
  \bibinfo{pages}{031602(R)} (\bibinfo{year}{2004}).

\bibitem[{\citenamefont{Shin et~al.}(2004)\citenamefont{Shin, Saba, Schirotzek,
  Pasquini, Leanhardt, Pritchard, and Ketterle}}]{shin_saba_04}
\bibinfo{author}{\bibfnamefont{Y.}~\bibnamefont{Shin}},
  \bibinfo{author}{\bibfnamefont{M.}~\bibnamefont{Saba}},
  \bibinfo{author}{\bibfnamefont{A.}~\bibnamefont{Schirotzek}},
  \bibinfo{author}{\bibfnamefont{T.~A.} \bibnamefont{Pasquini}},
  \bibinfo{author}{\bibfnamefont{A.~E.} \bibnamefont{Leanhardt}},
  \bibinfo{author}{\bibfnamefont{D.~E.} \bibnamefont{Pritchard}},
  \bibnamefont{and} \bibinfo{author}{\bibfnamefont{W.}~\bibnamefont{Ketterle}},
  \bibinfo{journal}{Phys. Rev. Lett.} \textbf{\bibinfo{volume}{92}},
  \bibinfo{pages}{150401} (\bibinfo{year}{2004}).

\bibitem[{\citenamefont{Stellmer et~al.}(2013)\citenamefont{Stellmer, Pasquiou,
  Grimm, and Schreck}}]{stellmer_pasquiou_13}
\bibinfo{author}{\bibfnamefont{S.}~\bibnamefont{Stellmer}},
  \bibinfo{author}{\bibfnamefont{B.}~\bibnamefont{Pasquiou}},
  \bibinfo{author}{\bibfnamefont{R.}~\bibnamefont{Grimm}}, \bibnamefont{and}
  \bibinfo{author}{\bibfnamefont{F.}~\bibnamefont{Schreck}},
  \bibinfo{journal}{Phys. Rev. Lett.} \textbf{\bibinfo{volume}{110}},
  \bibinfo{pages}{263003} (\bibinfo{year}{2013}).

\bibitem[{\citenamefont{Kosterlitz and
  Thouless}(1973)}]{kosterlitz_thouless_73}
\bibinfo{author}{\bibfnamefont{J.~M.} \bibnamefont{Kosterlitz}}
  \bibnamefont{and} \bibinfo{author}{\bibfnamefont{D.~J.}
  \bibnamefont{Thouless}}, \bibinfo{journal}{J. Phys. C.}
  \textbf{\bibinfo{volume}{6}}, \bibinfo{pages}{1181} (\bibinfo{year}{1973}).

\bibitem[{\citenamefont{Hadzibabic et~al.}(2006)\citenamefont{Hadzibabic,
  Kruger, Cheneau, Battelier, and Dalibard}}]{hadzibabic_kruger_06}
\bibinfo{author}{\bibfnamefont{Z.}~\bibnamefont{Hadzibabic}},
  \bibinfo{author}{\bibfnamefont{P.}~\bibnamefont{Kruger}},
  \bibinfo{author}{\bibfnamefont{M.}~\bibnamefont{Cheneau}},
  \bibinfo{author}{\bibfnamefont{B.}~\bibnamefont{Battelier}},
  \bibnamefont{and} \bibinfo{author}{\bibfnamefont{J.}~\bibnamefont{Dalibard}},
  \bibinfo{journal}{Nature} \textbf{\bibinfo{volume}{441}},
  \bibinfo{pages}{1118} (\bibinfo{year}{2006}).

\bibitem[{\citenamefont{Schweikhard et~al.}(2007)\citenamefont{Schweikhard,
  Tung, and Cornell}}]{schweikhard_tung_07}
\bibinfo{author}{\bibfnamefont{V.}~\bibnamefont{Schweikhard}},
  \bibinfo{author}{\bibfnamefont{S.}~\bibnamefont{Tung}}, \bibnamefont{and}
  \bibinfo{author}{\bibfnamefont{E.~A.} \bibnamefont{Cornell}},
  \bibinfo{journal}{Phys. Rev. Lett.} \textbf{\bibinfo{volume}{99}},
  \bibinfo{pages}{030401} (\bibinfo{year}{2007}).

\bibitem[{\citenamefont{Kr\"uger et~al.}(2007)\citenamefont{Kr\"uger,
  Hadzibabic, and Dalibard}}]{kruger_hadzibabic_07}
\bibinfo{author}{\bibfnamefont{P.}~\bibnamefont{Kr\"uger}},
  \bibinfo{author}{\bibfnamefont{Z.}~\bibnamefont{Hadzibabic}},
  \bibnamefont{and} \bibinfo{author}{\bibfnamefont{J.}~\bibnamefont{Dalibard}},
  \bibinfo{journal}{Phys. Rev. Lett.} \textbf{\bibinfo{volume}{99}},
  \bibinfo{pages}{040402} (\bibinfo{year}{2007}).

\bibitem[{\citenamefont{Clad\'e et~al.}(2009)\citenamefont{Clad\'e, Ryu,
  Ramanathan, Helmerson, and Phillips}}]{clade_ryu_09}
\bibinfo{author}{\bibfnamefont{P.}~\bibnamefont{Clad\'e}},
  \bibinfo{author}{\bibfnamefont{C.}~\bibnamefont{Ryu}},
  \bibinfo{author}{\bibfnamefont{A.}~\bibnamefont{Ramanathan}},
  \bibinfo{author}{\bibfnamefont{K.}~\bibnamefont{Helmerson}},
  \bibnamefont{and} \bibinfo{author}{\bibfnamefont{W.~D.}
  \bibnamefont{Phillips}}, \bibinfo{journal}{Phys. Rev. Lett.}
  \textbf{\bibinfo{volume}{102}}, \bibinfo{pages}{170401}
  (\bibinfo{year}{2009}).

\bibitem[{\citenamefont{Tung et~al.}(2010)\citenamefont{Tung, Lamporesi,
  Lobser, Xia, and Cornell}}]{tung_lamporesi_10}
\bibinfo{author}{\bibfnamefont{S.}~\bibnamefont{Tung}},
  \bibinfo{author}{\bibfnamefont{G.}~\bibnamefont{Lamporesi}},
  \bibinfo{author}{\bibfnamefont{D.}~\bibnamefont{Lobser}},
  \bibinfo{author}{\bibfnamefont{L.}~\bibnamefont{Xia}}, \bibnamefont{and}
  \bibinfo{author}{\bibfnamefont{E.~A.} \bibnamefont{Cornell}},
  \bibinfo{journal}{Phys. Rev. Lett.} \textbf{\bibinfo{volume}{105}},
  \bibinfo{pages}{230408} (\bibinfo{year}{2010}).

\bibitem[{\citenamefont{Hung et~al.}(2011)\citenamefont{Hung, Zhang, Gemekle,
  and Chin}}]{hung_zhang_11}
\bibinfo{author}{\bibfnamefont{C.-L.} \bibnamefont{Hung}},
  \bibinfo{author}{\bibfnamefont{X.}~\bibnamefont{Zhang}},
  \bibinfo{author}{\bibfnamefont{N.}~\bibnamefont{Gemekle}}, \bibnamefont{and}
  \bibinfo{author}{\bibfnamefont{C.}~\bibnamefont{Chin}},
  \bibinfo{journal}{Nature} \textbf{\bibinfo{volume}{470}},
  \bibinfo{pages}{236} (\bibinfo{year}{2011}).

\bibitem[{\citenamefont{Prokof'ev et~al.}(2001)\citenamefont{Prokof'ev,
  Ruebenacker, and Svistunov}}]{prokofev_ruebenacker_01}
\bibinfo{author}{\bibfnamefont{N.}~\bibnamefont{Prokof'ev}},
  \bibinfo{author}{\bibfnamefont{O.}~\bibnamefont{Ruebenacker}},
  \bibnamefont{and}
  \bibinfo{author}{\bibfnamefont{B.}~\bibnamefont{Svistunov}},
  \bibinfo{journal}{Phys. Rev. Lett.} \textbf{\bibinfo{volume}{87}},
  \bibinfo{pages}{270402} (\bibinfo{year}{2001}).

\bibitem[{\citenamefont{Simula and Blakie}(2006)}]{simula_blakie_06}
\bibinfo{author}{\bibfnamefont{T.~P.} \bibnamefont{Simula}} \bibnamefont{and}
  \bibinfo{author}{\bibfnamefont{P.~B.} \bibnamefont{Blakie}},
  \bibinfo{journal}{Phys. Rev. Lett.} \textbf{\bibinfo{volume}{96}},
  \bibinfo{pages}{020404} (\bibinfo{year}{2006}).

\bibitem[{\citenamefont{Holzmann and Krauth}(2008)}]{holzmann_krauth_08}
\bibinfo{author}{\bibfnamefont{M.}~\bibnamefont{Holzmann}} \bibnamefont{and}
  \bibinfo{author}{\bibfnamefont{W.}~\bibnamefont{Krauth}},
  \bibinfo{journal}{Phys. Rev. Lett.} \textbf{\bibinfo{volume}{100}},
  \bibinfo{pages}{190402} (\bibinfo{year}{2008}).

\bibitem[{\citenamefont{Bisset et~al.}(2009)\citenamefont{Bisset, Davis,
  Simula, and Blakie}}]{bisset_davis_09}
\bibinfo{author}{\bibfnamefont{R.~N.} \bibnamefont{Bisset}},
  \bibinfo{author}{\bibfnamefont{M.~J.} \bibnamefont{Davis}},
  \bibinfo{author}{\bibfnamefont{T.~P.} \bibnamefont{Simula}},
  \bibnamefont{and} \bibinfo{author}{\bibfnamefont{P.~B.}
  \bibnamefont{Blakie}}, \bibinfo{journal}{Phys. Rev. A}
  \textbf{\bibinfo{volume}{79}}, \bibinfo{pages}{033626}
  (\bibinfo{year}{2009}).

\bibitem[{\citenamefont{Cockburn and Proukakis}(2012)}]{cockburn_proukakis_12}
\bibinfo{author}{\bibfnamefont{S.~P.} \bibnamefont{Cockburn}} \bibnamefont{and}
  \bibinfo{author}{\bibfnamefont{N.~P.} \bibnamefont{Proukakis}},
  \bibinfo{journal}{Phys. Rev. A} \textbf{\bibinfo{volume}{86}},
  \bibinfo{pages}{033610} (\bibinfo{year}{2012}).

\bibitem[{\citenamefont{Hadzibabic and
  Dalibard}(2011)}]{hadzibabic_dalibard_12}
\bibinfo{author}{\bibfnamefont{Z.}~\bibnamefont{Hadzibabic}} \bibnamefont{and}
  \bibinfo{author}{\bibfnamefont{J.}~\bibnamefont{Dalibard}},
  \emph{\bibinfo{title}{Nano optics and atomics: transport of light and matter
  waves}} (\bibinfo{publisher}{Rivista del Nuovo Cimento},
  \bibinfo{year}{2011}), vol.~\bibinfo{volume}{34} of
  \emph{\bibinfo{series}{Proceedings of the International School of Physics
  "Enrico Fermi", Vol. CLXXIII}}, chap. \bibinfo{chapter}{Two Dimensional Bose
  Fluids: An Atomic Physics Perspective}, p. \bibinfo{pages}{389}.

\bibitem[{\citenamefont{Roumpos et~al.}(2012)\citenamefont{Roumpos, Lohse,
  Nitsche, Keeling, Szyma{\'n}ska, Littlewood, L{\"o}ffler, H{\"o}fling,
  Worschech, Forchel et~al.}}]{roumpos_lohse_12}
\bibinfo{author}{\bibfnamefont{G.}~\bibnamefont{Roumpos}},
  \bibinfo{author}{\bibfnamefont{M.}~\bibnamefont{Lohse}},
  \bibinfo{author}{\bibfnamefont{W.~H.} \bibnamefont{Nitsche}},
  \bibinfo{author}{\bibfnamefont{J.}~\bibnamefont{Keeling}},
  \bibinfo{author}{\bibfnamefont{M.~H.} \bibnamefont{Szyma{\'n}ska}},
  \bibinfo{author}{\bibfnamefont{P.~B.} \bibnamefont{Littlewood}},
  \bibinfo{author}{\bibfnamefont{A.}~\bibnamefont{L{\"o}ffler}},
  \bibinfo{author}{\bibfnamefont{S.}~\bibnamefont{H{\"o}fling}},
  \bibinfo{author}{\bibfnamefont{L.}~\bibnamefont{Worschech}},
  \bibinfo{author}{\bibfnamefont{A.}~\bibnamefont{Forchel}},
  \bibnamefont{et~al.}, \bibinfo{journal}{Proceedings of the National Academy
  of Sciences} \textbf{\bibinfo{volume}{109}}, \bibinfo{pages}{6467}
  (\bibinfo{year}{2012}).

\bibitem[{\citenamefont{Nitsche et~al.}(2014)\citenamefont{Nitsche, Kim,
  Roumpos, Schneider, Kamp, H\"ofling, Forchel, and Yamamoto}}]{nitsche_kim_14}
\bibinfo{author}{\bibfnamefont{W.~H.} \bibnamefont{Nitsche}},
  \bibinfo{author}{\bibfnamefont{N.~Y.} \bibnamefont{Kim}},
  \bibinfo{author}{\bibfnamefont{G.}~\bibnamefont{Roumpos}},
  \bibinfo{author}{\bibfnamefont{C.}~\bibnamefont{Schneider}},
  \bibinfo{author}{\bibfnamefont{M.}~\bibnamefont{Kamp}},
  \bibinfo{author}{\bibfnamefont{S.}~\bibnamefont{H\"ofling}},
  \bibinfo{author}{\bibfnamefont{A.}~\bibnamefont{Forchel}}, \bibnamefont{and}
  \bibinfo{author}{\bibfnamefont{Y.}~\bibnamefont{Yamamoto}},
  \bibinfo{journal}{Phys. Rev. B} \textbf{\bibinfo{volume}{90}},
  \bibinfo{pages}{205430} (\bibinfo{year}{2014}).

\bibitem[{\citenamefont{Petrov et~al.}(2001)\citenamefont{Petrov, Shlyapnikov,
  and Walraven}}]{petrov_shlyapnikov_01}
\bibinfo{author}{\bibfnamefont{D.~S.} \bibnamefont{Petrov}},
  \bibinfo{author}{\bibfnamefont{G.~V.} \bibnamefont{Shlyapnikov}},
  \bibnamefont{and} \bibinfo{author}{\bibfnamefont{J.~T.~M.}
  \bibnamefont{Walraven}}, \bibinfo{journal}{Phys. Rev. Lett.}
  \textbf{\bibinfo{volume}{87}}, \bibinfo{pages}{050404}
  (\bibinfo{year}{2001}).

\bibitem[{\citenamefont{Shvarchuck et~al.}(2002)\citenamefont{Shvarchuck,
  Buggle, Petrov, Dieckmann, Zielonkowski, Kemmann, Tiecke, von Klitzing,
  Shlyapnikov, and Walraven}}]{shvarchuck_buggle_02}
\bibinfo{author}{\bibfnamefont{I.}~\bibnamefont{Shvarchuck}},
  \bibinfo{author}{\bibfnamefont{C.}~\bibnamefont{Buggle}},
  \bibinfo{author}{\bibfnamefont{D.~S.} \bibnamefont{Petrov}},
  \bibinfo{author}{\bibfnamefont{K.}~\bibnamefont{Dieckmann}},
  \bibinfo{author}{\bibfnamefont{M.}~\bibnamefont{Zielonkowski}},
  \bibinfo{author}{\bibfnamefont{M.}~\bibnamefont{Kemmann}},
  \bibinfo{author}{\bibfnamefont{T.~G.} \bibnamefont{Tiecke}},
  \bibinfo{author}{\bibfnamefont{W.}~\bibnamefont{von Klitzing}},
  \bibinfo{author}{\bibfnamefont{G.~V.} \bibnamefont{Shlyapnikov}},
  \bibnamefont{and} \bibinfo{author}{\bibfnamefont{J.~T.~M.}
  \bibnamefont{Walraven}}, \bibinfo{journal}{Phys. Rev. Lett.}
  \textbf{\bibinfo{volume}{89}}, \bibinfo{pages}{270404}
  (\bibinfo{year}{2002}).

\bibitem[{\citenamefont{Hugbart et~al.}(2007)\citenamefont{Hugbart, Retter,
  Var\'on, Bouyer, Aspect, and Davis}}]{hugbart_retter_07}
\bibinfo{author}{\bibfnamefont{M.}~\bibnamefont{Hugbart}},
  \bibinfo{author}{\bibfnamefont{J.~A.} \bibnamefont{Retter}},
  \bibinfo{author}{\bibfnamefont{A.~F.} \bibnamefont{Var\'on}},
  \bibinfo{author}{\bibfnamefont{P.}~\bibnamefont{Bouyer}},
  \bibinfo{author}{\bibfnamefont{A.}~\bibnamefont{Aspect}}, \bibnamefont{and}
  \bibinfo{author}{\bibfnamefont{M.~J.} \bibnamefont{Davis}},
  \bibinfo{journal}{Phys. Rev. A} \textbf{\bibinfo{volume}{75}},
  \bibinfo{pages}{011602} (\bibinfo{year}{2007}).

\bibitem[{\citenamefont{Binney et~al.}(1992)\citenamefont{Binney, Dowrick,
  Fisher, and Newman}}]{binney_dowrick_92}
\bibinfo{author}{\bibfnamefont{J.~J.} \bibnamefont{Binney}},
  \bibinfo{author}{\bibfnamefont{N.~J.} \bibnamefont{Dowrick}},
  \bibinfo{author}{\bibfnamefont{A.~J.} \bibnamefont{Fisher}},
  \bibnamefont{and} \bibinfo{author}{\bibfnamefont{M.~E.~J.}
  \bibnamefont{Newman}}, \emph{\bibinfo{title}{{The Theory of Critical
  Phenomena: An Introduction to the Renormalization Group}}}
  (\bibinfo{publisher}{Oxford University Press, USA}, \bibinfo{year}{1992}).

\bibitem[{\citenamefont{Zinn-Justin}(2002)}]{Zinn-Justin2002}
\bibinfo{author}{\bibfnamefont{J.}~\bibnamefont{Zinn-Justin}},
  \emph{\bibinfo{title}{Quantum Field Theory and Critical Phenomena}}
  (\bibinfo{publisher}{Clarendon Press}, \bibinfo{address}{Oxford},
  \bibinfo{year}{2002}), \bibinfo{edition}{4th} ed.

\bibitem[{\citenamefont{Hohenberg and Halperin}(1977)}]{Hohenberg1977a}
\bibinfo{author}{\bibfnamefont{P.~C.} \bibnamefont{Hohenberg}}
  \bibnamefont{and} \bibinfo{author}{\bibfnamefont{B.~I.}
  \bibnamefont{Halperin}}, \bibinfo{journal}{Rev. Mod. Phys.}
  \textbf{\bibinfo{volume}{49}}, \bibinfo{pages}{435} (\bibinfo{year}{1977}).

\bibitem[{\citenamefont{Baym et~al.}(1999)\citenamefont{Baym, Blaizot,
  Holzmann, Lalo\"e, and Vautherin}}]{baym_blaizot_99}
\bibinfo{author}{\bibfnamefont{G.}~\bibnamefont{Baym}},
  \bibinfo{author}{\bibfnamefont{J.-P.} \bibnamefont{Blaizot}},
  \bibinfo{author}{\bibfnamefont{M.}~\bibnamefont{Holzmann}},
  \bibinfo{author}{\bibfnamefont{F.}~\bibnamefont{Lalo\"e}}, \bibnamefont{and}
  \bibinfo{author}{\bibfnamefont{D.}~\bibnamefont{Vautherin}},
  \bibinfo{journal}{Phys. Rev. Lett.} \textbf{\bibinfo{volume}{83}},
  \bibinfo{pages}{1703} (\bibinfo{year}{1999}).

\bibitem[{\citenamefont{Kashurnikov et~al.}(2001)\citenamefont{Kashurnikov,
  Prokof'ev, and Svistunov}}]{kashurnikov_prokofev_01}
\bibinfo{author}{\bibfnamefont{V.~A.} \bibnamefont{Kashurnikov}},
  \bibinfo{author}{\bibfnamefont{N.~V.} \bibnamefont{Prokof'ev}},
  \bibnamefont{and} \bibinfo{author}{\bibfnamefont{B.~V.}
  \bibnamefont{Svistunov}}, \bibinfo{journal}{Phys. Rev. Lett.}
  \textbf{\bibinfo{volume}{87}}, \bibinfo{pages}{120402}
  (\bibinfo{year}{2001}).

\bibitem[{\citenamefont{Arnold and Moore}(2001)}]{arnold_moore_01}
\bibinfo{author}{\bibfnamefont{P.}~\bibnamefont{Arnold}} \bibnamefont{and}
  \bibinfo{author}{\bibfnamefont{G.}~\bibnamefont{Moore}},
  \bibinfo{journal}{Phys. Rev. Lett.} \textbf{\bibinfo{volume}{87}},
  \bibinfo{pages}{120401} (\bibinfo{year}{2001}).

\bibitem[{\citenamefont{Gerbier et~al.}(2004)\citenamefont{Gerbier, Thywissen,
  Richard, Hugbart, Bouyer, and Aspect}}]{gerbier_thywissen_04}
\bibinfo{author}{\bibfnamefont{F.}~\bibnamefont{Gerbier}},
  \bibinfo{author}{\bibfnamefont{J.~H.} \bibnamefont{Thywissen}},
  \bibinfo{author}{\bibfnamefont{S.}~\bibnamefont{Richard}},
  \bibinfo{author}{\bibfnamefont{M.}~\bibnamefont{Hugbart}},
  \bibinfo{author}{\bibfnamefont{P.}~\bibnamefont{Bouyer}}, \bibnamefont{and}
  \bibinfo{author}{\bibfnamefont{A.}~\bibnamefont{Aspect}},
  \bibinfo{journal}{Phys. Rev. Lett.} \textbf{\bibinfo{volume}{92}},
  \bibinfo{pages}{030405} (\bibinfo{year}{2004}).

\bibitem[{\citenamefont{Davis and Blakie}(2006)}]{davis_blakie_06}
\bibinfo{author}{\bibfnamefont{M.~J.} \bibnamefont{Davis}} \bibnamefont{and}
  \bibinfo{author}{\bibfnamefont{P.~B.} \bibnamefont{Blakie}},
  \bibinfo{journal}{Phys. Rev. Lett.} \textbf{\bibinfo{volume}{96}},
  \bibinfo{pages}{060404} (\bibinfo{year}{2006}).

\bibitem[{\citenamefont{Smith et~al.}(2011)\citenamefont{Smith, Campbell,
  Tammuz, and Hadzibabic}}]{smith_campbell_11}
\bibinfo{author}{\bibfnamefont{R.~P.} \bibnamefont{Smith}},
  \bibinfo{author}{\bibfnamefont{R.~L.~D.} \bibnamefont{Campbell}},
  \bibinfo{author}{\bibfnamefont{N.}~\bibnamefont{Tammuz}}, \bibnamefont{and}
  \bibinfo{author}{\bibfnamefont{Z.}~\bibnamefont{Hadzibabic}},
  \bibinfo{journal}{Phys. Rev. Lett.} \textbf{\bibinfo{volume}{106}},
  \bibinfo{pages}{250403} (\bibinfo{year}{2011}).

\bibitem[{\citenamefont{Ritter et~al.}(2007)\citenamefont{Ritter, \"{O}ttl,
  Donner, Bourdel, K\"{o}hl, and Esslinger}}]{ritter_ottl_07}
\bibinfo{author}{\bibfnamefont{S.}~\bibnamefont{Ritter}},
  \bibinfo{author}{\bibfnamefont{A.}~\bibnamefont{\"{O}ttl}},
  \bibinfo{author}{\bibfnamefont{T.}~\bibnamefont{Donner}},
  \bibinfo{author}{\bibfnamefont{T.}~\bibnamefont{Bourdel}},
  \bibinfo{author}{\bibfnamefont{M.}~\bibnamefont{K\"{o}hl}}, \bibnamefont{and}
  \bibinfo{author}{\bibfnamefont{T.}~\bibnamefont{Esslinger}},
  \bibinfo{journal}{Phys. Rev. Lett.} \textbf{\bibinfo{volume}{98}},
  \bibinfo{pages}{090402} (\bibinfo{year}{2007}).

\bibitem[{\citenamefont{Donner et~al.}(2007)\citenamefont{Donner, Ritter,
  Bourdel, {\"O}ttl, K{\"o}hl, and Esslinger}}]{donner_ritter_07}
\bibinfo{author}{\bibfnamefont{T.}~\bibnamefont{Donner}},
  \bibinfo{author}{\bibfnamefont{S.}~\bibnamefont{Ritter}},
  \bibinfo{author}{\bibfnamefont{T.}~\bibnamefont{Bourdel}},
  \bibinfo{author}{\bibfnamefont{A.}~\bibnamefont{{\"O}ttl}},
  \bibinfo{author}{\bibfnamefont{M.}~\bibnamefont{K{\"o}hl}}, \bibnamefont{and}
  \bibinfo{author}{\bibfnamefont{T.}~\bibnamefont{Esslinger}},
  \bibinfo{journal}{Science} \textbf{\bibinfo{volume}{315}},
  \bibinfo{pages}{1556} (\bibinfo{year}{2007}).

\bibitem[{\citenamefont{Bezett and Blakie}(2009)}]{bezett_blakie_09}
\bibinfo{author}{\bibfnamefont{A.}~\bibnamefont{Bezett}} \bibnamefont{and}
  \bibinfo{author}{\bibfnamefont{P.~B.} \bibnamefont{Blakie}},
  \bibinfo{journal}{Phys. Rev. A} \textbf{\bibinfo{volume}{79}},
  \bibinfo{pages}{033611} (\bibinfo{year}{2009}).

\bibitem[{\citenamefont{Navon et~al.}(2015)\citenamefont{Navon, Gaunt, Smith,
  and Hadzibabic}}]{navon_gaunt_15}
\bibinfo{author}{\bibfnamefont{N.}~\bibnamefont{Navon}},
  \bibinfo{author}{\bibfnamefont{A.~L.} \bibnamefont{Gaunt}},
  \bibinfo{author}{\bibfnamefont{R.~P.} \bibnamefont{Smith}}, \bibnamefont{and}
  \bibinfo{author}{\bibfnamefont{Z.}~\bibnamefont{Hadzibabic}},
  \bibinfo{journal}{Science} \textbf{\bibinfo{volume}{347}},
  \bibinfo{pages}{167} (\bibinfo{year}{2015}).

\bibitem[{\citenamefont{Kibble}(1976)}]{kibble_76}
\bibinfo{author}{\bibfnamefont{T.~W.~B.} \bibnamefont{Kibble}},
  \bibinfo{journal}{J. Phys. A: Math. Gen.} \textbf{\bibinfo{volume}{9}},
  \bibinfo{pages}{1387} (\bibinfo{year}{1976}).

\bibitem[{\citenamefont{Landau and Lifshitz}(1980)}]{landau_lifshitz_book_80}
\bibinfo{author}{\bibfnamefont{L.~D.} \bibnamefont{Landau}} \bibnamefont{and}
  \bibinfo{author}{\bibfnamefont{E.~M.} \bibnamefont{Lifshitz}},
  \emph{\bibinfo{title}{Statistical Physics, Part 1}}
  (\bibinfo{publisher}{Butterworth--Heinemann}, \bibinfo{address}{Oxford, UK},
  \bibinfo{year}{1980}), \bibinfo{edition}{3rd} ed.

\bibitem[{\citenamefont{Zurek}(1985)}]{zurek_85}
\bibinfo{author}{\bibfnamefont{W.~H.} \bibnamefont{Zurek}},
  \bibinfo{journal}{Nature} \textbf{\bibinfo{volume}{317}},
  \bibinfo{pages}{505} (\bibinfo{year}{1985}).

\bibitem[{\citenamefont{Zurek}(1996)}]{zurek_96}
\bibinfo{author}{\bibfnamefont{W.~H.} \bibnamefont{Zurek}},
  \bibinfo{journal}{Physics Reports} \textbf{\bibinfo{volume}{276}},
  \bibinfo{pages}{177} (\bibinfo{year}{1996}).

\bibitem[{\citenamefont{Anglin and Zurek}(1999)}]{anglin_zurek_99}
\bibinfo{author}{\bibfnamefont{J.~R.} \bibnamefont{Anglin}} \bibnamefont{and}
  \bibinfo{author}{\bibfnamefont{W.~H.} \bibnamefont{Zurek}},
  \bibinfo{journal}{Phys. Rev. Lett} \textbf{\bibinfo{volume}{83}},
  \bibinfo{pages}{1707} (\bibinfo{year}{1999}).

\bibitem[{\citenamefont{Weiler et~al.}(2008)\citenamefont{Weiler, Neely,
  Scherer, Bradley, Davis, and Anderson}}]{weiler_neely_08}
\bibinfo{author}{\bibfnamefont{C.~N.} \bibnamefont{Weiler}},
  \bibinfo{author}{\bibfnamefont{T.~W.} \bibnamefont{Neely}},
  \bibinfo{author}{\bibfnamefont{D.~R.} \bibnamefont{Scherer}},
  \bibinfo{author}{\bibfnamefont{A.~S.} \bibnamefont{Bradley}},
  \bibinfo{author}{\bibfnamefont{M.~J.} \bibnamefont{Davis}}, \bibnamefont{and}
  \bibinfo{author}{\bibfnamefont{B.~P.} \bibnamefont{Anderson}},
  \bibinfo{journal}{Nature} \textbf{\bibinfo{volume}{455}},
  \bibinfo{pages}{948} (\bibinfo{year}{2008}).

\bibitem[{\citenamefont{Freilich et~al.}(2010)\citenamefont{Freilich, Bianchi,
  Kaufman, Langin, and Hall}}]{freilich_bianchi_10}
\bibinfo{author}{\bibfnamefont{D.~V.} \bibnamefont{Freilich}},
  \bibinfo{author}{\bibfnamefont{D.~M.} \bibnamefont{Bianchi}},
  \bibinfo{author}{\bibfnamefont{A.~M.} \bibnamefont{Kaufman}},
  \bibinfo{author}{\bibfnamefont{T.~K.} \bibnamefont{Langin}},
  \bibnamefont{and} \bibinfo{author}{\bibfnamefont{D.~S.} \bibnamefont{Hall}},
  \bibinfo{journal}{Science} \textbf{\bibinfo{volume}{329}},
  \bibinfo{pages}{1182} (\bibinfo{year}{2010}).

\bibitem[{\citenamefont{Gardiner and Davis}(2003)}]{gardiner_davis_03}
\bibinfo{author}{\bibfnamefont{C.~W.} \bibnamefont{Gardiner}} \bibnamefont{and}
  \bibinfo{author}{\bibfnamefont{M.~J.} \bibnamefont{Davis}},
  \bibinfo{journal}{J. Phys. B: At. Mol. Opt. Phys.}
  \textbf{\bibinfo{volume}{36}}, \bibinfo{pages}{4731} (\bibinfo{year}{2003}).

\bibitem[{\citenamefont{Blakie et~al.}(2008)\citenamefont{Blakie, Bradley,
  Davis, Ballagh, and Gardiner}}]{blakie_bradley_08}
\bibinfo{author}{\bibfnamefont{P.~B.} \bibnamefont{Blakie}},
  \bibinfo{author}{\bibfnamefont{A.~S.} \bibnamefont{Bradley}},
  \bibinfo{author}{\bibfnamefont{M.~J.} \bibnamefont{Davis}},
  \bibinfo{author}{\bibfnamefont{R.~J.} \bibnamefont{Ballagh}},
  \bibnamefont{and} \bibinfo{author}{\bibfnamefont{C.~W.}
  \bibnamefont{Gardiner}}, \bibinfo{journal}{Advances in Physics}
  \textbf{\bibinfo{volume}{57}}, \bibinfo{pages}{363} (\bibinfo{year}{2008}).

\bibitem[{\citenamefont{Proukakis and Jackson}(2008)}]{proukakis_jackson_08}
\bibinfo{author}{\bibfnamefont{N.~P.} \bibnamefont{Proukakis}}
  \bibnamefont{and} \bibinfo{author}{\bibfnamefont{B.}~\bibnamefont{Jackson}},
  \bibinfo{journal}{J. Phys. B: At. Mol. Opt.} \textbf{\bibinfo{volume}{41}},
  \bibinfo{pages}{203002} (\bibinfo{year}{2008}).

\bibitem[{\citenamefont{Cockburn and Proukakis}(2009)}]{cockburn_proukakis_09}
\bibinfo{author}{\bibfnamefont{S.~P.} \bibnamefont{Cockburn}} \bibnamefont{and}
  \bibinfo{author}{\bibfnamefont{N.~P.} \bibnamefont{Proukakis}},
  \bibinfo{journal}{Laser Phys.} \textbf{\bibinfo{volume}{19}},
  \bibinfo{pages}{558} (\bibinfo{year}{2009}).

\bibitem[{\citenamefont{Steel et~al.}(1998)\citenamefont{Steel, Olsen, Plimak,
  Drummond, Tan, Collett, Walls, and Graham}}]{steel_olsen_98}
\bibinfo{author}{\bibfnamefont{M.~J.} \bibnamefont{Steel}},
  \bibinfo{author}{\bibfnamefont{M.~K.} \bibnamefont{Olsen}},
  \bibinfo{author}{\bibfnamefont{L.~I.} \bibnamefont{Plimak}},
  \bibinfo{author}{\bibfnamefont{P.~D.} \bibnamefont{Drummond}},
  \bibinfo{author}{\bibfnamefont{S.~M.} \bibnamefont{Tan}},
  \bibinfo{author}{\bibfnamefont{M.~J.} \bibnamefont{Collett}},
  \bibinfo{author}{\bibfnamefont{D.~F.} \bibnamefont{Walls}}, \bibnamefont{and}
  \bibinfo{author}{\bibfnamefont{R.}~\bibnamefont{Graham}},
  \bibinfo{journal}{Phys. Rev. A} \textbf{\bibinfo{volume}{58}},
  \bibinfo{pages}{4824} (\bibinfo{year}{1998}).

\bibitem[{\citenamefont{Drummond and Corney}(1999)}]{drummond_corney_99}
\bibinfo{author}{\bibfnamefont{P.~D.} \bibnamefont{Drummond}} \bibnamefont{and}
  \bibinfo{author}{\bibfnamefont{J.~F.} \bibnamefont{Corney}},
  \bibinfo{journal}{Phys. Rev. A.} \textbf{\bibinfo{volume}{60}},
  \bibinfo{pages}{R2661} (\bibinfo{year}{1999}).

\bibitem[{\citenamefont{Chang et~al.}(2009)\citenamefont{Chang, Hamner, and
  Engels}}]{chiang_hamner_09}
\bibinfo{author}{\bibfnamefont{J.}~\bibnamefont{Chang}},
  \bibinfo{author}{\bibfnamefont{C.}~\bibnamefont{Hamner}}, \bibnamefont{and}
  \bibinfo{author}{\bibfnamefont{P.}~\bibnamefont{Engels}}
  (\bibinfo{year}{2009}), \bibinfo{note}{40th Annual Meeting of the APS
  Division of Atomic, Molecular and Optical Physics}.

\bibitem[{\citenamefont{Zurek}(2009)}]{zurek_09}
\bibinfo{author}{\bibfnamefont{W.~H.} \bibnamefont{Zurek}},
  \bibinfo{journal}{Phys. Rev. Lett.} \textbf{\bibinfo{volume}{102}},
  \bibinfo{pages}{105702} (\bibinfo{year}{2009}).

\bibitem[{\citenamefont{Damski and Zurek}(2010)}]{damski_zurek_10}
\bibinfo{author}{\bibfnamefont{B.}~\bibnamefont{Damski}} \bibnamefont{and}
  \bibinfo{author}{\bibfnamefont{W.~H.} \bibnamefont{Zurek}},
  \bibinfo{journal}{Phys. Rev. Lett.} \textbf{\bibinfo{volume}{104}},
  \bibinfo{pages}{160404} (\bibinfo{year}{2010}).

\bibitem[{\citenamefont{Witkowska et~al.}(2011)\citenamefont{Witkowska, Deuar,
  Gajda, and Rz\c{a}\.{z}ewski}}]{witkowska_deuar_11}
\bibinfo{author}{\bibfnamefont{E.}~\bibnamefont{Witkowska}},
  \bibinfo{author}{\bibfnamefont{P.}~\bibnamefont{Deuar}},
  \bibinfo{author}{\bibfnamefont{M.}~\bibnamefont{Gajda}}, \bibnamefont{and}
  \bibinfo{author}{\bibfnamefont{K.}~\bibnamefont{Rz\c{a}\.{z}ewski}},
  \bibinfo{journal}{Phys. Rev. Lett.} \textbf{\bibinfo{volume}{106}},
  \bibinfo{pages}{135301} (\bibinfo{year}{2011}).

\bibitem[{\citenamefont{del Campo et~al.}(2011)\citenamefont{del Campo,
  Retzker, and Plenio}}]{delcampo_retzker_11}
\bibinfo{author}{\bibfnamefont{A.}~\bibnamefont{del Campo}},
  \bibinfo{author}{\bibfnamefont{A.}~\bibnamefont{Retzker}}, \bibnamefont{and}
  \bibinfo{author}{\bibfnamefont{M.~B.} \bibnamefont{Plenio}},
  \bibinfo{journal}{New J. Phys.} \textbf{\bibinfo{volume}{13}},
  \bibinfo{pages}{083022} (\bibinfo{year}{2011}).

\bibitem[{\citenamefont{Lamporesi et~al.}(2013)\citenamefont{Lamporesi,
  Donadello, Serafini, Dalfovo, and Ferrari}}]{lamporesi_donadello_13}
\bibinfo{author}{\bibfnamefont{G.}~\bibnamefont{Lamporesi}},
  \bibinfo{author}{\bibfnamefont{S.}~\bibnamefont{Donadello}},
  \bibinfo{author}{\bibfnamefont{S.}~\bibnamefont{Serafini}},
  \bibinfo{author}{\bibfnamefont{F.}~\bibnamefont{Dalfovo}}, \bibnamefont{and}
  \bibinfo{author}{\bibfnamefont{G.}~\bibnamefont{Ferrari}},
  \bibinfo{journal}{Nat. Phys.} \textbf{\bibinfo{volume}{9}},
  \bibinfo{pages}{656} (\bibinfo{year}{2013}).

\bibitem[{\citenamefont{Donadello et~al.}(2014)\citenamefont{Donadello,
  Serafini, Tylutki, Pitaevskii, Dalfovo, Lamporesi, and Ferrari}}]{trento_prl}
\bibinfo{author}{\bibfnamefont{S.}~\bibnamefont{Donadello}},
  \bibinfo{author}{\bibfnamefont{S.}~\bibnamefont{Serafini}},
  \bibinfo{author}{\bibfnamefont{M.}~\bibnamefont{Tylutki}},
  \bibinfo{author}{\bibfnamefont{L.~P.} \bibnamefont{Pitaevskii}},
  \bibinfo{author}{\bibfnamefont{F.}~\bibnamefont{Dalfovo}},
  \bibinfo{author}{\bibfnamefont{G.}~\bibnamefont{Lamporesi}},
  \bibnamefont{and} \bibinfo{author}{\bibfnamefont{G.}~\bibnamefont{Ferrari}},
  \bibinfo{journal}{Phys. Rev. Lett.} \textbf{\bibinfo{volume}{113}},
  \bibinfo{pages}{065302} (\bibinfo{year}{2014}).

\bibitem[{\citenamefont{Chomaz et~al.}(2015)\citenamefont{Chomaz, Corman,
  Bienaim\'{e}, Desbuquois, Weitenberg, amd J\'{e}r\^{o}me~Beugnon, and
  Dalibard}}]{chomaz_corman_15}
\bibinfo{author}{\bibfnamefont{L.}~\bibnamefont{Chomaz}},
  \bibinfo{author}{\bibfnamefont{L.}~\bibnamefont{Corman}},
  \bibinfo{author}{\bibfnamefont{T.}~\bibnamefont{Bienaim\'{e}}},
  \bibinfo{author}{\bibfnamefont{R.}~\bibnamefont{Desbuquois}},
  \bibinfo{author}{\bibfnamefont{C.}~\bibnamefont{Weitenberg}},
  \bibinfo{author}{\bibfnamefont{S.~N.} \bibnamefont{amd
  J\'{e}r\^{o}me~Beugnon}}, \bibnamefont{and}
  \bibinfo{author}{\bibfnamefont{J.}~\bibnamefont{Dalibard}},
  \bibinfo{journal}{Nat. Comm.} \textbf{\bibinfo{volume}{6}},
  \bibinfo{pages}{6162} (\bibinfo{year}{2015}).

\bibitem[{\citenamefont{Corman et~al.}(2014)\citenamefont{Corman, Chomaz,
  Bienaim\'e, Desbuquois, Weitenberg, Nascimb\`ene, Dalibard, and
  Beugnon}}]{corman_chomaz_14}
\bibinfo{author}{\bibfnamefont{L.}~\bibnamefont{Corman}},
  \bibinfo{author}{\bibfnamefont{L.}~\bibnamefont{Chomaz}},
  \bibinfo{author}{\bibfnamefont{T.}~\bibnamefont{Bienaim\'e}},
  \bibinfo{author}{\bibfnamefont{R.}~\bibnamefont{Desbuquois}},
  \bibinfo{author}{\bibfnamefont{C.}~\bibnamefont{Weitenberg}},
  \bibinfo{author}{\bibfnamefont{S.}~\bibnamefont{Nascimb\`ene}},
  \bibinfo{author}{\bibfnamefont{J.}~\bibnamefont{Dalibard}}, \bibnamefont{and}
  \bibinfo{author}{\bibfnamefont{J.}~\bibnamefont{Beugnon}},
  \bibinfo{journal}{Phys. Rev. Lett.} \textbf{\bibinfo{volume}{113}},
  \bibinfo{pages}{135302} (\bibinfo{year}{2014}).

\bibitem[{\citenamefont{Das et~al.}(2012)\citenamefont{Das, Sabbatini, and
  Zurek}}]{das_sabbatini_12}
\bibinfo{author}{\bibfnamefont{A.}~\bibnamefont{Das}},
  \bibinfo{author}{\bibfnamefont{J.}~\bibnamefont{Sabbatini}},
  \bibnamefont{and} \bibinfo{author}{\bibfnamefont{W.~H.} \bibnamefont{Zurek}},
  \bibinfo{journal}{Sci. Rep.} \textbf{\bibinfo{volume}{2}},
  \bibinfo{pages}{352} (\bibinfo{year}{2012}).

\bibitem[{\citenamefont{Sadler et~al.}(2006)\citenamefont{Sadler, Higbie,
  Leslie, Vengalattore, and Stamper-Kurn}}]{sadler_higbie_06}
\bibinfo{author}{\bibfnamefont{L.~E.} \bibnamefont{Sadler}},
  \bibinfo{author}{\bibfnamefont{J.~M.} \bibnamefont{Higbie}},
  \bibinfo{author}{\bibfnamefont{S.~R.} \bibnamefont{Leslie}},
  \bibinfo{author}{\bibfnamefont{M.}~\bibnamefont{Vengalattore}},
  \bibnamefont{and} \bibinfo{author}{\bibfnamefont{D.~M.}
  \bibnamefont{Stamper-Kurn}}, \bibinfo{journal}{Nature}
  \textbf{\bibinfo{volume}{443}}, \bibinfo{pages}{312} (\bibinfo{year}{2006}).

\bibitem[{\citenamefont{De et~al.}(2014)\citenamefont{De, Campbell, Price,
  Putra, Anderson, and Spielman}}]{de_campbell_14}
\bibinfo{author}{\bibfnamefont{S.}~\bibnamefont{De}},
  \bibinfo{author}{\bibfnamefont{D.~L.} \bibnamefont{Campbell}},
  \bibinfo{author}{\bibfnamefont{R.~M.} \bibnamefont{Price}},
  \bibinfo{author}{\bibfnamefont{A.}~\bibnamefont{Putra}},
  \bibinfo{author}{\bibfnamefont{B.~M.} \bibnamefont{Anderson}},
  \bibnamefont{and} \bibinfo{author}{\bibfnamefont{I.~B.}
  \bibnamefont{Spielman}}, \bibinfo{journal}{Phys. Rev. A}
  \textbf{\bibinfo{volume}{89}}, \bibinfo{pages}{033631}
  (\bibinfo{year}{2014}).

\bibitem[{\citenamefont{Papp et~al.}(2008)\citenamefont{Papp, Pino, and
  Wieman}}]{papp_pino_08}
\bibinfo{author}{\bibfnamefont{S.~B.} \bibnamefont{Papp}},
  \bibinfo{author}{\bibfnamefont{J.~M.} \bibnamefont{Pino}}, \bibnamefont{and}
  \bibinfo{author}{\bibfnamefont{C.~E.} \bibnamefont{Wieman}},
  \bibinfo{journal}{Phys. Rev. Lett.} \textbf{\bibinfo{volume}{101}},
  \bibinfo{pages}{040402} (\bibinfo{year}{2008}).

\bibitem[{\citenamefont{McCarron et~al.}(2011)\citenamefont{McCarron, Cho,
  Jenkin, K\"oppinger, and Cornish}}]{mccarron_cho_11}
\bibinfo{author}{\bibfnamefont{D.~J.} \bibnamefont{McCarron}},
  \bibinfo{author}{\bibfnamefont{H.~W.} \bibnamefont{Cho}},
  \bibinfo{author}{\bibfnamefont{D.~L.} \bibnamefont{Jenkin}},
  \bibinfo{author}{\bibfnamefont{M.~P.} \bibnamefont{K\"oppinger}},
  \bibnamefont{and} \bibinfo{author}{\bibfnamefont{S.~L.}
  \bibnamefont{Cornish}}, \bibinfo{journal}{Phys. Rev. A}
  \textbf{\bibinfo{volume}{84}}, \bibinfo{pages}{011603}
  (\bibinfo{year}{2011}).

\bibitem[{\citenamefont{Liu et~al.}(2015)\citenamefont{Liu, Pattinson, Billam,
  Gardiner, Cornish, T.-M., Lin, Gou, Parker, and
  Proukakis}}]{liu_pattinson_15}
\bibinfo{author}{\bibfnamefont{I.-K.} \bibnamefont{Liu}},
  \bibinfo{author}{\bibfnamefont{R.~W.} \bibnamefont{Pattinson}},
  \bibinfo{author}{\bibfnamefont{T.~P.} \bibnamefont{Billam}},
  \bibinfo{author}{\bibfnamefont{S.~A.} \bibnamefont{Gardiner}},
  \bibinfo{author}{\bibfnamefont{S.~L.} \bibnamefont{Cornish}},
  \bibinfo{author}{\bibfnamefont{H.}~\bibnamefont{T.-M.}},
  \bibinfo{author}{\bibfnamefont{W.-W.} \bibnamefont{Lin}},
  \bibinfo{author}{\bibfnamefont{S.-C.} \bibnamefont{Gou}},
  \bibinfo{author}{\bibfnamefont{N.~G.} \bibnamefont{Parker}},
  \bibnamefont{and} \bibinfo{author}{\bibfnamefont{N.~P.}
  \bibnamefont{Proukakis}}, \bibinfo{journal}{arXiv:1408.0891v3}
  (\bibinfo{year}{2015}).

\bibitem[{\citenamefont{Sabbatini et~al.}(2011)\citenamefont{Sabbatini, Zurek,
  and Davis}}]{sabbatini_zurek_11}
\bibinfo{author}{\bibfnamefont{J.}~\bibnamefont{Sabbatini}},
  \bibinfo{author}{\bibfnamefont{W.~H.} \bibnamefont{Zurek}}, \bibnamefont{and}
  \bibinfo{author}{\bibfnamefont{M.~J.} \bibnamefont{Davis}},
  \bibinfo{journal}{Phys. Rev. Lett.} \textbf{\bibinfo{volume}{107}},
  \bibinfo{pages}{230402} (\bibinfo{year}{2011}).

\bibitem[{\citenamefont{{S}wis\l{}ocki
  et~al.}(2013)\citenamefont{{S}wis\l{}ocki, Witkowska, Dziarmaga, and
  Matuszewski}}]{swislocki_witkowska_13}
\bibinfo{author}{\bibfnamefont{T.}~\bibnamefont{{S}wis\l{}ocki}},
  \bibinfo{author}{\bibfnamefont{E.}~\bibnamefont{Witkowska}},
  \bibinfo{author}{\bibfnamefont{J.}~\bibnamefont{Dziarmaga}},
  \bibnamefont{and}
  \bibinfo{author}{\bibfnamefont{M.}~\bibnamefont{Matuszewski}},
  \bibinfo{journal}{Phys. Rev. Lett.} \textbf{\bibinfo{volume}{110}},
  \bibinfo{pages}{045303} (\bibinfo{year}{2013}).

\bibitem[{\citenamefont{Hofmann et~al.}(2014)\citenamefont{Hofmann, Natu, and
  Das~Sarma}}]{hofmann_natu_14}
\bibinfo{author}{\bibfnamefont{J.}~\bibnamefont{Hofmann}},
  \bibinfo{author}{\bibfnamefont{S.~S.} \bibnamefont{Natu}}, \bibnamefont{and}
  \bibinfo{author}{\bibfnamefont{S.}~\bibnamefont{Das~Sarma}},
  \bibinfo{journal}{Phys. Rev. Lett.} \textbf{\bibinfo{volume}{113}},
  \bibinfo{pages}{095702} (\bibinfo{year}{2014}).

\bibitem[{\citenamefont{Mathey et~al.}(2015)\citenamefont{Mathey, Gasenzer, and
  Pawlowski}}]{Mathey2014a.PhysRevA.92.023635}
\bibinfo{author}{\bibfnamefont{S.}~\bibnamefont{Mathey}},
  \bibinfo{author}{\bibfnamefont{T.}~\bibnamefont{Gasenzer}}, \bibnamefont{and}
  \bibinfo{author}{\bibfnamefont{J.~M.} \bibnamefont{Pawlowski}},
  \bibinfo{journal}{Phys. Rev. A} \textbf{\bibinfo{volume}{92}},
  \bibinfo{pages}{023635} (\bibinfo{year}{2015}).

\bibitem[{\citenamefont{{Pi\~neiro Orioli}
  et~al.}(2015)\citenamefont{{Pi\~neiro Orioli}, Boguslavski, and
  Berges}}]{Orioli:2015dxa}
\bibinfo{author}{\bibfnamefont{A.}~\bibnamefont{{Pi\~neiro Orioli}}},
  \bibinfo{author}{\bibfnamefont{K.}~\bibnamefont{Boguslavski}},
  \bibnamefont{and} \bibinfo{author}{\bibfnamefont{J.}~\bibnamefont{Berges}},
  \bibinfo{journal}{Phys. Rev. D} \textbf{\bibinfo{volume}{92}},
  \bibinfo{pages}{025041} (\bibinfo{year}{2015}).

\bibitem[{\citenamefont{Berges and Sexty}(2012)}]{Berges:2012us}
\bibinfo{author}{\bibfnamefont{J.}~\bibnamefont{Berges}} \bibnamefont{and}
  \bibinfo{author}{\bibfnamefont{D.}~\bibnamefont{Sexty}},
  \bibinfo{journal}{Phys. Rev. Lett.} \textbf{\bibinfo{volume}{108}},
  \bibinfo{pages}{161601} (\bibinfo{year}{2012}).

\bibitem[{\citenamefont{Nowak et~al.}(2011)\citenamefont{Nowak, Sexty, and
  Gasenzer}}]{Nowak:2010tm}
\bibinfo{author}{\bibfnamefont{B.}~\bibnamefont{Nowak}},
  \bibinfo{author}{\bibfnamefont{D.}~\bibnamefont{Sexty}}, \bibnamefont{and}
  \bibinfo{author}{\bibfnamefont{T.}~\bibnamefont{Gasenzer}},
  \bibinfo{journal}{Phys. Rev. B} \textbf{\bibinfo{volume}{84}},
  \bibinfo{pages}{020506(R)} (\bibinfo{year}{2011}).

\bibitem[{\citenamefont{Nowak et~al.}(2012)\citenamefont{Nowak, Schole, Sexty,
  and Gasenzer}}]{Nowak:2011sk}
\bibinfo{author}{\bibfnamefont{B.}~\bibnamefont{Nowak}},
  \bibinfo{author}{\bibfnamefont{J.}~\bibnamefont{Schole}},
  \bibinfo{author}{\bibfnamefont{D.}~\bibnamefont{Sexty}}, \bibnamefont{and}
  \bibinfo{author}{\bibfnamefont{T.}~\bibnamefont{Gasenzer}},
  \bibinfo{journal}{Phys. Rev. A} \textbf{\bibinfo{volume}{85}},
  \bibinfo{pages}{043627} (\bibinfo{year}{2012}).

\bibitem[{\citenamefont{Nowak et~al.}(2013)\citenamefont{Nowak, Erne, Karl,
  Schole, Sexty, and Gasenzer}}]{Nowak:2013juc}
\bibinfo{author}{\bibfnamefont{B.}~\bibnamefont{Nowak}},
  \bibinfo{author}{\bibfnamefont{S.}~\bibnamefont{Erne}},
  \bibinfo{author}{\bibfnamefont{M.}~\bibnamefont{Karl}},
  \bibinfo{author}{\bibfnamefont{J.}~\bibnamefont{Schole}},
  \bibinfo{author}{\bibfnamefont{D.}~\bibnamefont{Sexty}}, \bibnamefont{and}
  \bibinfo{author}{\bibfnamefont{T.}~\bibnamefont{Gasenzer}}, in
  \emph{\bibinfo{booktitle}{Proc. Int. School on Strongly Interacting Quantum
  Systems Out of Equilibrium, Les Houches, 2012 (to appear)}}
  (\bibinfo{year}{2013}), arXiv:1302.1448 [cond-mat.quant-gas].

\bibitem[{\citenamefont{Schole et~al.}(2012)\citenamefont{Schole, Nowak, and
  Gasenzer}}]{Schole:2012kt}
\bibinfo{author}{\bibfnamefont{J.}~\bibnamefont{Schole}},
  \bibinfo{author}{\bibfnamefont{B.}~\bibnamefont{Nowak}}, \bibnamefont{and}
  \bibinfo{author}{\bibfnamefont{T.}~\bibnamefont{Gasenzer}},
  \bibinfo{journal}{Phys. Rev. A} \textbf{\bibinfo{volume}{86}},
  \bibinfo{pages}{013624} (\bibinfo{year}{2012}).

\bibitem[{\citenamefont{Schmidt et~al.}(2012)\citenamefont{Schmidt, Erne,
  Nowak, Sexty, and Gasenzer}}]{Schmidt:2012kw}
\bibinfo{author}{\bibfnamefont{M.}~\bibnamefont{Schmidt}},
  \bibinfo{author}{\bibfnamefont{S.}~\bibnamefont{Erne}},
  \bibinfo{author}{\bibfnamefont{B.}~\bibnamefont{Nowak}},
  \bibinfo{author}{\bibfnamefont{D.}~\bibnamefont{Sexty}}, \bibnamefont{and}
  \bibinfo{author}{\bibfnamefont{T.}~\bibnamefont{Gasenzer}},
  \bibinfo{journal}{New J. Phys.} \textbf{\bibinfo{volume}{14}},
  \bibinfo{pages}{075005} (\bibinfo{year}{2012}).

\bibitem[{\citenamefont{{Karl} et~al.}(2013)\citenamefont{{Karl}, {Nowak}, and
  {Gasenzer}}}]{Karl:2013mn}
\bibinfo{author}{\bibfnamefont{M.}~\bibnamefont{{Karl}}},
  \bibinfo{author}{\bibfnamefont{B.}~\bibnamefont{{Nowak}}}, \bibnamefont{and}
  \bibinfo{author}{\bibfnamefont{T.}~\bibnamefont{{Gasenzer}}},
  \bibinfo{journal}{Sci. Rept.} \textbf{\bibinfo{volume}{3}},
  \bibinfo{eid}{2394} (\bibinfo{year}{2013}).

\bibitem[{\citenamefont{Karl et~al.}(2013)\citenamefont{Karl, Nowak, and
  Gasenzer}}]{Karl:2013kua}
\bibinfo{author}{\bibfnamefont{M.}~\bibnamefont{Karl}},
  \bibinfo{author}{\bibfnamefont{B.}~\bibnamefont{Nowak}}, \bibnamefont{and}
  \bibinfo{author}{\bibfnamefont{T.}~\bibnamefont{Gasenzer}},
  \bibinfo{journal}{Phys. Rev. A} \textbf{\bibinfo{volume}{88}},
  \bibinfo{pages}{063615} (\bibinfo{year}{2013}).

\bibitem[{\citenamefont{Gasenzer et~al.}(2014)\citenamefont{Gasenzer, McLerran,
  Pawlowski, and Sexty}}]{Gasenzer:2013era}
\bibinfo{author}{\bibfnamefont{T.}~\bibnamefont{Gasenzer}},
  \bibinfo{author}{\bibfnamefont{L.}~\bibnamefont{McLerran}},
  \bibinfo{author}{\bibfnamefont{J.~M.} \bibnamefont{Pawlowski}},
  \bibnamefont{and} \bibinfo{author}{\bibfnamefont{D.}~\bibnamefont{Sexty}},
  \bibinfo{journal}{Nucl. Phys. A} \textbf{\bibinfo{volume}{930}},
  \bibinfo{pages}{163} (\bibinfo{year}{2014}).

\bibitem[{\citenamefont{Damle et~al.}(1996)\citenamefont{Damle, Majumdar, and
  Sachdev}}]{damle_majumdar_96}
\bibinfo{author}{\bibfnamefont{K.}~\bibnamefont{Damle}},
  \bibinfo{author}{\bibfnamefont{S.~N.} \bibnamefont{Majumdar}},
  \bibnamefont{and} \bibinfo{author}{\bibfnamefont{S.}~\bibnamefont{Sachdev}},
  \bibinfo{journal}{Phys. Rev. A} \textbf{\bibinfo{volume}{54}},
  \bibinfo{pages}{5037} (\bibinfo{year}{1996}).

\bibitem[{\citenamefont{Connaughton et~al.}(2005)\citenamefont{Connaughton,
  Josserand, Picozzi, Pomeau, and Rica}}]{connaughton_josserand_05}
\bibinfo{author}{\bibfnamefont{C.}~\bibnamefont{Connaughton}},
  \bibinfo{author}{\bibfnamefont{C.}~\bibnamefont{Josserand}},
  \bibinfo{author}{\bibfnamefont{A.}~\bibnamefont{Picozzi}},
  \bibinfo{author}{\bibfnamefont{Y.}~\bibnamefont{Pomeau}}, \bibnamefont{and}
  \bibinfo{author}{\bibfnamefont{S.}~\bibnamefont{Rica}},
  \bibinfo{journal}{Phys. Rev. Lett.} \textbf{\bibinfo{volume}{95}},
  \bibinfo{pages}{263901} (\bibinfo{year}{2005}).

\bibitem[{\citenamefont{Nardin et~al.}(2009)\citenamefont{Nardin, Lagoudakis,
  Wouters, Richard, Baas, Andr\'e, Dang, Pietka, and
  Deveaud-Pl\'edran}}]{nardin_lagoudakis_09}
\bibinfo{author}{\bibfnamefont{G.}~\bibnamefont{Nardin}},
  \bibinfo{author}{\bibfnamefont{K.~G.} \bibnamefont{Lagoudakis}},
  \bibinfo{author}{\bibfnamefont{M.}~\bibnamefont{Wouters}},
  \bibinfo{author}{\bibfnamefont{M.}~\bibnamefont{Richard}},
  \bibinfo{author}{\bibfnamefont{A.}~\bibnamefont{Baas}},
  \bibinfo{author}{\bibfnamefont{R.}~\bibnamefont{Andr\'e}},
  \bibinfo{author}{\bibfnamefont{L.~S.} \bibnamefont{Dang}},
  \bibinfo{author}{\bibfnamefont{B.}~\bibnamefont{Pietka}}, \bibnamefont{and}
  \bibinfo{author}{\bibfnamefont{B.}~\bibnamefont{Deveaud-Pl\'edran}},
  \bibinfo{journal}{Phys. Rev. Lett.} \textbf{\bibinfo{volume}{103}},
  \bibinfo{pages}{256402} (\bibinfo{year}{2009}).

\bibitem[{\citenamefont{Belykh et~al.}(2013)\citenamefont{Belykh, Sibeldin,
  Kulakovskii, Glazov, Semina, Schneider, H\"ofling, Kamp, and
  Forchel}}]{belykh_sibeldin_13}
\bibinfo{author}{\bibfnamefont{V.~V.} \bibnamefont{Belykh}},
  \bibinfo{author}{\bibfnamefont{N.~N.} \bibnamefont{Sibeldin}},
  \bibinfo{author}{\bibfnamefont{V.~D.} \bibnamefont{Kulakovskii}},
  \bibinfo{author}{\bibfnamefont{M.~M.} \bibnamefont{Glazov}},
  \bibinfo{author}{\bibfnamefont{M.~A.} \bibnamefont{Semina}},
  \bibinfo{author}{\bibfnamefont{C.}~\bibnamefont{Schneider}},
  \bibinfo{author}{\bibfnamefont{S.}~\bibnamefont{H\"ofling}},
  \bibinfo{author}{\bibfnamefont{M.}~\bibnamefont{Kamp}}, \bibnamefont{and}
  \bibinfo{author}{\bibfnamefont{A.}~\bibnamefont{Forchel}},
  \bibinfo{journal}{Phys. Rev. Lett.} \textbf{\bibinfo{volume}{110}},
  \bibinfo{pages}{137402} (\bibinfo{year}{2013}).

\bibitem[{\citenamefont{Lagoudakis et~al.}(2011)\citenamefont{Lagoudakis,
  Manni, Pietka, Wouters, Liew, Savona, Kavokin, Andr\'e, and
  Deveaud-Pl\'edran}}]{lagoudakis_manni_11}
\bibinfo{author}{\bibfnamefont{K.~G.} \bibnamefont{Lagoudakis}},
  \bibinfo{author}{\bibfnamefont{F.}~\bibnamefont{Manni}},
  \bibinfo{author}{\bibfnamefont{B.}~\bibnamefont{Pietka}},
  \bibinfo{author}{\bibfnamefont{M.}~\bibnamefont{Wouters}},
  \bibinfo{author}{\bibfnamefont{T.~C.~H.} \bibnamefont{Liew}},
  \bibinfo{author}{\bibfnamefont{V.}~\bibnamefont{Savona}},
  \bibinfo{author}{\bibfnamefont{A.~V.} \bibnamefont{Kavokin}},
  \bibinfo{author}{\bibfnamefont{R.}~\bibnamefont{Andr\'e}}, \bibnamefont{and}
  \bibinfo{author}{\bibfnamefont{B.}~\bibnamefont{Deveaud-Pl\'edran}},
  \bibinfo{journal}{Phys. Rev. Lett.} \textbf{\bibinfo{volume}{106}},
  \bibinfo{pages}{115301} (\bibinfo{year}{2011}).

\bibitem[{\citenamefont{Klaers et~al.}(2010)\citenamefont{Klaers, Schmitt,
  Vewinger, and Weitz}}]{klaers_schmitt_10}
\bibinfo{author}{\bibfnamefont{J.}~\bibnamefont{Klaers}},
  \bibinfo{author}{\bibfnamefont{J.}~\bibnamefont{Schmitt}},
  \bibinfo{author}{\bibfnamefont{F.}~\bibnamefont{Vewinger}}, \bibnamefont{and}
  \bibinfo{author}{\bibfnamefont{M.}~\bibnamefont{Weitz}},
  \bibinfo{journal}{Nature} \textbf{\bibinfo{volume}{468}},
  \bibinfo{pages}{545} (\bibinfo{year}{2010}).

\bibitem[{\citenamefont{Demokritov et~al.}(2006)\citenamefont{Demokritov, V.~E.
  Demidov~an, Melkov, Serga, Hillebrands, and Slavin}}]{demokritov_demidov_06}
\bibinfo{author}{\bibfnamefont{S.~O.} \bibnamefont{Demokritov}},
  \bibinfo{author}{\bibfnamefont{O.~D.} \bibnamefont{V.~E. Demidov~an}},
  \bibinfo{author}{\bibfnamefont{G.~A.} \bibnamefont{Melkov}},
  \bibinfo{author}{\bibfnamefont{A.~A.} \bibnamefont{Serga}},
  \bibinfo{author}{\bibfnamefont{B.}~\bibnamefont{Hillebrands}},
  \bibnamefont{and} \bibinfo{author}{\bibfnamefont{A.~N.}
  \bibnamefont{Slavin}}, \bibinfo{journal}{Nature}
  \textbf{\bibinfo{volume}{443}}, \bibinfo{pages}{430} (\bibinfo{year}{2006}).

\end{thebibliography}
\bibliographystyle{apsrev-MJD}

\end{document}